%% file: f_theory-standard_model-v5.tex
\documentclass[11pt,a4paper]{article}                           

\usepackage[bookmarksopen, bookmarksnumbered, bookmarksopenlevel=2]{hyperref}
\usepackage{tikz}
\usepackage{tikz-3dplot}
\usetikzlibrary{calc}
\usetikzlibrary{decorations} %
\usepackage[UKenglish]{babel}
\usepackage{amsmath}
\usepackage{amssymb}
\usepackage{graphicx}
\usepackage{wrapfig}
\usepackage{hhline}
\usepackage[bf]{caption}
\usepackage{cite}
\usepackage[vcentermath]{youngtab}
\usepackage{geometry}
\usepackage{multirow}
\usepackage{longtable,array}

\makeatletter
\def\nobreakhline{%
  \noalign{\ifnum0=`}\fi
    \penalty\@M
    \futurelet\@let@token\LT@@nobreakhline}
\def\LT@@nobreakhline{%
  \ifx\@let@token\hline
    \global\let\@gtempa\@gobble
    \gdef\LT@sep{\penalty\@M\vskip\doublerulesep}% <-- change here
  \else
    \global\let\@gtempa\@empty
    \gdef\LT@sep{\penalty\@M\vskip-\arrayrulewidth}% <-- change here
  \fi
  \ifnum0=`{\fi}%
  \multispan\LT@cols
     \unskip\leaders\hrule\@height\arrayrulewidth\hfill\cr
  \noalign{\LT@sep}%
  \multispan\LT@cols
     \unskip\leaders\hrule\@height\arrayrulewidth\hfill\cr
  \noalign{\penalty\@M}%
  \@gtempa}
\makeatother
\hypersetup{
    pdftitle={Towards the Standard Model in F-theory},
    pdfauthor={Ling Lin, Timo Weigand, \ldots},
    pdfsubject={F-theory compactifications}
}
\numberwithin{equation}{section}
\numberwithin{table}{section}

\newcommand{\bea}{\begin{eqnarray}}
\newcommand{\eea}{\end{eqnarray}}
\newcommand{\T}[1]{\textmd{#1}}

\newcommand{\su}{\mathrm{u}}

\newcommand{\sv}{\mathrm{v}}
\newcommand{\sw}{\mathrm{w}}
\newcommand{\executeiffilenewer}[3]{%
 \ifnum\pdfstrcmp{\pdffilemoddate{#1}}%
 {\pdffilemoddate{#2}}>0%
 {\immediate\write18{#3}}\fi%
}

\usepackage{todonotes}
\begin{document}

\baselineskip=14pt
\parskip 5pt plus 1pt 

\interfootnotelinepenalty=10000

\vspace*{-1.5cm}
\begin{flushright}    % Publication numbers
  {\small

  }
\end{flushright}

\vspace*{2cm}
\begin{center}        % Main title
  {\LARGE Towards the Standard Model in F-theory}
\end{center}

\vspace*{0.75cm}
\begin{center}        % Authors
Ling Lin and Timo Weigand
\end{center}

\vspace*{0.15cm}
\begin{center}        % Institutes
\emph{Institut f\"ur Theoretische Physik, Ruprecht-Karls-Universit\"at, \\
             Philosophenweg 19, 69120 \\
             Heidelberg, Germany}
\end{center}

\vspace*{2cm}

%%%%%%%%%%%%%%%%%%%%%%%%%%%%%%%%%%%%%%%%%%%%%%%
%%%%%%%%%%%%%%%%%%%%%%%%%%%%%%%%%%%%%%%%%%%%%%%
%%%%%%%%%%%%%%%%%%%%%%%%%%%%%%%%%%%%%%%%%%%%%%%
%%%%%%%%%%%%%%%%%%%%%%%%%%%%%%%%%%%%%%%%%%%%%%%
%%%%%%%%%%%%%%%%%%%%%%%%%%%%%%%%%%%%%%%%%%%%%%%
%%%%%%%%%%%%%%%%%%%%%%%%%%%%%%%%%%%%%%%%%%%%%%%
%%%%%%%%%%%%%%%%%%%%%%%%%%%%%%%%%%%%%%%%%%%%%%%
%%%%%%%%%%%%%%%%%%%%%%%%%%%%%%%%%%%%%%%%%%%%%%%

\begin{abstract}

This article explores possible embeddings of the Standard Model gauge group and its matter representations into F-theory.
To this end we construct elliptic fibrations with gauge group $SU(3) \times SU(2) \times U(1) \times U(1)$ as suitable restrictions of a ${\rm Bl}_2{\mathbb P}^2$-fibration with rank-two Mordell-Weil group. We analyse the five inequivalent toric enhancements to gauge group $SU(3) \times SU(2)$ along two independent divisors $W_3$ and $W_2$ in the base.
For each of the resulting smooth fibrations, the representation spectrum generically consists of a bifundamental $({\bf 3, \bf 2})$, three types of $({\bf 1,\bf 2})$ representations and five types of $({\bf 3},{\bf 1})$ representations (plus conjugates), in addition to charged singlet states. The precise spectrum of zero-modes in these representations depends on the 3-form background.
We analyse the geometrically realised Yukawa couplings among all these states and find complete agreement with field theoretic expectations based on their $U(1)$ charges.
We classify possible identifications of the found representations with the Standard Model field content extended by right-handed neutrinos and extra singlets. The linear combination of the two abelian gauge group factors orthogonal to hypercharge acts as a selection rule which, depending on the specific model, can forbid dangerous dimension-four and -five proton decay operators.

\end{abstract}
 
\vspace*{\fill}

\thispagestyle{empty}
\clearpage
\setcounter{page}{1}

%%%%%%%%%%%%%%%%%%%%%%%%%%%%%%%%%%%%%%%%%%%%%%%
%%%%%%%%%%%%%%%%%%%%%%%%%%%%%%%%%%%%%%%%%%%%%%%
%%%%%%%%%%%                 %%%%%%%%%%%%%%%%%%%
%%%%%%%%%%%  DOCUMENT BODY  %%%%%%%%%%%%%%%%%%%
%%%%%%%%%%%                 %%%%%%%%%%%%%%%%%%%
%%%%%%%%%%%%%%%%%%%%%%%%%%%%%%%%%%%%%%%%%%%%%%%
%%%%%%%%%%%%%%%%%%%%%%%%%%%%%%%%%%%%%%%%%%%%%%%
%%%%%%%%%%%%%%%%%%%%%%%%%%%%%%%%%%%%%%%%%%%%%%%

\newpage

\tableofcontents
%%%%%%%%%%%%%%%%%%%%%%%%%%%%%%%%%%%%%%%%%%%%%%%%%%%%%%%%%%%%%%%%%%%%%%%%%%%%%%%%%%%%%%%%%%%%%%%%%%%%%%%%%%%%%%%%%%%%%%%%%%%%%%%%%%%%%%%%%%%%%%%

\input{section1_introduction_v3.tex}

\input{section2_F-theory_with_two_U1s_v3.tex}

\input{section3_models_with_SU2_v3.tex}

\input{section4_models_with_SU3_v3.tex}

\input{section5_toric_SU3xSU2_v3.tex}

\input{section6_F-theory_MSSM_v3.tex}

\input{section7_F-theory_Fluxes_v3.tex}

\input{section8_Conclusions_v3.tex}

\input{section_appendix_v3.tex}

\newpage
\bibliography{papers}  
\bibliographystyle{custom1}

\end{document}

%% file: section1_introduction_v3.tex
\section{Introduction}
%%%%%%%%%%%%%%%%%%%%%%%%%%%%%%%%%%%%%%%%%%%%%%%%%%%%%%%%%%%%%%%%%%%%%%%%%%%%%%%%%%%%%%%%%%%%%%%%%%%%%%%%%%%%%%%%%%%%%%%%%%%%%%%%%%%%%%%%%%%%%%%

If string theory is indeed the correct ultra-violet completion of Quantum Field Theory in presence of gravity, it must be possible to derive the observed Standard Model of Particle Physics as one of the consistent four-dimensional vacuum solutions to the string equations of motion.
A very efficient approach to embedding the Standard Model into string theory is  via Grand Unified Theories (GUTs) of particle physics. 
In this context, exceptional symmetry based on the Lie groups $E_6$, $E_7$ or $E_8$ is known to play a distinguished role. In fact, the success of particle physics model building in heterotic $E_8 \times E_8$ string theory, recent examples being \cite{Bouchard:2005ag,Braun:2005nv,Lebedev:2006kn,Anderson:2011ns} and references therein, 
is largely due to the ease with which exceptional symmetry arises here. %For recent constructions of heterotic Standard Model building see e.g. \cite{Bouchard:2005ag,Braun:2005nv,Lebedev:2006kn,Anderson:2011ns} and references therein.

%More recently, this GUT-based approach has been extended to studying string vacua with branes in the non-perturbative regime, most notably in F-theory

In string vacua with branes, exceptional symmetry requires non-perturbative ingredients such as $[p,q]$-strings in strongly coupled Type IIB theory.
In order to extend the GUT paradigm to string theories with branes one therefore has to leave the perturbative regime. Indeed 
 considerable effort  has been invested recently into exploiting such non-perturbative effects in the context of GUT phenomenology  in F-theory as initiated in  \cite{Donagi:2008ca,oai:arXiv.org:0802.3391,Beasley:2008kw,Donagi:2008kj} (see e.g the reviews \cite{Heckman:2010bq,Weigand:2010wm,Maharana:2012tu} for more references).

On the other hand, in string compactifications with branes a rather different route towards the Standard Model suggests itself which directly yields the observed gauge group $SU(3) \times SU(2) \times U(1)_Y$ without any detour via a GUT group such as $SU(5)$ or $SO(10)$. This approach has traditionally been pursued in a perturbative context with intersecting D-branes in Type IIA or Type IIB orientifold theories \cite{Blumenhagen:2000wh,Aldazabal:2000cn,
Cvetic:2001nr}, where stacks of multiple D-branes in general position give rise to gauge group $U(N)$ and thus the building block of the Standard Model gauge group. 
The apparent unification of the gauge couplings, which is particularly well motivated in models with low-energy supersymmetry, translates into a relation on the volume of the D-branes on which the Standard Model is realised \cite{Blumenhagen:2003jy}. 
The construction of Standard-like models via such intersecting branes in Type II orientifolds has been pursued in a rich literature reviewed for instance in \cite{Lust:2004ks,Blumenhagen:2006ci,Marchesano:2007de,Cvetic:2011vz,Ibanez:2012zz}.\footnote{For example, most recently 
Standard-like vacua have been constructed in toroidal orientifolds in \cite{Honecker:2012qr,Honecker:2012jd} and in RCFT orientifolds in \cite{Dijkstra:2004cc,Anastasopoulos:2006da}. See \cite{Dolan:2011qu,Cicoli:2012vw} and references therein for recent advances in realistic model building with branes at singularities.}

A distinctive organising principle for the couplings between the charged particles of such brane vacua is provided by abelian selection rules.
In perturbative Type II orientifolds, as pointed out already, the gauge group on a stack of $N$ multiple branes is $U(N) = SU(N) \times U(1)/{\mathbb Z_N}$. The diagonal abelian gauge group factor typically receives 
a St\"uckelberg mass induced by the coupling to the closed string axions \cite{Ibanez:1998qp}, but remains as a perturbative selection rule (`massive $U(1)$') constraining the structure of perturbatively allowed couplings.
These selection rules can only be broken non-perturbatively by D-brane instantons \cite{Blumenhagen:2006xt,Ibanez:2006da,Haack:2006cy,Florea:2006si,Blumenhagen:2009qh}, oftentimes to an exact discrete $\mathbb Z_k$ symmetry \cite{BerasaluceGonzalez:2011wy,BerasaluceGonzalez:2012vb,Ibanez:2012wg,Anastasopoulos:2012zu,Honecker:2013hda}. 
Such selection rules can be a curse or a blessing:
As one of their advantages, they can forbid undesirable, dangerous couplings, e.g. interactions that would induce unacceptably rapid proton decay.
In this sense, for instance the origin of baryon or lepton number in the Standard Model can be traced to perturbatively exact symmetries from a string theory perspective. 
Similarly, extra $U(1)$ symmetries may conspire to perturbatively forbid hierarchically suppressed couplings, which in turn are generated only non-perturbatively, thereby explaining their smallness. A systematic analysis of the phenomenology of such effects in perturbative MSSM quivers can be found in \cite{Ibanez:2008my,Cvetic:2009yh,Cvetic:2009ez}.

On the other hand, extra $U(1)$ symmetries can also forbid desirable couplings. Consider a realisation of the Standard Model via
brane stacks of the form $U(3)_a \times U(2)_b \times U(1)_c$ plus possible extra $U(1)$ branes.
The left-handed quark $Q$ resides in representation $({\bf 3},{\bf 2})_{1_a, -1_b}$ (with subscripts denoting the $U(1)$ charges) and the up-type Higgs $H_u$, say, in representation $({\bf 1},{\bf 2})_{1_b,-1_c}$. If the right-handed up-quarks $u_R^c$ are realised as the antisymmetric representation of $U(3)$, i.e. $u_R^c = ({\bf \overline 3}, {\bf 1})_{-2_a}$,
the $U(1)$ charges prohibit a perturbative Yukawa coupling $Q \, u_R^c \, H_u$ even though this coupling is allowed by the Standard Model gauge symmetry itself.

This example is similar to the well-known \cite{Blumenhagen:2001te,Blumenhagen:2007zk} absence of a perturbative coupling ${\bf 10} \, {\bf 10} \,  {\bf 5} $ in attempts to realise an $SU(5)$ GUT with perturbative intersecting branes via $U(5)_a \times U(1)_b$. 
% Here the ${\bf 5}_{1_a, - 1_b}$ contains the Standard Model up-Higgs and ${\bf 10}_{2_a}$ contains the fields $Q$, $u_R^c$ and $e_R^c$.
This problem is beautifully solved in F-theory models, where such a coupling arises at a non-perturbative point of $E_6$ enhancement \cite{Donagi:2008ca,oai:arXiv.org:0802.3391,Beasley:2008kw,Donagi:2008kj}. 
Such enhancement points signal that the brane configuration is not smoothly connected in moduli space to a well-defined Type IIB limit \cite{Donagi:2009ra,Krause:2012yh} and one should think of the matter states as arising from multi-pronged $[p,q]$-strings, which are perturbatively absent.  
Equivalently, the underlying reason for existence of a ${\bf 10} \, {\bf 10} \,  {\bf 5}$ coupling in F-theory is that the structure of $U(1)$ gauge symmetries in F-theory can be much more general than in the subclass of perturbative Type IIB models. 
Indeed, in a more general F-theory model without a smooth Type IIB limit, $SU(N)$ gauge symmetries are not necessarily accompanied by extra diagonal $U(1)$ factors.\footnote{See, however, \cite{Grimm:2010ez,Grimm:2011tb,Braun:2014nva,Anderson:2014yva} for the description of such overall massive $U(1)$s in F-theory.} The structure of abelian gauge symmetries in F-theory is a particularly rich topic with beautiful connections to algebraic geometry and has been investigated in great detail in the very recent literature \cite{Grimm:2010ez,Braun:2011zm,Krause:2011xj,Grimm:2011fx,oai:arXiv.org:1202.3138,Morrison:2012ei,oai:arXiv.org:1210.6034,Mayrhofer:2012zy,Braun:2013yti,Borchmann:2013jwa,Cvetic:2013nia,Braun:2013nqa,Cvetic:2013uta,Borchmann:2013hta,Cvetic:2013jta,Cvetic:2013qsa,Morrison:2014era,Martini:2014iza,Bizet:2014uua}, motivated in part by the need for extra $U(1)$ selection rules in the context of F-theory GUT model building  \cite{Marsano:2009wr,Hayashi:2010zp,Dolan:2011iu,Dolan:2011aq,Marsano:2011nn,
Maharana:2012tu,Krippendorf:2014xba}.

Given the crucial role of $U(1)$ selection rules for Standard Model building on the one hand and the striking differences between the structure of $U(1)$ selection rules in perturbative and non-perturbative models on the other hand, 
it is an obvious question to what extent direct, non-GUT realisations of the Standard Model in F-theory differ from their perturbative, well-studied counter-parts. 
Motivated by our interest in extending our knowledge of realistic vacua to new, unexplored regions of the landscape,
we therefore investigate in this work the possibilities of directly constructing string vacua with Standard Model gauge group and matter representations in F-theory. 
For phenomenological reasons we focus on string vacua which in addition to the Standard Model gauge group allow for one abelian group factor. This extra $U(1)$ will eventually acquire a St\"uckelberg mass term upon switching on gauge flux (as required to generate a chiral spectrum) and remain as a perturbative selection rule, which will help avoid in principle dangerous dimension-four (and higher) proton decay operators.

Our approach is therefore to construct elliptic fibrations for F-theory compactifications with gauge group $SU(3) \times SU(2) \times U(1)_1 \times U(1)_2$ such that one linear combination of $U(1)_1$ and $U(1)_2$ will play the role of hypercharge $U(1)_Y$ and stay exactly massless even in presence of gauge fluxes. 
To this end we start with an elliptic fibration with gauge group $U(1)_1 \times U(1)_2$ and further constrain the complex structure moduli such as to create the non-abelian gauge group factors $SU(3) \times SU(2)$.
Elliptic fibrations with gauge group $U(1)_1 \times U(1)_2$ have been analysed in detail in \cite{Borchmann:2013jwa,Cvetic:2013nia,Cvetic:2013uta,Borchmann:2013hta,Cvetic:2013jta}. In section \ref{sec_U1U1} we summarise their main properties with special emphasis on the structure of Yukawa couplings among the charged singlet states. 
We then analyse the subclass of all possible gauge group enhancements to $SU(3) \times SU(2) \times U(1)_1 \times U(1)_2$ which can be achieved torically. This amounts to studying the $SU(2)$ and $SU(3)$ tops  \cite{Candelas:1996su,Candelas:1997eh} over the $U(1)_1 \times U(1)_2$-fibrations, which generate the corresponding singularities in the fibre over two independent divisors $W_2$ and $W_3$.
The construction of the Standard Model gauge group in F-theory has previously been approached in \cite{Choi:2013hua,Choi:2010su,Choi:2010nf} via a geometric deformation of an $SU(5)$ singularity along a single divisor such that the $SU(3)$ and $SU(2)$ singularities arise over homologous divisors. This is different to our approach, where both divisors are generically unrelated.

For our $U(1)_1 \times U(1)_2$ fibration, there are three tops for $SU(2)$ and $SU(3)$ each \cite{Bouchard:2003bu}. In sections \ref{sec:SU(2)-tops} and \ref{sec:SU(3)-tops} we compute in turn the structure of matter curves and the geometrically realised Yukawa couplings in models with an extra $SU(2)$ and $SU(3)$ gauge symmetry, respectively. We exemplify our computations for one of the two possible tops and relegate the details of the remaining analysis to appendices \ref{SU2DetailsApp} and \ref{app-SU3}. 
In section \ref{sec_3211} we combine these results into a single elliptic fibration with gauge group $SU(3) \times SU(2) \times U(1)_1 \times U(1)_2$. Of the a priori $3 \times 3$ resulting types of fibrations, only five turn out to be inequivalent. In each of the five inequivalent fibrations,
the structure of matter fields charged only under $SU(3)$ or $SU(2)$ (and/or the abelian gauge groups) carries over, but a new field arises at the intersection of the two non-abelian stacks. In generic situations, to which we restrict ourselves, this field transforms in the $({\bf 3}, {\bf 2})$-representation and  plays the role of the left-handed quarks in the Standard Model. We analyse all possible Yukawa couplings involving this state. 
Interestingly, one of the couplings turns out to correspond to a non-standard Kodaira fibre.  
Our approach is to analyse the elliptic fibre in full generality. For a sufficiently generic base space ${\cal B}$, these fibrations define a smooth elliptically fibred Calabi-Yau fourfold suitable for F-theory compactification.
% In one of the  nine possible combinations of tops  enhancement beyond Kodaira fibre type arises in one point, which must be absent in order for the elliptic fibration to describe an F-theory compactification, at least in the usual sense as understood as of this writing. 

In section \ref{sec_SM}, we match the various matter representations of the five inequivalent types of fibrations with the Standard Model fields, working in the context of the MSSM with extra singlets and an a priori unspecified supersymmetry breaking scale.
 Since the fibrations under consideration give rise to five different types of $({\bf 3},{\bf 1})$-fields and three different classes of $({\bf 1},{\bf 2})$-fields, each localised on a different matter curve and with different $U(1)$ charges, a plethora of possible identifications exists, which we list in appendix \ref{appsec:huge_table}. 
In principle, different generations of matter can be distributed over different curves. This way some of the generations may enjoy a perturbative Yukawa coupling, while others remain perturbatively massless. 
An important distinctive property of the so-obtained MSSM candidate setups is the spectrum of perturbatively allowed dimension-four and -five couplings. We list all possible such couplings.
For a low supersymmetry breaking scale, certain combinations of these couplings lead to unacceptable proton-decay, while in scenarios with  intermediate or high-scale supersymmetry breaking the constraints are more relaxed.
With an eye on the possibility of intermediate scale supersymmetry breaking, we do not exclude any models based on such dimension-four and -five couplings. A more detailed phenomenological assessment will appear in future work. 
 
The philosophy behind our classification of `toric Standard Models' is that gauge fluxes ensure the correct spectrum of MSSM matter. In particular, the remaining fundamental and singlet fields which are exotic states from an MSSM perspective are assumed to be absent at the massless level by virtue of a suitable choice of fluxes.
In section \ref{sec_fluxes} we summarise the constraints on these fluxes, especially from the requirement that hypercharge remain massless, leaving a more systematic treatment of gauge fluxes for the future. Such an analysis will be required to determine which of the Standard Model fibrations can also encompass the physically required number of zero-modes for the various fields, in particular in which cases no chiral exotics are forced upon us.
We conclude in section \ref{sec_Conclusions} with  an outlook and open questions.

%% file: section2_F-theory_with_two_U1s_v3.tex
\section[F-theory with \texorpdfstring{\boldmath $U(1) \times U(1)$}{U(1) x U(1)} Gauge Group]{F-theory with  \texorpdfstring{\boldmath $U(1) \times U(1)$}{U(1) x U(1)} Gauge Group} \label{sec_U1U1}
%%%%%%%%%%%%%%%%%%%%%%%%%%%%%%%%%%%%%%%%%%%%%%%%%%%%%%%%%%%%%%%%%%%%%%%%%%%%%%%%%

We are interested in engineering F-theory vacua with Standard Model gauge group $SU(3) \times SU(2) \times U(1)_Y$ and one additional abelian gauge group factor. 
Our starting point are therefore elliptic fibrations which allow for two abelian gauge groups $U(1)_1$ and $U(1)_2$; a further restriction of the complex structure of such fibrations will then induce the non-abelian factors $SU(3) \times SU(2)$. One linear combination of $U(1)_1$ and $U(1)_2$ will correspond to $U(1)_Y$, while the remaining combination must 
become massive by a flux-induced St\"uckelberg mechanism and act as an extra selection rule on the couplings of the model.

F-theory compactifications with two abelian gauge groups are based on elliptic fibrations with Mordell-Weil group of rank two. Such elliptic fibrations allow for a description as the vanishing locus of the hypersurface equation \cite{Borchmann:2013jwa,Cvetic:2013nia,Cvetic:2013uta,Borchmann:2013hta,Cvetic:2013jta}
\begin{align}
  \label{eq:hypersurface-equation}
  \begin{split}
    P_T =  &\sv \, \sw (c_1 \, \sw \, s_1 + c_2 \, \sv \, s_0) + \su \, (b_0 \, \sv^2 \, s_0^2 + b_1 \, \sv \, \sw \, s_0 \, s_1 + b_2 \, \sw^2 \, s_1^2) + \\
    &\su^2 (d_0 \, \sv \, s_0^2 \, s_1 + d_1 \, \sw \, s_0 \, s_1^2 + d_2 \, \su \, s_0^2 \, s_1^2)
  \end{split}
\end{align}
inside a $\text{Bl}_2 \mathbb{P}^2$-fibration. Such types of fibration had previously been considered also in \cite{Klemm:1996hh}. 
The ambient space $\text{Bl}_2 \mathbb{P}^2$ of the elliptic fibre is a toric space which has the toric diagram represented by polygon 5 in the classification \cite{Bouchard:2003bu} (see also figure \ref{fig:base_polygon} in the appendix). 
Here and in the sequel we will stick to the notation of \cite{Borchmann:2013jwa,Borchmann:2013hta} and denote the coordinates of $\text{Bl}_2 \mathbb{P}^2$ by $\mathrm{u},\mathrm{v},\mathrm{w},s_0,s_1$, where $s_0$ and $s_1$ correspond to two blow-up $\mathbb{P}^1$s inside $\mathbb P^2$ with homogeneous coordinates $[\su : \sw : \sv]$. The Stanley-Reisner ideal is generated by $\{ \su\,\sv , \su\,\sw , s_0\,\sw, s_1\,\sv, s_0 \, s_1 \}$ and the divisor classes associated with the fibre ambient space coordinates are given as follows:
\begin{align}\label{tab:divisor-classes-small}
 \begin{array}{c|ccccc}
    \hphantom{U} & \su & \sv & \sw & s_0 & s_1 \\ \hline
    {U} & 1 & 1 & 1 & \cdot & \cdot \\
    S_0 & \cdot & \cdot & 1 & 1& \cdot  \\
    S_1 & \cdot & 1 & \cdot & \cdot & 1
  \end{array}
\end{align}
The quantities $b_i, \, c_j, \, d_k$ transform as sections of certain line bundles over the base ${\cal B}$ of the fibration, whose class is determined by the requirement that the fibration is Calabi-Yau. These classes are collected in table \ref{coeff} (taken from \cite{Borchmann:2013hta}; see also \cite{Cvetic:2013nia,Cvetic:2013uta,Cvetic:2013jta}).
\begin{table}[t]
  \centering
  \begin{tabular}{c|c|c|c|c|c|c|c}% {c@{$\mspace{6mu}$}cc@{$\mspace{4mu}$}c@{$\mspace{4mu}$}c@{$\mspace{5mu}$}c@{$\mspace{5mu}$}c@{$\mspace{5mu}$}c}
 $ b_{0}$  & $  b_{1} $ & $b_{2}$ & $  c_{1}  $ & $  c_{2}  $&$d_{0}$&$d_{1}$&$d_{2}$\\
\hline
 $\alpha-\beta+ \bar {\cal K}$&$\bar {\cal K}$&$-\alpha +\beta+\bar {\cal K}$&$-\alpha+\bar {\cal K}$&$-\beta + \bar {\cal K}$&$\alpha + \bar {\cal K}$&$\beta+\bar {\cal K}$&$\alpha+\beta+\bar {\cal K}$\\
\end{tabular}
\caption{Classes of the sections appearing in (\ref{eq:hypersurface-equation}) for $\alpha$ and $\beta$ pullback of `arbitrary' classes of ${\cal B}$ and $\overline{\mathcal{K}} = \pi^{-1} \overline{\mathcal{K}}_{\cal B}$.}\label{coeff}
\end{table}
For suitable 3-dimensional base spaces ${\cal B}$ the hypersurface (\ref{eq:hypersurface-equation}) then describes a smooth elliptically fibred Calabi-Yau 4-fold
\bea
\pi: Y_4 \rightarrow {\cal B}.
\eea
It exhibits three independent rational sections 
\bea
U = \{u\}, \qquad  S_0=\{s_0\}, \qquad S_1= \{s_1\}.
\eea
Throughout the article we use, unless stated otherwise, $\{ \ldots \}$ as a short-hand notation for $\{ \ldots = 0 \}$.
One of these sections, e.g. $S_0$, can be interpreted as the zero-section.\footnote{The fact that all three independent sections are rational (as opposed to holomorphic) is an artefact of the representation of the fibration as a hypersurface. Indeed, the fibration is birationally equivalent to a complete intersection which does exhibit a
holomorphic zero-section \cite{Borchmann:2013hta}. The pre-image of this section under the birational map can be identified with the zero section \cite{Borchmann:2013jwa,Borchmann:2013hta}. Alternatively, one can define an F-theory compactification with a rational zero-section as in \cite{Cvetic:2013nia,Grimm:2013oga,Cvetic:2013uta}.}  
The image of the remaining two independent sections under the Shioda map \cite{Shioda:1989,Wazir:2001,Park:2011ji}  then identifies the generators of two independent $U(1)$ gauge groups  as \cite{Borchmann:2013jwa,Cvetic:2013nia,Cvetic:2013uta,Borchmann:2013hta}
\begin{align}
  \label{eq:U(1)-generators}
  \begin{split}
    \omega_1 &= S_1 - S_0 - \overline{\mathcal{K}}, \\
    \omega_2 &= U - S_0 - \overline{\mathcal{K}} - [ c_1 ],
  \end{split}
\end{align}
where $\overline{\mathcal{K}} = \pi^{-1} \overline{\mathcal{K}}_{\cal B}$ is the pre-image of the anti-canonical bundle of the base ${\cal B}$. Here and in the sequel our notation does not distinguish between a divisor (class) and its dual 2-form.

In our analysis of the matter representations of F-theory compactified on $Y_4$ we will  also need the form of the singular Weierstrass model which is birationally equivalent to the blow-down of the hypersurface (\ref{eq:hypersurface-equation}) (i.e. the hypersurface inside the $\mathbb{P}^2$-fibration over ${\cal B}$ achieved by setting $s_0 = s_1 =1$ in (\ref{eq:hypersurface-equation})).
The explicit map of this blow-down to Weierstrass form
\bea \label{Weier}
y^2 = x^3 + f \, x \, z^4 + g \, z^6 
\eea
has been worked out in the physics literature in \cite{Borchmann:2013jwa,Cvetic:2013nia,Cvetic:2013uta,Borchmann:2013hta}.
In our subsequent analysis we will make use of the expression for the Weierstrass sections $f$ and $g$ in terms of the defining sections $b_i, c_j, d_k$ appearing in (\ref{eq:hypersurface-equation}) as given after equation (2.38) and (2.39) in \cite{Borchmann:2013jwa}, which we recall here for completeness,

\begin{equation} \label{sec-map1}
 f=-\tfrac13\T d^2 + \T c\, \T e\qquad\textmd{and}\qquad g=-  f\,\left(\tfrac13\T d\right) -\left(\tfrac{1}{3}\T d\right)^3  + \T c^2\, \T k,
\end{equation}
where
\begin{equation}
 \begin{split} \label{sec-map2}
  \T d &=b_1^2 + 8\,b_0\,b_2 - 4\,c_1\,d_0 - 4\,c_2\,d_1,\\
  \T c &=-\frac{4}{c_1}(b_0\,b_2^2 - b_2\,c_1\,d_0 + c_1^2\,d_2),\\
  \T e &=\frac{2  c_1 \left( b_0 \left( b_1  c_1  d_1-b_1^2 b_2+2  b_2  c_1  d_0+2  b_2  c_2
    d_1-2  c_1^2  d_2\right)\right)}{ b_0  b_2^2+ c_1 ( c_1  d_2- b_2  d_0)}+\\&\qquad\qquad+\frac{2  c_1 \left(-2  b_0^2  b_2^2+ c_2 ( b_1  b_2  d_0+ b_1
    c_1  d_2-2  b_2  c_2  d_2-2  c_1  d_0
    d_1)\right)}{ b_0  b_2^2+ c_1 ( c_1  d_2- b_2  d_0)},\\
  \T k &=\frac{ c_1^2 ( b_0  b_1  b_2- b_0  c_1  d_1- b_2  c_2
    d_0+ c_1  c_2  d_2)^2}{\left( b_0  b_2^2+ c_1 ( c_1
    d_2- b_2 d_0)\right)^2}.
 \end{split}
\end{equation}

\subsection{Charged Singlets and Their Yukawa Couplings}\label{sec:singlets-prime-ideals}

The fibration gives rise to six types of charged singlet states  $ \mathbf{1}^{(i)}$  localised on curves on ${\cal B}$, which have already been analysed in \cite{Borchmann:2013jwa,Cvetic:2013nia,Cvetic:2013uta,Borchmann:2013hta,Cvetic:2013jta}. Here we continue with the analysis of \cite{Borchmann:2013jwa,Borchmann:2013hta} (alternatively, see \cite{Cvetic:2013nia,Cvetic:2013uta}), which derives the curves as loci in the base over which the fibre of the blown-down version of (\ref{eq:hypersurface-equation}) exhibits a conifold singularity. These loci are given as the union of the set of solutions to each of the following three pairs of equations,
\begin{equation}\label{eq:singlet-locus-1} 
\begin{split}
0 & = d_0\, c_2^2 + b_0^2 \,c_1 - b_0\, b_1\, c_2\,,\\
0 & = d_1\, b_0\, c_2 - b_0^2\, b_2 - c_2^2\, d_2 \, ,
\end{split}
\end{equation}
and
\begin{equation}\label{eq:singlet-locus-2} 
\begin{split}
0 & = d_0\, b_2\, c_1 - b_0\, b_2^2 - c_1^2\, d_2\,,\\
0 & = d_1\, c_1^2 - b_1\, b_2\, c_1 + b_2^2\, c_2\,,
\end{split}
\end{equation}
and
\begin{equation}\label{eq:singlet-locus-3} 
\begin{split}
0 & = d_0\, c_1^3\, c_2^2 + b_0^2\, c_1^4 - b_0\, b_1\, c_1^3\, c_2 - 
  c_2^3 (b_2^2\, c_2 -b_1\, b_2\, c_1 + c_1^2 d_1)\,,\\
0 & = d_2\, c_1^4\, c_2^2 + (b_0\, c_1^2 + c_2\, (-b_1\, c_1 + b_2\, c_2)) (b_0\, b_2\, c_1^2 + c_2 (-b_1\, b_2\, c_1 + b_2^2\, c_2 + c_1^2\, d_1))\,.
\end{split}
\end{equation}
In \cite{Borchmann:2013jwa,Borchmann:2013hta} three singlet curves were identified as complete intersections: $C^{(1)} = \{b_0\} \cap \{c_2\}$ solves both (\ref{eq:singlet-locus-1}) and (\ref{eq:singlet-locus-3}), $C^{(3)} = \{b_2\} \cap \{ c_1\}$ solves (\ref{eq:singlet-locus-2}) and (\ref{eq:singlet-locus-3}), and $C^{(5)} = \{c_1\} \cap \{ c_2\}$ solves (\ref{eq:singlet-locus-3}). If one inserts these equations into the hypersurface equation (\ref{eq:hypersurface-equation}) one confirms that the fibre factorises into two $\mathbb P^1$s and can identify the singlet states as M2-branes wrapping one of the fibre components. The remaining three curves were represented in \cite{Borchmann:2013jwa,Borchmann:2013hta} as (\ref{eq:singlet-locus-1}) with $b_0 \neq 0 \neq c_2$ ($C^{(2)}$), (\ref{eq:singlet-locus-2}) with $b_2 \neq 0 \neq c_1$ ($C^{(4)}$), and (\ref{eq:singlet-locus-3}) with $b_0 \neq 0 \neq b_2$ and $c_1 \neq 0 \neq c_2$ ($C^{(6)}$). Plugging these more lengthy expressions into the hypersurface 
equation also leads to a factorisation of the fibre, i.e.~the appearance of charged singlets. Their location and charges are summarised as follows:
\begin{align}
  \label{tab:singlets-charges}
  \begin{array}{c|c| r @{} l }
    \multirow{2}{*}{\text{singlet}} & \multirow{2}{*}{\text{locus}} & \multicolumn{2}{c}{(U(1)_1, U(1)_2) \text{-}} \\
    & & \multicolumn{2}{c}{\text{charges}} \\ \hline \rule{0pt}{3ex}
    \mathbf{1}^{(1)} / \overline{\mathbf{1}}^{(1)} & \{b_0\} \cap \{c_2 \} & (1,-1) &/ (-1 ,1) \\
    \mathbf{1}^{(2)} / \overline{\mathbf{1}}^{(2)} & C^{(2)} & (1, 0) &/ (-1 , 0) \\
    \mathbf{1}^{(3)} / \overline{\mathbf{1}}^{(3)}  & \{b_2\} \cap \{c_1 \} & (1,2) &/ (-1,-2) \\
    \mathbf{1}^{(4)} / \overline{\mathbf{1}}^{(4)}  & C^{(4)} & (1, 1) &/ (-1,-1) \\
    \mathbf{1}^{(5)} / \overline{\mathbf{1}}^{(5)}  & \{c_1\} \cap \{c_2 \} & (0,2) &/ (0,-2) \\
    \mathbf{1}^{(6)} / \overline{\mathbf{1}}^{(6)}  & C^{(6)} & (0, 1) &/ (0,-1)
  \end{array} 
\end{align}
From the charge assignment we expect six types of Yukawa couplings. Of these, the couplings $\mathbf{1}^{(1)} \,  \overline{\mathbf{1}}^{(4)} \, \mathbf{1}^{(5)}$ over $b_0 = c_1 = c_2 = 0$, $\mathbf{1}^{(2)} \, \overline{\mathbf{1}}^{(3)} \, \mathbf{1}^{(5)}$ over $b_2 = c_1 = c_2 = 0$, and $\mathbf{1}^{(2)} \, \overline{\mathbf{1}}^{(4)} \, \mathbf{1}^{(6)}$ over $C^{(2)} \cap C^{(4)} \cap C^{(6)}$ can in fact be directly seen \cite{Borchmann:2013jwa, Cvetic:2013nia} when plugging in the corresponding equations of the curves into the hypersurface equation. However the charges also allow for couplings $\mathbf{1}^{(1)} \, \overline{\mathbf{1}}^{(2)} \, \mathbf{1}^{(6)}$, $\overline{\mathbf{1}}^{(3)} \, \mathbf{1}^{(4)} \, \mathbf{1}^{(6)}$ and $\overline{\mathbf{1}}^{(5)} \, \mathbf{1}^{(6)} \, \mathbf{1}^{(6)}$, which due to the form of the curves $C^{(2)}$, $C^{(4)}$, $C^{(6)}$ are more complicated to analyse. The difficulty is that the set of solutions to equations (\ref{eq:singlet-locus-1}) to (\ref{eq:singlet-locus-3}) consists of several irreducible components which intersect each other precisely at the interesting Yukawa points. To find an appropriate form of the singlet curves, we apply a classic method in algebraic geometry (e.g.~\cite{hartshorne:alggeo,cox:alggeo,cox:ideals}), which in the context of F-theory has been first presented in \cite{Cvetic:2013uta,Cvetic:2013jta} (`technique using prime ideals', see also \cite{Piragua}).

The general idea is that the expressions on the right-hand sides of equations (\ref{eq:singlet-locus-1}) -- (\ref{eq:singlet-locus-3}) are elements of the polynomial ring $R = \mathbb{C}[b_i,c_j,d_k]$. If we formally treat the sections $b_i,c_j,d_k$ as independent variables of these polynomials, then basic algebraic geometry tells us that the common zero locus $V(\{f_n\})$ of a set of polynomials $f_n \in R$ is the same as the common zero locus of the ideal generated by $\{f_n\}$, $V(\{f_n\}) = V(\langle \{ f_n \} \rangle)$. Intersections of zero loci are described by the formula $V(I_1) \cap V(I_2) = V(I_1 + I_2)$, where $I_1 + I_2$ is the sum of ideals. Each (proper) ideal $I \varsubsetneq R$ has a so-called primary decomposition $I = \bigcap_{i=1}^{r<\infty} J_i$, where $J_i$ are termed primary ideals; by definition, the radical $\sqrt{J_i}$ is a prime ideal (in fact the smallest containing $J_i$), more precisely it is the (minimal) associated prime ideal. Translated into geometry this means that the vanishing locus is 
decomposed into components $V(I) = \bigcup_{i=1}^{r<\infty} V(J_i) = \bigcup_{i=1}^{r<\infty} V(\sqrt{J_i})$, where the last equality is a consequence of the famous `Hilbert's Nullstellensatz' (see e.g.~\cite{cox:ideals}). The components $V(\sqrt{J_i})$ are precisely the irreducible components of $V(I)$. Furthermore, one can compute the codimension of $V(I)$ algebraically via the so-called Krull dimension $\dim_K$ of the quotient ring $R/I$: $\text{codim} V(I) = \dim_K R - \dim_K R/I$.

After this short mathematical interlude, the procedure to find the singlet curves and also the Yukawa points becomes clear: The right-hand sides of equations (\ref{eq:singlet-locus-1}), (\ref{eq:singlet-locus-2}) and (\ref{eq:singlet-locus-3}) define three ideals generated by two polynomials. Their associated prime ideals with such Krull dimension that their vanishing locus has codimension two correspond to curves in the three-dimensional base -- the matter curves. Yukawa points arise at intersections of three matter curves; correspondingly we have to form the sum of prime ideals associated to the matter curves involved. In general, when we compute the associated prime ideals of such a sum and calculate their Krull dimension, we will find that the intersection locus has a number of irreducible components with different codimensions. On a generic three-dimensional base, Yukawa points correspond to the codimension-three components, while all higher-codimension components are absent; hence if we compute the 
intersection of 
certain curves and only find codimension-four or higher components, we conclude that these curves cannot meet in a generic three-dimensional base and form Yukawa couplings.

We have carried out the calculations using the computer algebra system \textsc{Singular}\footnote{We would like to thank Hernan Piragua for introducing us to the `prime ideal technique' in \textsc{Singular}. See also \cite{Piragua}.} \cite{singular} and indeed find the six associated prime ideals listed in \cite{Cvetic:2013uta,Cvetic:2013jta}. Three of them have only two generators, $I^{(1)} = \langle b_0,  c_2 \rangle$, $I^{(3)} = \langle b_2 , c_1 \rangle$, $I^{(5)} = \langle c_1 , c_2 \rangle$; their corresponding vanishing loci are the complete intersection curves listed in (\ref{tab:singlets-charges}). The other three associated prime ideals $I^{(2)}$, $I^{(4)}$ and $I^{(6)}$, which are associated prime ideals of (\ref{eq:singlet-locus-1}), (\ref{eq:singlet-locus-2}) and (\ref{eq:singlet-locus-3}), respectively, now correspond to the curves $C^{(2)}$, $C^{(4)}$ and $C^{(6)}$ in our previous notation. They have more generators, which are quite lengthy polynomial expressions in $b_i, c_j, d_k$; our 
findings for their explicit form 
coincide with the results presented in \cite{Cvetic:2013uta,Cvetic:2013jta}. 

With this technique we can now analyse the Yukawa couplings $\mathbf{1}^{(1)} \, \overline{\mathbf{1}}^{(2)} \, \mathbf{1}^{(6)}$, $\overline{\mathbf{1}}^{(3)} \, \mathbf{1}^{(4)} \, \mathbf{1}^{(6)}$ and $\overline{\mathbf{1}}^{(5)} \, \mathbf{1}^{(6)} \, \mathbf{1}^{(6)}$. To this end we first calculate the associated prime ideals of the sum of the ideals corresponding to the curves. In each case we indeed find one prime ideal corresponding to a codimension-three zero locus, confirming the existence of the intersection points of those triplets of singlet curves. All three codimension-three intersection loci are in fact complete intersections,
\begin{align}
\begin{split}\label{eq:intersection-point1}
& V(I^{(1)}) \cap V(I^{(2)}) \cap V(I^{(6)}) = \\
& \{b_0 \} \cap \{c_2\} \cap \{b_2^2\,d_0^2-b_1\,b_2\,d_0\,d_1+c_1\,d_0\,d_1^2+b_1^2\,b_2\,d_2-2\,b_2\,c_1\,d_0\,d_2-b_1\,c_1\,d_1\,d_2+c_1^2\,d_2^2\} \, ,
\end{split} \\
\begin{split}\label{eq:intersection-point2}
& V(I^{(3)}) \cap V(I^{(4)}) \cap V(I^{(6)}) = \\
& \{b_2\} \cap \{c_1\} \cap \{ b_0^2 \, d_1^2 - b_0\,b_1\,d_0\,d_1 + c_2\,d_1\,d_0^2 + b_0\,b_1^2\,d_2 - b_1\,c_2\,d_0\,d_2 - 2\,b_0\,c_2\,d_1\,d_2 + c_2^2\,d_2^2 \} \, ,
\end{split}\\
\begin{split}\label{eq:intersection-point3}
& V(I^{(5)}) \cap V(I^{(6)}) \cap V(I^{(6)}) = \\
& \{c_1 \} \cap \{c_2\} \cap \{b_1\,d_0\,d_1-b_2\,d_0^2-b_0\,d_1^2-b_1^2\,d_2+4\,b_0\,b_2\,d_2\} \, .
\end{split}
\end{align}
Interestingly, the last set of Yukawa points (\ref{eq:intersection-point3}) coincides with the singular locus of $V(I^{(6)})$. Due to the complicated form of $V(I^{(6)})$ (it has 39 generators which themselves are complicated polynomials), we have not determined the type of the singularity, but the form of the Yukawa coupling involving two $\mathbf{1}^{(6)}$-states suggests that it is a point of self-intersection of the $\mathbf{1}^{(6)}$-curve where also the $\mathbf{1}^{(5)}$-curve passes through.

The final proof for the existence of the Yukawa couplings comes by inspecting the fibre over the intersection points. The couplings $\mathbf{1}^{(1)} \, \overline{\mathbf{1}}^{(2)} \, \mathbf{1}^{(6)}$ and $\overline{\mathbf{1}}^{(3)} \, \mathbf{1}^{(4)} \, \mathbf{1}^{(6)}$ have already been argued to exist geometrically in \cite{Cvetic:2013uta} using the `prime ideal technique', and independently in \cite{Borchmann:2013hta} in an indirect manner by exploiting their formal relation to the chiral index of certain $G_4$-fluxes. 
Here we therefore focus on the remaining $\overline{\mathbf{1}}^{(5)} \, \mathbf{1}^{(6)} \, \mathbf{1}^{(6)}$ coupling. If we solve the last equation in (\ref{eq:intersection-point3}) for $b_1=(d_0 \, d_1 \pm \sqrt{d_0^2 - 4 \, b_0 \, d_2} \,  \sqrt{d_1^2 - 4 \, b_2 \, d_2})/(2 \, d_2)$, we see that the complete intersection locus (\ref{eq:intersection-point3}) really consists of two sets of points defined by each sign. Note that as far as codimension-three loci are concerned the appearance of the  square root or of $d_2$ in the denominator does not pose any problems.
Plugging this together with $c_1 = c_2 =0$ into the hypersurface equation (\ref{eq:hypersurface-equation}) yields, after some tedious algebra, the factorisation
\begin{align}\label{eq:factorisation-singlet-yukawa}
\begin{split}
&P_T |_{c_1=c_2=0 , \, b_1=(d_0 \, d_1 \pm \sqrt{d_0^2 - 4 \, b_0 \, d_2} \,  \sqrt{d_1^2 - 4 \, b_2 \, d_2})/(2 \, d_2)} \, = \frac{1}{4\,d_2} \,  \su \\
& \times \left[ 2 \, d_2 \, s_0 \, s_1 \su + \left( d_0 - \sqrt{d_0^2 - 4\, b_0 \, d_2} \right) s_0 \, \sv + \left(d_1 \pm \sqrt{d_1^2 - 4\,b_2\,d_2} \right) s_1 \sw \right] \\
& \times \left[ 2 \, d_2 \, s_0 \, s_1 \su + \left( d_0 + \sqrt{d_0^2 - 4\, b_0 \, d_2} \right) s_0 \, \sv + \left(d_1 \mp \sqrt{d_1^2 - 4\,b_2\,d_2} \right) s_1 \sw \right] \, ,
\end{split}
\end{align}
which is well-defined since no fibre coordinate appears under the square root. The fibre component defined by the factor $\su$ corresponds to the singlet state $\overline{\mathbf{1}}^{(5)}$, as explicit calculation of the intersection numbers with the $U(1)$ generators (\ref{eq:U(1)-generators}) using the Stanley-Reissner ideal and the divisor table (\ref{tab:divisor-classes-small}) quickly shows. The other two components are obviously in the same divisor class of the fibre ambient space and must have the same intersection numbers; indeed their intersection numbers with the $U(1)$ generators reveal that both correspond to $\mathbf{1}^{(6)}$-states. Furthermore, each component intersects the others exactly once, giving rise to an affine $SU(3)$ diagram. Similar calculations also verify the analogous fibre structure enhancement over the other two Yukawa points (\ref{eq:intersection-point1}) and (\ref{eq:intersection-point2}). More details may be found in \cite{philipp:BA}.

% A similar, although slightly longer calculation (hence omitted here, but may be found in \cite{philipp:BA}) also confirms the existence of the Yukawa coupling $\mathbf{1}^{(1)} - \overline{\mathbf{1}}^{(2)} - \mathbf{1}^{(6)}$. Currently we are also working \cite{philipp:BA} on a method to explicitly see the appearance of the Yukawa coupling $\overline{\mathbf{1}}^{(3)} - \mathbf{1}^{(4)} - \mathbf{1}^{(6)}$ over the more complicated codimension 3 loci $V(I^{(3)}) \cap V(I^{(4)}) \cap V(I^{(6)})$.

\subsection{Introducing Non-Abelian Symmetry}

Non-abelian gauge symmetry arises  if the sections  $g_m \in \{b_i, c_j, d_k\}$ appearing in (\ref{eq:hypersurface-equation})   take a non-generic form such that the fibration acquires a singularity in the fibre
 over one or several divisors in the base.
  A special type of such gauge enhancement is realised by restricting every section $g_m \in \{b_i, c_j, d_k\}$ such that it exhibits a certain vanishing order $k$ along a divisor $W = \{w\}$ in the base.
 In other words, one restricts
 \bea \label{gmkform}
g_m = g_{m,k} w^k,
 \eea
 where now $g_{m,k}$ is a generic section of class $[g_m] - k [W]$ in the base; in particular, it does not vanish identically on $W$.\footnote{Other types of non-abelian singularities would involve non-trivial relations between the $g_m$
rather than factorisations of the type (\ref{gmkform}), as studied recently in the context of fibrations with Mordell-Weil group of rank one in \cite{Mayrhofer:2012zy,Kuntzler:2014ila}.}

The complex structure restrictions of type (\ref{gmkform}) compatible with gauge group $G$ along $W$ can be determined with the help of toric geometry and are encoded in the construction of toric tops \cite{Candelas:1996su,Candelas:1997eh}. The possible tops for all sixteen hypersurface realisations of genus-one fibrations have been classified in \cite{Bouchard:2003bu}. The top construction directly gives the resolution of the singularities in the fibre over $w=0$. 
The toric resolution of a singularity associated with gauge group $G$ of rank $r$ introduces blow-up coordinates $e_i, i=1, \ldots,r$ together with new scaling relations in the toric ambient space of the fibre.
The hypersurface equation is replaced by a hypersurface in the blown-up ambient space in which each $g_m$ is replaced by $g_{m,k} \, e_0^k \, e_1^{l_1} \, e_2^{l_2} \ldots e_r^{l_r}$ for suitable powers $l_i$. For details on how to read off these $l_i$ from the toric tops we also refer to \cite{Borchmann:2013hta}.

The vanishing set $\{e_0\} \cap \{P_T\}$ is a divisor in the resolved fourfold which can be identified with the fibration of the original (singular) fibre \textit{without} the singular point over $W$, i.e.~the generic fibre has the topology of a $\mathbb{P}^1$.
Each of the sets $\{e_i\} \cap \{P_T\}, i>0$ is a \textit{resolution divisor} $E_i$, which --  over a generic point in $W$ -- introduces one further $\mathbb{P}^1$ to resolve the singularity of the fibre. The intersection diagram of these $\mathbb{P}^1$s \textit{over a generic point} in $W$ is the affine Dynkin diagram of the non-abelian gauge group. The resolution divisors $E_{i,i>0}$ correspond to (minus) the simple roots of the gauge Lie algebra. As in the case of the singlets, the $\mathbb{P}^1$s in the fibre can split over special curves and points (\mbox{codimension-two and -three}, respectively) in the base, leading to the appearance of matter states charged under the non-abelian gauge group and Yukawa couplings involving these states and also the charged singlets. Note that an alternative procedure to detect such matter via a deformation (as opposed to resolution) of the singular fibres has been described recently in \cite{Grassi:2013kha,Grassi:2014sda}.

The complex structure moduli restrictions (\ref{gmkform}) affect the precise location of the charged singlets in the base (but not their charges). In general the loci displayed in (\ref{tab:singlets-charges}) will contain components with non-abelian matter, which we have to disregard when we focus on the singlets. For the singlets  $ \mathbf{1}^{(i)}$ with $i = 1, 3, 5$ this is simply done by replacing the coefficients $g_m$ that are supposed to vanish by their factors $g_{m,k}$ that do not vanish identically on $W$. For the other singlets we can again use \textsc{Singular} to determine the prime ideals associated with the curves.

%% file: section3_models_with_SU2_v3.tex
\section[Toric Fibrations with Additional \texorpdfstring{\boldmath $SU(2)$}{SU(2)} Symmetry]{Toric Fibrations with Additional \texorpdfstring{\boldmath $SU(2)$}{SU(2)} Symmetry}\label{sec:SU(2)-tops}

In this section we analyse in detail toric realisations of gauge group $SU(2)$ along a base divisor 
\bea
W_2: \{w_2 = 0\}
\eea
in elliptic fibrations of type (\ref{eq:hypersurface-equation}). As detailed in appendix \ref{app:tops}, the fibres of such toric models are described by the three $A_1$-tops over polygon 5 \cite{Bouchard:2003bu}. 
The resolution of an $SU(2)$ singularity over a divisor $W_2$ requires one resolution divisor $E_1$ corresponding to the single (simple) root $-\alpha$. Over a generic point on $W_2$, the fibre splits into two $\mathbb{P}^1$-components described by 
\bea
\mathbb{P}^1_i = \{e_i\} \cap \{P_T\} \cap Y_a \cap Y_b  
\eea
for  $i=0,1$, where $Y_{a,b}$ are two generic divisors in the base. These two $\mathbb{P}^1$s intersect in the affine $SU(2)$ diagram and will split into further $\mathbb{P}^1$s over matter curves and Yukawa points. %Each of the three inequivalent tops gives rise to a different matter content and pattern of Yukawa couplings.

\subsection[\texorpdfstring{$SU(2)$}{SU(2)}-I Top]{\texorpdfstring{{\boldmath $SU(2)$}}{SU(2)}-I Top}

The first $A_1$-top, depicted in figure \ref{fig:tops}
 in appendix \ref{app:tops}, corresponds to the following restrictions of the coefficients of the hypersurface equation (\ref{eq:hypersurface-equation}),
\begin{align}\label{eq:SU(2)-I-coeffs}
 b_0 = b_{0,1} \, e_0 , \quad b_2 = b_{2,0} \, e_1 , \quad c_1 = c_{1,0} \, e_1 , \quad d_0 = d_{0,1} \, e_0 , \quad d_2 = d_{2,1} \, e_0,
\end{align}
while the other coefficients remain unrestricted. Concretely, the hypersurface describing the resolved elliptic fibration is given by
\begin{align}\label{eq:SU(2)-I-hypersurface-equation}
  \begin{split}
    P_T =  &\sv \, \sw (c_{1,0}\,e_1 \, \sw \, s_1 + c_2 \, \sv \, s_0) + \su \, (b_{0,1}\,e_0 \, \sv^2 \, s_0^2 + b_1 \, \sv \, \sw \, s_0 \, s_1 + b_{2,0}\,e_1 \, \sw^2 \, s_1^2) + \\
    &\su^2 (d_{0,1}\,e_0 \, \sv \, s_0^2 \, s_1 + d_1 \, \sw \, s_0 \, s_1^2 + d_{2,1}\,e_0 \, \su \, s_0^2 \, s_1^2).
  \end{split}
\end{align}
This is the blow-up of a singular fibration with an $A_1$-singular fibre over the base divisor $W_2=\{w_2\}$ with $\pi^{-1} W_2 = E_0 \, E_1$. The singular fibration is obtained by setting $e_1=1$ and identifying $e_0$ with $w_2$.
One can map this blow-down to Weierstrass form (\ref{Weier}) and confirm a Kodaira fibre of (split) type $I_2$ over $\{w_2\}$ from the vanishing orders $(0,0,2)$ of $(f,g,\Delta)$.

The top allows two different triangulations. For definiteness, we choose one of these triangulations, for which the Stanley-Reisner-ideal (SR-ideal) is generated by
\begin{align}\label{eq:SU(2)-I-SR-ideal}
  \su \, \sv , \su \, \sw , \sw \, s_0 , \sv \, s_1 , s_0 \, s_1 , e_0 \, \sw , e_1 \, s_0 , e_1 \, \su .
\end{align}

From the top one can further read off the scaling relations among the coordinates and their corresponding divisor classes in the ambient space, which are summarised in the following table:
\begin{align}\label{tab:SU(2)-I-divisor-classes}
  \begin{array}{c|cccccc|c}
    \hphantom{U} & \su & \sv & \sw & s_0 & s_1 & e_1 & e_0 \\ \hline
    {U} & 1 & 1 & 1 & \cdot & \cdot & \cdot & \cdot \\
    S_0 & \cdot & \cdot & 1 & 1& \cdot & \cdot & \cdot \\
    S_1 & \cdot & 1 & \cdot & \cdot & 1 & \cdot & \cdot \\
    E_1 & \cdot & \cdot & -1 & \cdot & \cdot & 1 & -1
  \end{array}
\end{align}

In the presence of non-abelian symmetry the $U(1)$ generators (\ref{eq:U(1)-generators}) need to be corrected such that the $SU(2)$ root has zero $U(1)$ charge. The resulting $U(1)$ generators take the form
\begin{align}\label{eq:SU(2)-I-U(1)-generators}
  \begin{split}
    \omega_1^\text{I} &= S_1 - S_0 - \overline{\mathcal{K}} + \frac{1}{2} E_1 ,\\
    \omega_2^\text{I} &= U - S_0 - \overline{\mathcal{K}} - [c_{1,0}] .
  \end{split}
\end{align}
Note that the charges (\ref{tab:singlets-charges}) of the singlets are not affected as these states are not charged under the $SU(2)$ root.

\subsubsection*{Matter Curves}

The Kodaira type of the resolved fibre changes in codimension-two, i.e.~over curves along the divisor $W_2$ in the base. 
These loci can be found by analysing the vanishing order of the discriminant of the singular blow-down of (\ref{eq:SU(2)-I-hypersurface-equation}) along with the Weierstrass sections $f$ and $g$
which define the birationally equivalent Weierstrass model (\ref{Weier}).
One finds
\bea
\Delta  \simeq w_2^2 \Big(  c_{2} \, (c_{1,0}^2 \, d_1 - b_1 \, b_{2,0} \, c_{1,0} + b_{2,0}^2 \, c_2) \, \ell_3 \,  (b_1^2 - 4 c_2 d_1)^2 + {\cal O}(w_2)  \Big)
\eea
with $\ell_3$ a complicated expression given in table (\ref{tab:SU(2)-I-matter}).
A straightforward analysis of the Weierstrass sections $f$ and $g$ reveals that the fibre
over the curves $\{w_2 \} \cap \{c_2\}$,  $\{w_2 \} \cap \{c_{1,0}^2 \, d_1 - b_1 \, b_{2,0} \, c_{1,0} + b_{2,0}^2 \, c_2\} $ and  $\{w_2 \} \cap \{\ell_3\}$ is of split Kodaira type $I_3$, corresponding to vanishing orders $(0,0,3)$ for $(f,g,\Delta)$.  This  indicates an enhancement of the singularity type from $A_1$ to $A_2$  due to the splitting of one of the fibre components such that the fibre over the curves forms the affine Dynkin diagram of $SU(3)$.
The curves therefore host massless matter multiplets in $SU(2)$ representation ${\bf 2}_{(q_1,q_2)}$ plus their conjugates, with the subscripts denoting the $U(1)$ charges.\footnote{
 Note that the anti-fundamental representation of $SU(2)$ is equivalent to the fundamental, but in the present context it has the opposite $U(1)$ charges and will therefore be denoted by $\overline{\mathbf{2}}$. 
} We will explicitly analyse the fibre and compute the $U(1)$ charges momentarily.
By contrast, along $\{w_2\} \cap \{ b_1^2 - 4 c_2 d_1\} $ the fibre is of Kodaira type $III$, corresponding to vanishing orders $(1,2,3)$ for $(f,g,\Delta)$. Since the singularity type remains $A_1$, no charged matter representations arise over this curve, consistent in particular with the results of \cite{Grassi:2011hq}.  The matter curves and $U(1)$ charges are summarised in table \ref{tab:SU(2)-I-matter}.
\begin{table}[ht]
\begin{center}
  \begin{tabular}{c|c|c|c|c}
    \multirow{2}{*}{matter} & \multirow{2}{*}{locus = $W_2 \cap \ldots$} & splitting of fibre & $U(1)-$ & highest weight\\
    & & components & charges & states\\ \hline\hline \rule{0pt}{3ex}
   $\mathbf{2}^\mathrm{I}_1$ & $\{c_{2}\}$ & $\mathbb{P}^1_0 \rightarrow \mathbb{P}^1_{0s_1} \! + \mathbb{P}^1_{0A}$ & $(\frac{1}{2}, -1)$ & $\mathbf{2} \! : \mathbb{P}^1_{0 A} , \, \overline{\mathbf{2}} \! : \mathbb{P}^1_{0 s_1}$ \\ [.5ex] \hline \rule{0pt}{3ex}
    $\mathbf{2}^\mathrm{I}_2$ & $\{ c_{1,0}^2 \, d_1 - b_1 \, b_{2,0} \, c_{1,0} + b_{2,0}^2 \, c_2 \}$ & $\mathbb{P}^1_0 \rightarrow \mathbb{P}^1_{0B} + \mathbb{P}^1_{0C}$ & $(\frac{1}{2},1)$ &  $\mathbf{2} \! : \mathbb{P}^1_{0 C} , \, \overline{\mathbf{2}} \! : \mathbb{P}^1_{0 B}$\\ [.5ex] \hline \rule{0pt}{3ex}
    \multirow{3}{*}{$\mathbf{2}^\mathrm{I}_3$}& $\{\ell_3\} := \{  b_{0,1}^2\,d_1^2$ & & \\
    & $+ b_{0,1}\,(b_1^2\,d_{2,1} - b_1\,d_{0,1}\,d_1- 2\,c_2\,d_1\,d_{2,1})$ & $\mathbb{P}^1_1 \rightarrow \mathbb{P}^1_{1A} + \mathbb{P}^1_{1B}$ & $(\frac{1}{2}, 0)$ &  $\mathbf{2} \! : \mathbb{P}^1_{1 B} , \, \overline{\mathbf{2}} \! : \mathbb{P}^1_{1 A}$ \\
    & $+ c_2\,(d_{0,1}^2\,d_1 - b_1\,d_{0,1}\,d_{2,1} + c_2\,d_{2,1}^2)\}$ & &
  \end{tabular}
\end{center}
\caption{Matter states and their charges in the $SU(2)$-I top. Note that for legibility we have omitted the conjugate $\overline{\mathbf{2}}$-states and their charges, which simply come with the opposite sign as the shown charges.}
\label{tab:SU(2)-I-matter}
\end{table}

The splitting process in the fibre is due to the factorisation of the hypersurface equation over the enhancement loci. For the first curve, the factorisation is straightforward to see after setting $c_2 = e_0 = 0$ in (\ref{eq:SU(2)-I-hypersurface-equation}),
\begin{align}\label{eq:SU(2)-I-splitting-first-curve}
  {P_T}|_{(e_0=0, \, c_2=0)} = s_1\,\sw \left( c_{1,0}\,e_1\,\sv\,\sw + b_1\,s_0\,\su\,\sv + b_{2,0}\,e_1\,s_1\,\sw\,\su + d_1\,s_0\,s_1\,\su^2 \right).
\end{align}
Since $e_0 \, \sw$ is in the SR-ideal, $\sw$ cannot vanish so that the zero locus of (\ref{eq:SU(2)-I-splitting-first-curve}) splits into the zero locus of $s_1$ and of the expression in brackets, defining the components $\mathbb{P}_{0s_1}$ and $\mathbb{P}_{0A}$. One can further calculate the intersections between these components and $\mathbb{P}^1_1$ (which does not split and remains the root of $SU(2)$) and easily verify the structure to be an affine $SU(3)$ diagram. Explicit calculations identify $\mathbb{P}^1_{0 A}$ with the highest weight state of the $\mathbf{2}$-representation with $U(1)$ charges $(\frac{1}{2},-1)$, whose states we denote by $\mathbf{2}^\mathrm{I}_1$. Correspondingly $\mathbb{P}^1_{0 s_1}$ is the highest weight state of the conjugate representation $\overline{\mathbf{2}}^\mathrm{I}_1$, whose states have $U(1)$ charges $(-\frac{1}{2},1)$.

For the second curve defined by $W_2 \cap \{c_{1,0}^2 \, d_1 - b_1 \, b_{2,0} \, c_{1,0} + b_{2,0}^2 \, c_2\}$, one could solve the equation for $b_1$, $c_2$ or $d_1$ and plug the expressions into (\ref{eq:SU(2)-I-hypersurface-equation}) to detect a factorisation. However any of these expressions will involve division by $c_{1,0}$ or $b_{2,0}$, which for the analysis of Yukawa points  below turns out to be disadvantageous. Instead we factorise 
\bea  \label{eq:SU(2)-I-quadratic-curve}
 c_{1,0}^2 \, d_1 - b_1 \, b_{2,0} \, c_{1,0} + b_{2,0}^2 \, c_2 = \frac{1}{d_1}  \, {\mathcal{C}_+}  \,  {\mathcal{C}_-}  \quad {\rm with} \quad {\mathcal{C}_\pm} =  c_{1,0}\,d_1 - b_{2,0} \left( \frac{b_1}{2} \pm \sqrt{\frac{b_1^2}{4} - c_2 \, d_1} \right),
 \eea
 % \begin{align}\label{eq:SU(2)-I-quadratic-curve} 
%   \frac{1}{d_1} \underbrace{\left[ c_{1,0}\,d_1 - b_{2,0} \left( \frac{b_1}{2} + \sqrt{\frac{b_1^2}{4} - c_2 \, d_1} \right) \! \right]}_{\mathcal{C}_+} \underbrace{\left[ c_{1,0}\,d_1 - b_{2,0} \left( \frac{b_1}{2} - \sqrt{ \frac{b_1^2}{4} - c_2 \, d_1} \right) \! \right]}_{\mathcal{C}_-} = 0,
% \end{align}
corresponding to a splitting of the curve into two components $W_2 \cap \{\mathcal{C}_\pm = 0\}$. Note that the square root introduces a branch cut in the base along which the two components are interchanged. Therefore the whole locus $W_2 \cap \{ c_{1,0}^2 \, d_1 - b_1 \, b_{2,0} \, c_{1,0} + b_{2,0}^2 \, c_2 \}$ is still one irreducible curve.
Furthermore the above factorisation is valid for generic points for which $d_1 \neq 0$. Since, as it turns out, at $d_1=0$ no Yukawa points are localised, this is sufficient for our purposes.

 The factorisation (\ref{eq:SU(2)-I-quadratic-curve}) now allows us to solve $\mathcal{C}_{\pm}=0$ for $c_{1,0}$ and substitute it into the hypersurface equation. With this substitution we can see that over each part of the curve, $\mathbb{P}^1_0$ splits into two components,
\begin{align}\label{eq:SU(2)-I-splitting-quadratic-curve}
  \begin{split}
    & P_T|_{(e_0=0, \, {\mathcal C}_\pm=0)} \\%c_{1,0} = \frac{b_{2,0}}{d_1} ( \frac{b_1}{2} \pm \sqrt{ \frac{b_1^2}{4} - c_2 \, d_1} ) )} \\
    = \,& \frac{1}{d_1} \underbrace{\left[ d_1\,s_1\,\su + \sv \left( \frac{b_1}{2} \pm \sqrt{ \frac{b_1^2}{4} - c_2 \, d_1} \right) \! \right]}_{\mathbb{P}^1_{0B}} \underbrace{\left[ b_{2,0}\,e_1\,s_1 + d_1\,s_0\,s_1\,\su + s_0\,\sv \left( \frac{b_1}{2} \mp \sqrt{ \frac{b_1^2}{4} - c_2 \, d_1} \right) \! \right]}_{\mathbb{P}^1_{0C}} ,
  \end{split}
\end{align}
where we have set $\sw=1$ using the SR-ideal. First note that there is no fibre coordinate appearing under the square roots. Therefore the factorisation defines two irreducible fibre components over each part $W_2 \cap {\cal C}_\pm$. At the branch cut the first/second component over one part of the curve is identified with the first/second component over the other part so there is no monodromy acting on the fibre components, making $\mathbb{P}^1_{0B}$ and $\mathbb{P}^1_{0C}$ well-defined on the whole curve. Explicit calculations show that $\mathbb{P}^1_{0B}$, $\mathbb{P}^1_{0C}$ and $\mathbb{P}^1_{1}$ (which again does not split) intersect each other in the affine $SU(3)$ diagram. $\mathbb{P}^1_{0C}$ is the highest weight state of $\mathbf{2}^\mathrm{I}_2$ with charges $(\frac{1}{2},1)$, and $\mathbb{P}^1_{0B}$ is that of $\overline{\mathbf{2}}^\mathrm{I}_2$ with charges $(-\frac{1}{2},-1)$.

For the third curve, we apply a similar factorisation method; the defining equation can be written as 
\begin{align}\label{eq:SU(2)-I-complicated-quadratic-curve}
\ell_3 = 1/d_1^2 \, \mathcal{D_+} \, \mathcal{D}_-    \quad {\rm with} \quad   \mathcal{D}_\pm = b_{0,1}\,d_1^2 - \left[ c_2\,d_1\,d_{2,1} + (d_{0,1}\,d_1-b_1\,d_{2,1}) \left( \frac{b_1}{2} \pm \sqrt{ \frac{b_1^2}{4} - c_2\,d_1} \right)\!\right] .
\end{align}
Again, there is a branch cut in the base coming from the square root which identifies  the two parts $W_2 \cap \{\mathcal{D}_\pm=0\}$ at the branch locus. The fibre enhancement over each part can be deduced (after some calculation) by solving $\mathcal{D}_\pm=0$ for $b_{0,1}$ and inserting the expression into the hypersurface equation (\ref{eq:SU(2)-I-hypersurface-equation}). We find that $\mathbb{P}^1_1$ splits into two components,
\begin{align}\label{eq:SU(2)-I-splitting-complicated-quadratic-curve}
  \begin{split}
    & P_T|_{(e_1=0, \, \mathcal{D}_\pm=0)} = \\
    & \frac{1}{d_1} \underbrace{\left[ d_1\,s_1 + \left( \frac{b_1}{2} \pm \sqrt{\frac{b_1^2}{4} - c_2\,d_1} \right)\!\sv \right]}_{\mathbb{P}^1_{1A}} \\
    & \times \underbrace{\left[ d_{2,1}\,e_0\,s_1 + \left[ d_{0,1} - \frac{d_{2,1}}{d_1} \left( \frac{b_1}{2} \pm \sqrt{ \frac{b_1^2}{4} - c_2\,d_1 } \right)\!\right]\!e_0\,\sv + d_1\,\sw\,s_1 + \left( \frac{b_1}{2} \mp \sqrt{\frac{b_1^2}{4} - c_2\,d_1} \right)\!\sv\,\sw \right]}_{\mathbb{P}^1_{1B}} . 
  \end{split}
\end{align}
Analogous to the situation over the second curve, the factors are not interchanged by any monodromy when passing the branch locus in the base, making the splitting $\mathbb{P}^1_1 \rightarrow \mathbb{P}^1_{1A} + \mathbb{P}^1_{1B}$ well-defined over the whole curve. The intersection structure together with $\mathbb{P}^1_0$ (which remains irreducible) turns out to be again the affine $SU(3)$ diagram. $\mathbb{P}^1_{1B}$ is the highest weight state of $\mathbf{2}^\mathrm{I}_3$ with charges $(\frac{1}{2},0)$ and $\mathbb{P}^1_{1A}$ is that of $\overline{\mathbf{2}}^\mathrm{I}_3$ with charges $(-\frac{1}{2},0)$.

\subsubsection*{Yukawa Points}

$SU(2)$ matter and singlet curves intersect at codimension-three loci in the base to form gauge invariant Yukawa couplings of the form $\mathbf{2}-\overline{\mathbf{2}}- \mathbf{1}/\overline{\mathbf{1}}$ and $\mathbf{2}-\mathbf{2}-\mathbf{1}/\overline{\mathbf{1}}$. We list all such couplings in table \ref{tab:SU(2)-I-Yukawas}.

\begin{table}[ht]
\begin{align*}
  \begin{array}{c|c|c}
    \text{coupling} & \text{locus} = W_2 \cap \ldots & \text{splitting of fibre components} \\ \hline\hline \rule{0pt}{3ex}
    \mathbf{2}^\mathrm{I}_1 - \mathbf{2}^\mathrm{I}_2 - \overline{\mathbf{1}}^{(2)} & \{c_2\} \cap \{c_{1,0}\,d_1 - b_1\,b_{2,0} \} & \mathbb{P}^1_0 \rightarrow \mathbb{P}^1_{0 s_1 C} + \mathbb{P}^1_{0AB} + \mathbb{P}^1_{0AC} \\[.5ex] \hline \rule{0pt}{3ex}
    \mathbf{2}^\mathrm{I}_1 - \overline{\mathbf{2}}^\mathrm{I}_2 - \mathbf{1}^{(5)} & \{c_2\} \cap \{c_{1,0}\} & \mathbb{P}^1_0 \rightarrow \mathbb{P}^1_{0s_1 B} + \mathbb{P}^1_{0AB'} + \mathbb{P}^1_{0AC'} \\[.5ex] \hline\hline \rule{0pt}{3ex}
    \mathbf{2}^\mathrm{I}_1 - \mathbf{2}^\mathrm{I}_3 - \overline{\mathbf{1}}^{(1)} & \{c_2\} \cap \{b_{0,1}\} & \mathbb{P}^1_0 \rightarrow \mathbb{P}^1_{0s_1} \!+ \mathbb{P}^1_{0A}, \, \mathbb{P}^1_1 \rightarrow \mathbb{P}^1_{1A} + \mathbb{P}^1_{1B} \\[.5ex] \hline \rule{0pt}{3ex}
    \mathbf{2}^\mathrm{I}_1 - \overline{\mathbf{2}}^\mathrm{I}_3 - \mathbf{1}^{(6)} & \{c_2\} \cap \{b_1^2\,d_{2,1} - b_1\,d_{0,1}\,d_1 + b_{0,1}\,d_1^2\} & \mathbb{P}^1_0 \rightarrow \mathbb{P}^1_{0s_1} \!+ \mathbb{P}^1_{0A}, \, \mathbb{P}^1_1 \rightarrow \mathbb{P}^1_{1A} + \mathbb{P}^1_{1B} \\[.5ex] \hline\hline \rule{0pt}{3ex}
    \mathbf{2}^\mathrm{I}_2 - \mathbf{2}^\mathrm{I}_3 - \overline{\mathbf{1}}^{(4)} & \left(\{\mathcal{C}_+\} \cap \{\mathcal{D}_+\}\right) \cup \left(\{\mathcal{C}_-\} \cap \{\mathcal{D}_-\}\right) & \mathbb{P}^1_0 \rightarrow \mathbb{P}^1_{0B} + \mathbb{P}^1_{0C}, \, \mathbb{P}^1_1 \rightarrow \mathbb{P}^1_{1A} + \mathbb{P}^1_{1B} \\[.5ex] \hline \rule{0pt}{3ex}
    \mathbf{2}^\mathrm{I}_2 - \overline{\mathbf{2}}^\mathrm{I}_3 - \overline{\mathbf{1}}^{(6)} & \left(\{\mathcal{C}_+\} \cap \{\mathcal{D}_-\}\right) \cup \left(\{\mathcal{C}_-\} \cap \{\mathcal{D}_+\}\right) & \mathbb{P}^1_0 \rightarrow \mathbb{P}^1_{0B} + \mathbb{P}^1_{0C}, \, \mathbb{P}^1_1 \rightarrow \mathbb{P}^1_{1A} + \mathbb{P}^1_{1B} \\[.5ex] \hline\hline \rule{0pt}{3ex}
    \mathbf{2}^\mathrm{I}_2 - \mathbf{2}^\mathrm{I}_2 - \overline{\mathbf{1}}^{(3)} & \{b_{2,0}\} \cap \{c_{1,0}\} & \mathbb{P}^1_0 \rightarrow \mathbb{P}^1_{0B} + \mathbb{P}^1_{0Cs_0} + \mathbb{P}^1_{0C'} \\[.5ex] \hline \rule{0pt}{3ex}
    \mathbf{2}^\mathrm{I}_3 - \mathbf{2}^\mathrm{I}_3 - \overline{\mathbf{1}}^{(2)} & \{b_{0,1}\,d_1 - c_2\,d_{2,1}\} \cap \{b_1\,d_{2,1} - d_{0,1}\,d_1\} & \mathbb{P}^1_1 \rightarrow \mathbb{P}^1_{1A} + \mathbb{P}^1_{1B'} + \mathbb{P}^1_{1B''}
  \end{array}
\end{align*}
\caption{Yukawa couplings in the $SU(2)$-I top.}\label{tab:SU(2)-I-Yukawas}
\end{table}

To derive these, one first checks explicitly that none of the possible Yukawa couplings lies at $d_1=0$ so that the factorisations (\ref{eq:SU(2)-I-quadratic-curve}) and (\ref{eq:SU(2)-I-complicated-quadratic-curve}) are applicable.
The first two couplings arise over the intersection locus of the $\mathbf{2}^\mathrm{I}_1$- and $\mathbf{2}^\mathrm{I}_2$-curves. This locus splits into two sets of points, which can be identified with the intersection of $\{c_2\}$ with $\{\mathcal{C}_+\}$ and $\{d_1 \mathcal{C}_-\}$, respectively.
The intersection with $\{\mathcal{C}_+\}$ leads to the first set of Yukawa points, over which the fibre of the divisor $E_0$ splits into three components $\mathbb{P}^1_{0 s_1 C}$, $\mathbb{P}^1_{0AB}$ and $\mathbb{P}^1_{0AC}$. The intersection structure of these three components and $\mathbb{P}^1_1$ (which does not split) forms an affine $SU(4)$ diagram. 
The splitting of the fibre components  (\ref{tab:SU(2)-I-matter}) over the respective $\mathbf{2}$-curves arises as follows:
\begin{itemize}
  \item Approaching the Yukawa points along the $\mathbf{2}^\mathrm{I}_1$-curve, $\mathbb{P}^1_{0 s_1} \rightarrow \mathbb{P}^1_{0 s_1C}$ remains irreducible while $\mathbb{P}^1_{0A}$ splits into two components, $\mathbb{P}^1_{0AB} + \mathbb{P}^1_{0AC}$.
  \item From the perspective of the  $\mathbf{2}^\mathrm{I}_2$-curve the Yukawa points lie on the $\{\mathcal{C}_+\}$-part, where the fibre components $\mathbb{P}^1_{0B}$ and $\mathbb{P}^1_{0C}$ are defined in (\ref{eq:SU(2)-I-splitting-quadratic-curve}) with `+'-sign for $\mathbb{P}^1_{0B}$ and `$-$'-sign for $\mathbb{P}^1_{0C}$. When we approach the Yukawas by setting $c_2= 0$, $\mathbb{P}^1_{0B} \rightarrow \mathbb{P}^1_{0AB}$ remains irreducible while the equation for $\mathbb{P}^1_{0C}$ splits off a factor $s_1$, giving the splitting $\mathbb{P}^1_{0C} \rightarrow \mathbb{P}^1_{0 s_1 C} + \mathbb{P}^1_{0AC}$.
\end{itemize}

The second set of Yukawa points $W_2 \cap \{c_2\} \cap \{c_{1,0}\}$ can be viewed as the intersection of the $\{d_1 \mathcal{C}_-\}$-part of $\mathbf{2}^\mathrm{I}_2$ with $\mathbf{2}^\mathrm{I}_1$.
Again $E_0$ splits into three components, $\mathbb{P}^1_{0 s_1 B}$, $\mathbb{P}^1_{0AB'}$ and $\mathbb{P}^1_{0AC'}$ (the primes denote that these have different charges under the $SU(2)$ root and the $U(1)$ generators, corresponding to the conjugate $\mathbf{2}^\mathrm{I}_2$-state and a different singlet), which together with $\mathbb{P}^1_1$ form the affine $SU(4)$ diagram. The splitting processes are as follows:
\begin{itemize}
  \item Along $\mathbf{2}^\mathrm{I}_1$, again the component $\mathbb{P}^1_{0s_1}$ remains irreducible and $\mathbb{P}^1_{0A} \rightarrow \mathbb{P}^1_{0AB'} + \mathbb{P}^1_{0AC'}$ splits. The primes denote that the charge of the components under the $SU(2)$ root and the $U(1)$ generators are different than over the first Yukawa point.
  \item Along $\mathbf{2}^\mathrm{I}_2$, we have to look at the fibre components $\mathbb{P}^1_{0B}$ and $\mathbb{P}^1_{0C}$ over $\{\mathcal{C}_-\}$, which are defined through (\ref{eq:SU(2)-I-splitting-quadratic-curve}) with the second sign choice; setting $c_2=0$ now leaves $\mathbb{P}^1_{0C}$ irreducible, while the equation for $\mathbb{P}^1_{0B}$ just becomes $d_1\,s_1\,\su$, defining the components $\mathbb{P}^1_{0s_1 B}$ (where $s_1=0$) and $\mathbb{P}^1_{0AB'}$ (where $\su=0$).
\end{itemize}

The third and fourth sets of Yukawa points are the intersection points of the $\mathbf{2}^\mathrm{I}_1$-curve over $\{c_2\}$ with $\mathbf{2}^\mathrm{I}_3$ over $\{ d_1 \mathcal{D}_-\}$ and  $\{\mathcal{D}_+\}$, respectively. The splitting is straightforward and gives the same $\mathbb{P}^1$s with identical charges over both points. What differs, though, is the intersection \textit{pattern}. Over the third set, the pattern is $\mathbb{P}^1_{0s_1} - \mathbb{P}^1_{0A} - \mathbb{P}^1_{1A} - \mathbb{P}^1_{1B} (- \mathbb{P}^1_{0s_1})$; M2-branes can wrap the combination $\mathbb{P}^1_{0s_1}+\mathbb{P}^1_{1B}$ giving rise to $\overline{\mathbf{1}}^{(1)}$-states, but not $\mathbb{P}^1_{0s_1} + \mathbb{P}^1_{1A}$, which is needed for $\mathbf{1}^{(6)}$-states. Accordingly, the intersection pattern over the fourth set is $\mathbb{P}^1_{0s_1} - \mathbb{P}^1_{0A} - \mathbb{P}^1_{1B} - \mathbb{P}^1_{1A} (- \mathbb{P}^1_{0s_1})$, allowing ${\mathbf{1}}^{(6)}$- but not $\
\overline{\mathbf{1}}^{(1)}$-states. Of course, both patterns have the same structure as the affine $SU(4)$ diagram.

The fifth and sixth types of couplings arise over the intersection points between the $\mathbf{2}^\mathrm{I}_2$- and $\mathbf{2}^\mathrm{I}_3$-curve. With the factorisations (\ref{eq:SU(2)-I-quadratic-curve}) and (\ref{eq:SU(2)-I-complicated-quadratic-curve}), these points group into the four intersection loci of $\{\mathcal{C}_\pm\}$ and $\{\mathcal{D}_\pm\}$. Analogously to the situation over the previous two types of Yukawa points, one finds for both the fifth and sixth coupling the same $\mathbb{P}^1$s with the same charges, but different intersection patterns, leading to either $\overline{\mathbf{1}}^{(4)}$-states over $\{\mathcal{C}_\pm\} \cap \{\mathcal{D}_\pm\}$ or $\overline{\mathbf{1}}^{(6)}$-states over $\{\mathcal{C}_\pm\} \cap \{\mathcal{D}_\mp\}$. The intersection structure is an affine $SU(4)$ diagram in both cases.

The last two Yukawa points are self-intersection points of $\mathbf{2}^\mathrm{I}_2$ resp.~$\mathbf{2}^\mathrm{I}_3$. For $\mathbf{2}^\mathrm{I}_2$ it is the intersection point $W_2 \cap \{b_{2,0}\} \cap \{c_{1,0}\}$ of $\{\mathcal{C}_+\}$ with $\{\mathcal{C}_-\}$ which also lies on the singlet curve $\mathbf{1}^{(3)}$.\footnote{$\{\mathcal{C}_+\}$ and $\{\mathcal{C}_-\}$ also have the codimension-three locus $W_2 \cap \{b_1^2-4\,c_2\,d_1\} \cap \{c_1\,d_1-1/2\,b_1\,b_2\}$ in common, which is just the branch locus of the square root. However this set of points does not lie on any singlet curve; consistently there is no further enhancement in the fibre.} From the factorisation of the hypersurface equation (\ref{eq:SU(2)-I-splitting-quadratic-curve}) one easily sees that, irrespective of along which part we approach the point, $\mathbb{P}^1_{0B}$ remains irreducible, while the equation of $\mathbb{P}^1_{0C}$ splits off a factor $s_0$ as we set $b_{2,0}=0$, thus $\mathbb{P}^1_{0C} \
\rightarrow \mathbb{P}^1_{0Cs_0} + \mathbb{P}^1_{0C'}$. The last coupling is over 
the point of $\{\mathcal{D}_+\} \cap \{\mathcal{D}_-\}$ which lies on $\mathbf{1}^{(2)}$. The splitting process here is not obvious from (\ref{eq:SU(2)-I-splitting-complicated-quadratic-curve}), but straightforward calculation reveals that $\mathbb{P}^1_{1B}$ splits into two components. Again, the intersection structure is an affine $SU(4)$ diagram over both self-intersection points.

Note that, while for the discussion of the Yukawas above we have only used the loci of the $\mathbf{2}$-curves to determine the Yukawa points,  we have used \textsc{Singular} to verify that indeed all the Yukawa points (\ref{tab:SU(2)-I-Yukawas})  also lie on the corresponding singlet curve. This is consistent with the appearance of the associated singlet states in the split fibres as discussed above.

\subsection[\texorpdfstring{$SU(2)$}{SU(2)}-II and -III Tops]{\texorpdfstring{\boldmath $SU(2)$}{SU(2)}-II and -III Tops}\label{subsec:SU(2)-tops-short}

The analysis of the remaining two $A_1$-tops as depicted in figure \ref{fig:tops} is very analogous and is carried out in appendix \ref{SU2DetailsApp}, to which we refer for more details.
Here we merely collect the massless spectrum and the associated Yukawa couplings as these will be needed for our construction of Standard-Model-like F-theory compactifications.

\subsubsection*{\texorpdfstring{\boldmath $SU(2)$}{SU(2)}-II Top}

%Since we will need them  for our construction of Standard-Model-like F-theory compactifications, we here merely collect the spectra and Yukawa couplings:

The second $A_1$-top, called  $SU(2)$-II top in appendix \ref{SU2App1},
% \begin{align}\label{eq:SU(2)-II-coeffs}
 %  b_0 = b_{0,1} \, e_0, \quad b_2 = b_{2,0} \, e_1, \quad c_2 = c_{2,1} \, e_0 , \quad d_1 = d_{1,0} \, e_1 , \quad d_2 = d_{2,0} \, e_1. 
% \end{align}
leads to an $SU(2)$-charged matter spectrum of the following form:
\begin{align}\label{tab:SU(2)-II-matter-short}
  \begin{array}{c|c}
    \text{matter}   & U(1)-\text{charges} \\ \hline \rule{0pt}{3ex}
    \mathbf{2}^{\mathrm{II}}_1  & (\frac{1}{2}, \frac{3}{2}) \\[.5ex]\hline \rule{0pt}{3ex}
    \mathbf{2}^{\mathrm{II}}_2  & (\frac{1}{2},-\frac{1}{2}) \\[.5ex]\hline \rule{0pt}{3ex}
   \mathbf{2}^\mathrm{II}_3 &   (\frac{1}{2}, \frac{1}{2}) \\
    \end{array}
\end{align}
All Yukawa couplings allowed by the $U(1)$ charges are realised geometrically.  More precisely the set of Yukawas is given by
\begin{align}
\begin{split}
&  \mathbf{2}_1^{\mathrm{II}} - \mathbf{2}^{\mathrm{II}}_2 - \overline{\mathbf{1}}^{(4)}, \qquad 
   \mathbf{2}^{\mathrm{II}}_1 - \overline{\mathbf{2}}^{\mathrm{II}}_2 - \overline{\mathbf{1}}^{(5)}, \qquad 
   \mathbf{2}^{\mathrm{II}}_1 - \mathbf{2}^{\mathrm{II}}_3 - \overline{\mathbf{1}}^{(3)},  \qquad 
  \mathbf{2}^{\mathrm{II}}_1 - \overline{\mathbf{2}}^{\mathrm{II}}_3 - \overline{\mathbf{1}}^{(6)}, \\
&  \mathbf{2}^{\mathrm{II}}_2 - \mathbf{2}^{\mathrm{II}}_3 - \overline{\mathbf{1}}^{(2)}, \qquad 
       \mathbf{2}^{\mathrm{II}}_2 - \overline{\mathbf{2}}^{\mathrm{II}}_3 - \mathbf{1}^{(6)},  \qquad 
         \mathbf{2}^{\mathrm{II}}_2 - \mathbf{2}^{\mathrm{II}}_2 - \overline{\mathbf{1}}^{(1)}, \qquad 
          \mathbf{2}^{\mathrm{II}}_3 - \mathbf{2}^{\mathrm{II}}_3 - \overline{\mathbf{1}}^{(4)}. 
\end{split}
\end{align}

In fact the $SU(2)$-II top is equivalent to the $SU(2)$-I top. On way to see this is to notice that upon identifying the $U(1)$ charges in the two tops as
\begin{align}\label{eq:U1-charge-transformation}
\begin{split}
U(1)_1^{\rm II} &= -U(1)_1^{\rm I} \, , \\
U(1)_2^{\rm II} &= U(1)_2^{\rm I} - U(1)_1^{\rm I}  \, ,
\end{split}
\end{align}
the spectrum and Yukawa structure is exactly the same if one identifies the states ${\bf 2}^{\rm II}_i \leftrightarrow \overline{\bf 2}^{\rm I}_i, i=1,2,3$ and exchanges the singlets ${\bf 1}^{(1)} \leftrightarrow \overline{\bf 1}^{(3)}, {\bf 1}^{(2)} \leftrightarrow \overline{\bf 1}^{(4)}$. One can also arrive at this identification from the symmetries of the tops. More details can be found in appendix \ref{app:tops}.

Although both models are the same when considering only the gauge group $SU(2) \times U(1)_1 \times U(1)_2$, they will give rise to different models when combining them with an $SU(3)$-top, as we will discuss below in section \ref{sec_3211}.

\subsubsection*{\texorpdfstring{\boldmath $SU(2)$}{SU(2)}-III Top}

The last $A_1$-top is the $SU(2)$-III top with matter content 
\begin{align}\label{tab:SU(2)-III-matter-short}
  \begin{array}{c|c}
    \text{matter}  & U(1)-\text{charges} \\ \hline \rule{0pt}{3ex}
    \mathbf{2}^{\mathrm{III}}_1  & (1,0) \\ \rule{0pt}{3ex}
    \mathbf{2}^{\mathrm{III}}_2  & (1,1)\\ \rule{0pt}{3ex}
    \mathbf{2}^{\mathrm{III}}_3 & (0,1)
  \end{array}
\end{align}

In this top, in addition to the three fundamental matter curves and a notorious type $III$ enhancement locus with no additional matter, one finds a change of fibre type to \emph{non-split $I_3$} type over yet another curve. As explained in appendix \ref{SU2DetailsApp}, the non-split fibre type can either be seen from the Weierstrass data or be explicitly confirmed by analysing the monodromies along the curve in question. As a result of the monodromy, this locus does not carry massless matter.

The geometrically realised  Yukawa couplings
\begin{align}
\begin{split}
& \mathbf{2}^{\mathrm{III}}_1 - \overline{\mathbf{2}}^{\mathrm{III}}_2 - \mathbf{1}^{(6)}, \qquad 
    \mathbf{2}^{\mathrm{III}}_1 - \mathbf{2}^{\mathrm{III}}_3 - \overline{\mathbf{1}}^{(4)}, \qquad  
      \mathbf{2}^{\mathrm{III}}_1 - \overline{\mathbf{2}}^{\mathrm{III}}_3 - \overline{\mathbf{1}}^{(1)}, \\
& \mathbf{2}^{\mathrm{III}}_2 - \mathbf{2}^{\mathrm{III}}_3 - \overline{\mathbf{1}}^{(3)},  \qquad  
      \mathbf{2}^{\mathrm{III}}_2 - \overline{\mathbf{2}}^{\mathrm{III}}_3 - \overline{\mathbf{1}}^{(2)}, \qquad  
           \mathbf{2}^{\mathrm{III}}_3 - \mathbf{2}^{\mathrm{III}}_3 - \overline{\mathbf{1}}^{(5)} 
\end{split}           ´
\end{align}
exhaust again all gauge invariant combinations.

This top is inequivalent to the first two tops. Under the transformation $U(1)'_1 \equiv - U(1)_1, \, U(1)'_2 \equiv U(1)_2 - U(1)_1$ analogous to 
(\ref{eq:U1-charge-transformation}) (with the same identification of the singlets), the spectrum and Yukawa structure of the $SU(2)$-III top is mapped to itself as ${\bf 2}^{\rm III}_1 \leftrightarrow \overline{\bf 2}^{\rm III}_2$ and ${\bf 1}^{(1)} \leftrightarrow \overline{\bf 1}^{(3)}, {\bf 1}^{(2)} \leftrightarrow \overline{\bf 1}^{(4)}$.

%% file: section4_models_with_SU3_v3.tex
\section[Toric Fibrations with Additional \texorpdfstring{\boldmath $SU(3)$}{SU(3)} Symmetry]{Toric Fibrations with Additional \texorpdfstring{\boldmath $SU(3)$}{SU(3)} Symmetry}\label{sec:SU(3)-tops}

The construction of $SU(3)$ gauge symmetry via tops is analogous to the $SU(2)$ cases.
The resolution of the $A_2$-singularity over a divisor $W_3: w_3=0$ in the base introduces  three toric divisors $F_0, F_1, F_2$ given by the vanishing locus of the coordinates $f_0, f_1, f_2$.
 Each $F_i$ is a  $\mathbb{P}^1$-fibration over $W_3$, and $\pi^{-1} W_3 = F_0 F_1 F_2$. Over a generic point on $W_3$ the intersection structure of the $\mathbb{P}^1$-fibres reproduces the affine $SU(3)$ diagram. We choose the root assignment $F_1 \leftrightarrow -\alpha_1, \, F_2 \leftrightarrow -\alpha_2, \, F_0 \leftrightarrow \alpha_1+\alpha_2$. There exist three $SU(3)$ tops, which we will now present in detail.

\subsection[\textit{SU}(3)-A Top]{\texorpdfstring{{\boldmath $SU(3)$}}{SU(3)}-A Top}

The first top corresponds to the following restriction of the hypersurface coefficients,
\begin{align}\label{eq:SU(3)-A-coeffs}
  \begin{split}
    b_0 &= b_{0,1}\,f_0 ,\quad b_2 = b_{2,0}\,f_1\,f_2 ,\quad c_1 = c_{1,0}\,f_2 ,\quad c_2 = c_{2,1}\,f_0\,f_2 , \\
    d_0 &= d_{0,1}\,f_0\,f_1 ,\quad d_1 = d_{1,0}\,f_1 ,\quad d_2 = d_{2,1}\,f_0\,f_1^2 ,
  \end{split}
\end{align}
where only $b_1$ remains unchanged. Out of the four different triangulations we choose the one whose SR-ideal is generated by  
\begin{align}\label{eq:SU(3)-A-SR-ideal}
  \su\,\sv , \su\,\sw , \sw\,s_0 , \sv\,s_1 , s_0\,s_1 , f_0\,\sw , f_0\,s_1 , f_1\,s_0 , f_1\,\sv , f_2\,s_0 , f_2\,s_1 , f_2\,\su .
\end{align}

The coordinates and their corresponding divisor classes are summarised in the following table:
\begin{align}\label{tab:SU(3)-A-divisor-classes}
  \begin{array}{c|ccccccc|c}
    \hphantom{U} & \su & \sv & \sw & s_0 & s_1 & f_1 & f_2 & f_0 \\ \hline
    \mathrm{U} & 1 & 1 & 1 & \cdot & \cdot & \cdot & \cdot & \cdot \\
    S_0 & \cdot & \cdot & 1 & 1& \cdot & \cdot & \cdot & \cdot \\
    S_1 & \cdot & 1 & \cdot & \cdot & 1 & \cdot & \cdot & \cdot \\
    F_1 & \cdot & 1 & \cdot & \cdot & \cdot & 1 & \cdot & -1 \\
    F_2 & \cdot & \cdot & -1 & \cdot & \cdot & \cdot & 1 & -1
  \end{array}
\end{align}

The $U(1)$ generators (\ref{eq:U(1)-generators}) need to be corrected such that the roots of $SU(3)$ have zero $U(1)$ charge. The generators with this property take the form
\begin{align}\label{eq:SU(3)-A-U(1)-generators}
  \begin{split}
    \omega_1^\text{A} &= S_1 - S_0 - \overline{\mathcal{K}} + \frac{2}{3} F_1 + \frac{1}{3} F_2 ,\\
    \omega_2^\text{A} &= U - S_0 - \overline{\mathcal{K}} - [c_{1,0}] + \frac{2}{3} F_1 + \frac{1}{3} F_2.
  \end{split}
\end{align}
The charges (\ref{tab:singlets-charges}) of the singlets are not affected as they are not charged under the roots of $SU(3)$.

\subsubsection*{Matter curves}

The matter curves are found by analysing the discriminant of the associated Weierstrass model
\bea
\Delta \simeq w^3 \Big(   b_{0,1}   \,  c_{1,0} \,   (b_{0,1}\,c_{1,0} - b_1\,c_{2,1}) \, (b_1\,b_{2,0} - c_{1,0}\,d_{1,0}) \,   (b_{0,1}\,d_{1,0}^2 - b_1\,d_{0,1}\,d_{1,0} + b_1^2\,d_{2,1} ) \, b_{1}^3    + {\cal O}(w_3) \Big). \nonumber
\eea
As can be read off from the Weierstrass sections $f$ and $g$, the fibre along the five curves associated with the first five factors in the bracket is of split Kodaira type $I_4$. The fibre components intersect as in the affine $SU(4)$ diagram, corresponding to $SU(3)$ matter in the fundamental representation (plus conjugate). Their $U(1)$ charges together with the geometric data can be found in table \ref{tab:SU(3)-A-matter}.

\begin{table}[ht]
\begin{center}
  \begin{tabular}{c|c|c|c|l}
  \multirow{2}{*}{matter} & \multirow{2}{*}{$\text{locus} = W_3 \cap \ldots$} & splitting of fibre & $U(1)-$ & \multirow{2}{*}{highest weight states of...}\\
    &  & components & charges & \\ \hline \rule{0pt}{3ex}
    $\mathbf{3}^{\mathrm{A}}_1$ & $\{b_{0,1}\}$ & $\mathbb{P}^1_1 \rightarrow \mathbb{P}^1_{1\sw} + \mathbb{P}^1_{1A}$ & $(\frac{2}{3}, -\frac{1}{3})$ & $\mathbf{3} \! : \mathbb{P}^1_{1 \sw} + \mathbb{P}^1_0 + \mathbb{P}^1_2, \, \overline{\mathbf{3}} \! : \mathbb{P}^1_{1A} + \mathbb{P}^1_0 $\\ \rule{0pt}{3ex}
   $ \mathbf{3}^{\mathrm{A}}_2$ & $\{c_{1,0}\}$ & $\mathbb{P}^1_0 \rightarrow \mathbb{P}^1_{0\su} + \mathbb{P}^1_{0A}$ & $(-\frac{1}{3}, -\frac{4}{3})$ & $\mathbf{3} \! : \mathbb{P}^1_{0 \su} , \, \overline{\mathbf{3}} \! : \mathbb{P}^1_{0 A}$\\ \rule{0pt}{3ex}
   $\mathbf{3}^{\mathrm{A}}_3$ & $\{b_{0,1}\,c_{1,0} - b_1\,c_{2,1}\}$ & $\mathbb{P}^1_1 \rightarrow  \mathbb{P}^1_{1B} + \mathbb{P}^1_{1C}$ & $(-\frac{1}{3}, \frac{2}{3})$ & $\mathbf{3} \! : \mathbb{P}^1_{1C} + \mathbb{P}^1_{0} + \mathbb{P}^1_{2} , \, \overline{\mathbf{3}} \!: \mathbb{P}^1_{1 B} + \mathbb{P}^1_0$\\ \rule{0pt}{3ex}
    $\mathbf{3}^{\mathrm{A}}_4$ & $\{b_1\,b_{2,0} - c_{1,0}\,d_{1,0}\}$ & $\mathbb{P}^1_0 \rightarrow \mathbb{P}^1_{0B} + \mathbb{P}^1_{0C}$ & $(\frac{2}{3}, \frac{2}{3})$ & $\mathbf{3} \! : \mathbb{P}^1_{0B} , \, \overline{\mathbf{3}} \! : \mathbb{P}^1_{0C}$ \\ \rule{0pt}{3ex}
    \multirow{2}{*}{$\mathbf{3}^{\mathrm{A}}_5$} & $\{b_{0,1}\,d_{1,0}^2 - b_1\,d_{0,1}\,d_{1,0}$ & \multirow{2}{*}{$\mathbb{P}^1_2 \rightarrow \mathbb{P}^1_{2A} + \mathbb{P}^1_{2B}$} & \multirow{2}{*}{$(-\frac{1}{3},-\frac{1}{3})$} & \multirow{2}{*}{$\mathbf{3} \! : \mathbb{P}^1_{2A} + \mathbb{P}^1_0 , \, \overline{\mathbf{3}} \! : \mathbb{P}^1_{2B} + \mathbb{P}^1_0 + \mathbb{P}^1_1$}\\ \rule{0pt}{3ex}
    &  $+ b_1^2\,d_{2,1}\}$ & & &
  \end{tabular}
\caption{Matter states in the $SU(3)$-A top.}
\label{tab:SU(3)-A-matter}
\end{center}
\end{table}

The splitting process over the first four curves can be straightforwardly verified. For the fifth curve, we proceed as for the $SU(2)$ tops und use expressions with square roots to split the curve into two parts; because $d_{2,1}\neq0$ for all the Yukawa points below, we factorise the quadratic term such that we can solve for $b_1$ (analogously to e.g.~(\ref{eq:SU(2)-I-quadratic-curve}), where $c_{1,0}$ played the role of $b_1$ here). Inserting the resulting expressions (one for each of the two parts of the curve) for $b_1$ into the hypersurface equation, we find that $\mathbb{P}^1_2$ splits into two components on both parts of the curve; similarly to the $SU(2)$-tops, there is no monodromy interchanging the components when passing from one part of the curve to the other and back.

In addition, over the curve $\{w_3\} \cap \{b_{1}\}$ the Weierstrass data $(f,g,\Delta)$ vanish to orders $(2,2,4)$. This indicates a Kodaira type $IV$ fibre in which no extra $\mathbb P^1$ splits off, but rather the three fibre components intersect in a single point, as can be checked explicitly. Over this curve, no extra matter representation arises.

\subsubsection*{Yukawa Points}

The $SU(3)$ matter and the singlet curves intersect at certain codimension-three loci in the base to form gauge invariant Yukawa couplings $\mathbf{3}-\overline{\mathbf{3}}-\mathbf{1}/\overline{\mathbf{1}}$. In this case the fibre is enhanced to the affine $SU(5)$-diagram, and the realised couplings are in 1-to-1 correspondence with the gauge theoretic selection rules.
In addition one can also form gauge invariant couplings of the type $\mathbf{3}-\mathbf{3}-\mathbf{3}$. As it turns out all such gauge invariant couplings are indeed realised geometrically, and  the fibre structure is the affine $SO(8)$-diagram, cf.~table \ref{tab:SU(3)-A-Yukawas}.

\begin{table}[ht]
\begin{align*}
  \begin{array}{c|c|c}
    \text{coupling} & \text{locus}=W_3 \cap \ldots & \text{splitting of fibre components} \\ \hline \rule{0pt}{3ex}
    \mathbf{3}^{\mathrm{A}}_1 - \overline{\mathbf{3}}^{\mathrm{A}}_2 - \overline{\mathbf{1}}^{(4)} & \{b_{0,1}\} \cap \{c_{1,0}\} & \mathbb{P}^1_0 \rightarrow \mathbb{P}^1_{0\su} + \mathbb{P}^1_{0A}, \, \mathbb{P}^1_1 \rightarrow \mathbb{P}^1_{1\sw}+\mathbb{P}^1_{1A} \\ \hline \rule{0pt}{3ex}
    \mathbf{3}^{\mathrm{A}}_1 - \overline{\mathbf{3}}^{\mathrm{A}}_3 - \mathbf{1}^{(1)} & \{b_{0,1}\} \cap \{c_{2,1}\} & \mathbb{P}^1_1 \rightarrow \mathbb{P}^1_{1\sw C} + \mathbb{P}^1_{1AB}\! +\mathbb{P}^1_{1AC} \\ \hline \rule{0pt}{3ex}
    \mathbf{3}^{\mathrm{A}}_1 - \overline{\mathbf{3}}^{\mathrm{A}}_4 - \mathbf{1}^{(6)} & \{b_{0,1}\} \cap \{b_1\,b_{2,0} - c_{1,0}\,d_{1,0}\} & \mathbb{P}^1_0 \rightarrow \mathbb{P}^1_{0B} + \mathbb{P}^1_{0C}, \, \mathbb{P}^1_1 \rightarrow \mathbb{P}^1_{1\sw}+\mathbb{P}^1_{1A} \\ \hline \rule{0pt}{3ex}
    \mathbf{3}^{\mathrm{A}}_1 - \overline{\mathbf{3}}^{\mathrm{A}}_5 - \overline{\mathbf{1}}^{(2)} & \{b_{0,1}\} \cap \{b_1\,d_{2,1}-d_{0,1}\,d_{1,0}\} & \mathbb{P}^1_1 \rightarrow \mathbb{P}^1_{1\sw} + \mathbb{P}^1_{1D}, \, \mathbb{P}^1_2 \rightarrow \mathbb{P}^1_{2A}+\mathbb{P}^1_{2B} \\ \hline \rule{0pt}{3ex}
    \mathbf{3}^{\mathrm{A}}_2 - \overline{\mathbf{3}}^{\mathrm{A}}_3 - \mathbf{1}^{(5)} & \{c_{1,0}\} \cap \{c_{2,0}\} & \mathbb{P}^1_0 \rightarrow \mathbb{P}^1_{0\su} + \mathbb{P}^1_{0A}, \, \mathbb{P}^1_1 \rightarrow \mathbb{P}^1_{1B}+\mathbb{P}^1_{1C} \\ \hline \rule{0pt}{3ex}
    \mathbf{3}^{\mathrm{A}}_2 - \overline{\mathbf{3}}^{\mathrm{A}}_4 - \mathbf{1}^{(3)} & \{c_{1,0}\} \cap \{b_{2,0}\} & \mathbb{P}^1_0 \rightarrow \mathbb{P}^1_{0\su A} \! + \mathbb{P}^1_{0AB} +\mathbb{P}^1_{0AC} \\ \hline \rule{0pt}{3ex}
    \mathbf{3}^{\mathrm{A}}_2 - \overline{\mathbf{3}}^{\mathrm{A}}_5 - \mathbf{1}^{(6)} & \{c_{1,0}\} \cap \left(\text{$\mathbf{3}^\mathrm{A}_5$-locus}\right)  & \mathbb{P}^1_0 \rightarrow \mathbb{P}^1_{0\su} + \mathbb{P}^1_{0A}, \, \mathbb{P}^1_2 \rightarrow \mathbb{P}^1_{2A}+\mathbb{P}^1_{2B} \\ \hline \rule{0pt}{3ex}
    \mathbf{3}^{\mathrm{A}}_3 - \overline{\mathbf{3}}^{\mathrm{A}}_4 - \mathbf{1}^{(2)} & \left(\mathbf{3}^\mathrm{A}_3 \right) \cap \left( \mathbf{3}^\mathrm{A}_4 \right) \setminus \left(\{c_{1,0}\} \cap \{b_1\} \right) & \mathbb{P}^1_0 \rightarrow \mathbb{P}^1_{0B} + \mathbb{P}^1_{0C} , \, \mathbb{P}^1_1 \rightarrow \mathbb{P}^1_{1B} + \mathbb{P}^1_{1C} \\ \hline \rule{0pt}{3ex}
    \mathbf{3}^{\mathrm{A}}_3 - \overline{\mathbf{3}}^{\mathrm{A}}_5 - \overline{\mathbf{1}}^{(6)} & \left(\mathbf{3}^\mathrm{A}_3\right) \cap \left(\mathbf{3}^\mathrm{A}_5\right) \! \setminus \! \left( \vphantom{\mathbf{3}^\mathrm{A}_5} \{b_{0,1}\} \cap \{b_1\} \right) & \mathbb{P}^1_1 \rightarrow \mathbb{P}^1_{1B} + \mathbb{P}^1_{1C}, \, \mathbb{P}^1_2 \rightarrow \mathbb{P}^1_{2A}+\mathbb{P}^1_{2B} \\ \hline \rule{0pt}{3ex}
    \mathbf{3}^{\mathrm{A}}_4 - \overline{\mathbf{3}}^{\mathrm{A}}_5 - \overline{\mathbf{1}}^{(4)} & \left(\mathbf{3}^\mathrm{A}_4\right) \cap \left(\mathbf{3}^\mathrm{A}_5\right) \! \setminus \! \left( \vphantom{\mathbf{3}^\mathrm{A}_5} \{d_{1,0}\} \cap \{b_1\} \right) & \mathbb{P}^1_0 \rightarrow \mathbb{P}^1_{0B} + \mathbb{P}^1_{0C}, \, \mathbb{P}^1_2 \rightarrow \mathbb{P}^1_{2A}+\mathbb{P}^1_{2B} \\ \hline \hline  \rule{0pt}{3ex}
    \mathbf{3}^{\mathrm{A}}_1 - \mathbf{3}^{\mathrm{A}}_3 - \mathbf{3}^{\mathrm{A}}_5 & \{b_{0,1}\} \cap \{b_1\} & \mathbb{P}^1_1 \rightarrow \mathbb{P}^1_{1\sw B} + \mathbb{P}^1_{1AB'} + \mathbb{P}^1_{AC'}, \, \mathbb{P}^1_2 \rightarrow \mathbb{P}^1_{2A} + \mathbb{P}^1_{2B} , \\ 
    & & \mathbb{P}^1_{AB'}=\mathbb{P}^1_{2A} \\ \hline \rule{0pt}{3ex}
    \mathbf{3}^{\mathrm{A}}_2 - \mathbf{3}^{\mathrm{A}}_3 - \mathbf{3}^{\mathrm{A}}_4 & \{c_{1,0}\} \cap \{b_1\} & \mathbb{P}^1_0 \rightarrow \mathbb{P}^1_{0\su C} + \mathbb{P}^1_{0AB'} + \mathbb{P}^1_{0AC'}, \, \mathbb{P}^1_1 \rightarrow \mathbb{P}^1_{1B} + \mathbb{P}^1_{1C} , \\ 
    & & \mathbb{P}^1_{0AC} = \mathbb{P}^1_{1C} \\ \hline \rule{0pt}{3ex}
    \mathbf{3}^{\mathrm{A}}_4 - \mathbf{3}^{\mathrm{A}}_5 - \mathbf{3}^{\mathrm{A}}_5 & \{d_{1,0}\} \cap \{b_1\} & \mathbb{P}^1_0 \rightarrow \mathbb{P}^1_{0B} + \mathbb{P}^1_{0C}, \, \mathbb{P}^1_2 \rightarrow \mathbb{P}^1_{2A} + \mathbb{P}^1_{2B0} + \mathbb{P}^1_{2B'} , \\
    & & \mathbb{P}^1_{0B} = \mathbb{P}^1_{2B0}
  \end{array}
\end{align*}
\caption{Yukawa couplings in the $SU(3)$-A top.}
\label{tab:SU(3)-A-Yukawas}
\end{table}

There, in the last column, the subscripts should help to visualise $\mathbb{P}^1$s' splitting process. E.g.~if we approach the second Yukawa point along the $\mathbf{3}^\mathrm{A}_1$-curve, then we find that $\mathbb{P}^1_{1\sw}$ remains a single component and $\mathbb{P}^1_{1A}$ splits into two, $\mathbb{P}^1_{1AB}$ and $\mathbb{P}^1_{1AC}$; $\mathbb{P}^1_{1AB}$ is the component $\mathbb{P}^1_{1B}$ from the $\mathbf{3}^\mathrm{A}_3$-curve that does not split over the Yukawa point, while $\mathbb{P}^1_{1C}$ splits into two components, of which one is identified with $\mathbb{P}^1_{1AC}$ and the other one coincides with $\mathbb{P}^1_{1\sw}$, hence the notation $\mathbb{P}^1_{1\sw C}$.

Over the last three Yukawa points with the coupling type $\mathbf{3}-\mathbf{3}-\mathbf{3}$, two of the three divisors $F_{0,1,2}$ each split off the same $\mathbb{P}^1$-component, which therefore is a multiplicity 2 component; this corresponds to the central node of the affine $SO(8)$-diagram with dual Coxeter label 2.

\subsection[\texorpdfstring{$SU(3)$}{SU(3)}-B and -C Top]{\texorpdfstring{\boldmath $SU(3)$}{SU(3)}-B and -C Top}

Let us briefly list the main results from the analysis of the remaining two $SU(3)$ tops, with more details relegated to appendix \ref{app-SU3}.

\subsubsection*{\texorpdfstring{\boldmath $SU(3)$}{SU(3)}-B Top}

The $SU(3)$-B top gives rise to five ${\bf 3}$-matter curves with the following $U(1)$ charges:
\begin{align}\label{tab:SU(3)-B-matter-short}
  \begin{array}{c|c|| c | c}
    \text{matter} & U(1)-\text{charges}& \text{matter} & U(1)-\text{charges}  \\ \hline \rule{0pt}{3ex}
    \mathbf{3}^{\mathrm{B}}_1  & (-\frac{2}{3}, -\frac{4}{3}) & \mathbf{3}^{\mathrm{B}}_4 &(\frac{1}{3}, \frac{2}{3})   \\ \rule{0pt}{3ex}
    \mathbf{3}^{\mathrm{B}}_2  & (-\frac{2}{3}, \frac{2}{3})  &     \mathbf{3}^{\mathrm{B}}_5  &(\frac{1}{3},-\frac{1}{3})   \\ \rule{0pt}{3ex}
    \mathbf{3}^{\mathrm{B}}_3 & (-\frac{2}{3}, -\frac{1}{3}) & & %
 %   \mathbf{3}^{\mathrm{B}}_4 & (\frac{1}{3}, \frac{2}{3}) & & \\ \rule{0pt}{3ex}
%
%    \mathbf{3}^{\mathrm{B}}_5  & (\frac{1}{3},-\frac{1}{3})
  \end{array}
\end{align}

These enjoy a rich spectrum of geometrically realised Yukawa couplings with the singlets,  
\begin{align}
\begin{split}
&  \mathbf{3}^{\mathrm{B}}_1 - \overline{\mathbf{3}}^{\mathrm{B}}_2 - \mathbf{1}^{(5)}, \qquad 
   \mathbf{3}^{\mathrm{B}}_1 - \overline{\mathbf{3}}^{\mathrm{B}}_3 - \mathbf{1}^{(6)}, \qquad 
    \mathbf{3}^{\mathrm{B}}_1 - \overline{\mathbf{3}}^{\mathrm{B}}_4 - \mathbf{1}^{(3)}, \qquad 
  \mathbf{3}^{\mathrm{B}}_1 - \overline{\mathbf{3}}^{\mathrm{B}}_5 - \mathbf{1}^{(4)},   \\ 
 &  \mathbf{3}^{\mathrm{B}}_2 - \overline{\mathbf{3}}^{\mathrm{B}}_3 - \overline{\mathbf{1}}^{(6)}, \qquad  
     \mathbf{3}^{\mathrm{B}}_2 - \overline{\mathbf{3}}^{\mathrm{B}}_4 - \mathbf{1}^{(2)}, \qquad  
  \mathbf{3}^{\mathrm{B}}_2 - \overline{\mathbf{3}}^{\mathrm{B}}_5 - \mathbf{1}^{(1)}, \qquad 
   \mathbf{3}^{\mathrm{B}}_3 - \overline{\mathbf{3}}^{\mathrm{B}}_4 - \mathbf{1}^{(4)}, \\ 
&   \mathbf{3}^{\mathrm{B}}_3 - \overline{\mathbf{3}}^{\mathrm{B}}_5 - \mathbf{1}^{(2)}, \qquad 
   \mathbf{3}^{\mathrm{B}}_4 - \overline{\mathbf{3}}^{\mathrm{B}}_5 - \overline{\mathbf{1}}^{(6)} \qquad
\end{split}
\end{align}
  and among one another,
  \bea
   \mathbf{3}^{\mathrm{B}}_3 - \mathbf{3}^{\mathrm{B}}_4 - \mathbf{3}^{\mathrm{B}}_5, \qquad   
   \mathbf{3}^{\mathrm{B}}_1 - \mathbf{3}^{\mathrm{B}}_4 - \mathbf{3}^{\mathrm{B}}_4, \qquad 
   \mathbf{3}^{\mathrm{B}}_2 - \mathbf{3}^{\mathrm{B}}_5 - \mathbf{3}^{\mathrm{B}}_5,
\eea
which is in 1-to-1 correspondence with the gauge theoretic selection rules.

Under the transformation $U(1)'_1 \equiv -U(1)_1, \, U(1)'_2 \equiv U(1)_2 - U(1)_1$, the spectrum and Yukawa couplings remains invariant with the identification ${\bf 3}^{\rm B}_1 \leftrightarrow \overline{\bf 3}^{\rm B}_2, \, {\bf 3}^{\rm B}_4 \leftrightarrow \overline{\bf 3}^{\rm B}_5$ and ${\bf 1}^{(1)} \leftrightarrow \overline{\bf 1}^{(3)} , \, {\bf 1}^{(2)} \leftrightarrow \overline{\bf 1}^{(4)}$.

\subsubsection*{\texorpdfstring{\boldmath $SU(3)$}{SU(3)}-C Top}

The third top $SU(3)$-C gives rise to $SU(3)$-charged states with the following $U(1)$ charges:

\begin{align}\label{tab:SU(3)-C-matter-short}
  \begin{array}{c|c||c|c}
    \text{matter}  & U(1)-\text{charges} &     \text{matter}  & U(1)-\text{charges}  \\ \hline \rule{0pt}{3ex}
    \mathbf{3}^{\mathrm{C}}_1  & (-\frac{2}{3}, -1) &  \mathbf{3}^{\mathrm{C}}_4  & (-\frac{2}{3}, 0)  \\ \rule{0pt}{3ex}
    \mathbf{3}^{\mathrm{C}}_2 & (\frac{1}{3}, -1)&   \mathbf{3}^{\mathrm{C}}_5  & (\frac{1}{3},0) \\ \rule{0pt}{3ex}
    \mathbf{3}^{\mathrm{C}}_3 & (\frac{1}{3}, 1) 
%
 %    \mathbf{3}^{\mathrm{C}}_4  & (-\frac{2}{3}, 0) \\ \rule{0pt}{3ex}
%
%     \mathbf{3}^{\mathrm{C}}_5  & (\frac{1}{3},0)
  \end{array}
\end{align}
The couplings
\begin{align}
\begin{split}
& \mathbf{3}^{\mathrm{C}}_1 - \overline{\mathbf{3}}^{\mathrm{C}}_2 - \mathbf{1}^{(2)}, \qquad 
  \mathbf{3}^{\mathrm{C}}_1 - \overline{\mathbf{3}}^{\mathrm{C}}_3 - \mathbf{1}^{(3)} , \qquad 
    \mathbf{3}^{\mathrm{C}}_1 - \overline{\mathbf{3}}^{\mathrm{C}}_4 - \mathbf{1}^{(6)} , \qquad 
      \mathbf{3}^{\mathrm{C}}_1 - \overline{\mathbf{3}}^{\mathrm{C}}_5 - \mathbf{1}^{(4)},  \\ 
&    \mathbf{3}^{\mathrm{C}}_2 - \overline{\mathbf{3}}^{\mathrm{C}}_3 - \mathbf{1}^{(5)}, \qquad 
      \mathbf{3}^{\mathrm{C}}_2 - \overline{\mathbf{3}}^{\mathrm{C}}_4 - \overline{\mathbf{1}}^{(1)} , \qquad 
      \mathbf{3}^{\mathrm{C}}_2 - \overline{\mathbf{3}}^{\mathrm{C}}_5 - \mathbf{1}^{(6)} , \qquad 
      \mathbf{3}^{\mathrm{C}}_3 - \overline{\mathbf{3}}^{\mathrm{C}}_4 - \overline{\mathbf{1}}^{(4)},\\
&        \mathbf{3}^{\mathrm{C}}_3 - \overline{\mathbf{3}}^{\mathrm{C}}_5 - \overline{\mathbf{1}}^{(6)}, \qquad 
        \mathbf{3}^{\mathrm{C}}_4 - \overline{\mathbf{3}}^{\mathrm{C}}_5 - \mathbf{1}^{(2)}
\end{split}
\end{align}
and      
      \bea
        \mathbf{3}^{\mathrm{C}}_1 - \mathbf{3}^{\mathrm{C}}_3 - \mathbf{3}^{\mathrm{C}}_5, \qquad 
          \mathbf{3}^{\mathrm{C}}_2 - \mathbf{3}^{\mathrm{C}}_3 - \mathbf{3}^{\mathrm{C}}_4, \qquad 
            \mathbf{3}^{\mathrm{C}}_4 - \mathbf{3}^{\mathrm{C}}_5 - \mathbf{3}^{\mathrm{C}}_5
          \eea
are in agreement with gauge theoretic expectations.

Similar to the situation with the $SU(2)$ tops, the $SU(3)$-C top is in fact equivalent to the $SU(3)$-A top (cf.~appendix \ref{app:tops}). Analogous to the $U(1)$ charges identification (\ref{eq:U1-charge-transformation}), we find with
\begin{align}\label{eq:U1-charge-transformation-SU3}
\begin{split}
	U(1)_1^{\rm C} &= -U(1)_1^{\rm A} \, , \\
U(1)_2^{\rm C} &= U(1)_2^{\rm A} - U(1)_1^{\rm A} 
\end{split}
\end{align}
that the spectrum agrees by identifying the states ${\bf 3}^{\rm A}_i \leftrightarrow {\bf 3}^{\rm C}_i, i=1,...,5$ and ${\bf 1}^{(1)} \leftrightarrow \overline{\bf 1}^{(3)}, {\bf 1}^{(2)} \leftrightarrow \overline{\bf 1}^{(4)}$. Again one needs both the $SU(3)$-A and -C top to construct all inequivalent toric $SU(3) \times SU(2) \times U(1)_1 \times U(1)_2$ models. 

%% file: section5_toric_SU3xSU2_v3.tex
\section[Toric \texorpdfstring{\boldmath $SU(3) \times SU(2) \times U(1) \times U(1)$}{SU(3) x SU(2) x U(1) x U(1)} Realisations]{Toric \texorpdfstring{\boldmath $SU(3) \times SU(2) \times U(1)_1 \times U(1)_2$}{SU(3)xSU(2)xU(1)xU(1)} Realisations} \label{sec_3211}

%\subsection{\texorpdfstring{{Models with \boldmath $SU(3) \times SU(2) \times U(1)_1 \times U(1)_2$ Symmetry}}{Models with SU(3) x SU(2) x U(1) x U(1) symmetry}}

%To construct a F-theory model with non-abelian gauge symmetry with group $SU(3) \times SU(2)$, we follow the spirit of type IIB constructions \TODO{Referenz!} where one realises each simple factor of the gauge group on a different stack of 7-branes such that we find the bi-fundamental matter at the intersection of the stacks. 

We are now ready to construct F-theory compactifications with gauge group $SU(3) \times SU(2)$, together with two abelian factors. 
To this end we start with our elliptic fibration realised as the hypersurface (\ref{eq:hypersurface-equation}) within a $\text{Bl}_2 \mathbb{P}^2$-fibration over ${\cal B}$ and realise an $SU(2)$ and an $SU(3)$ singularity over two independent base divisors $W_2$ and $W_3$, respectively. We focus here on the torically realisable singularities and their resolutions enforced by the tops described in the previous sections.
The base sections $g_m \in \{b_i,c_j,d_k\}$ in (\ref{eq:hypersurface-equation}) must now be of the form 
\bea
g_m = g_{m;k,l} \, w_2^k \, w_3^l
\eea
 with $\{w_n = 0\} = W_n$ and $g_{m;k,l}$ generic sections in the class $[g_m] - k [w_2] - l [w_3]$. The powers $k$ and $l$ depend on which of the three $SU(2)$ and $SU(3)$ tops are combined. However, as we will discuss shortly, only 5 of the $3 \times 3 = 9$ possible tops leading to the gauge group $SU(3) \times SU(2) \times U(1)_1 \times U(1)_2$ are inequivalent.

The discriminant of the fibration takes the form
\bea
\Delta = w_2^2 \, w_3^3 \,( P + {\cal O}(w_2) + {\cal O}(w_3))
\eea
 with $P$ a section of the base that 
does not vanish identically along $W_2$ and $W_3$.
For generic choice of $w_2$ and $w_3$, the fibration over $w_2$ and $w_3$ then looks like the three individual $SU(2)$ and $SU(3)$ fibrations with vanishing orders $k$ and $l$ over $w_2$ and $w_3$, respectively, but with an additional enhancement over the intersection curve $\{w_2\} \cap \{w_3 \}$ of the two non-abelian loci. Here, the generic choice of $w_2$ and $w_3$ in particular means that this intersection locus is assumed not to coincide with any of the matter curves of the individual tops. 

% Independent of the vanishing orders $l$ of $w_3$, the exponents $k$ of $w_2$ are such that if we rewrite $g_{m,\{k,l\}} \,w_2^k \,w_3^l = \tilde{g}_{m,k} \, w_2^k$, the terms of the hypersurface equation have the correct vanishing orders in $w_2$ for an $SU(2)$ singularity over $W_2$. The discriminant has the form $\Delta = w_2^2 \, \tilde{P} + ...$, where $\tilde{P}$ does not vanish identically along $W_2$. For $w_3 \neq 0$ the loci $\tilde{P}=0=w_2$ are the codimension 2 enhancement loci we would find if we had studied a fibration with only the $SU(2)$ singularity, where we would assume $\tilde{g}_{m,k}$ are \textit{generic} sections of the base of the appropriate bundle. If we replace $w_2$ with $w_3$ in that line of argument, we see that independent of the $SU(2)$ singularity and enhancement loci, we can engineer an $SU(3)$ singularity over $W_3$. Since the discriminant has the form 
% \bea
% \Delta = w_2^2 \, w_3^3 \,( P + {\cal O}(w_2) + {\cal O}(w_3))
% \eea
 % with $P$ a section of the base that 
% does not vanish identically along $W_2$ and $W_3$, there is a further codimension 2 enhancement over $w_2 = 0 =w_3$.

In a toric construction, this means that the toric diagram for the complete five-dimensional ambient space consists of two tops over the polygon for the fibre ambient space. The two tops extend in two mutually orthogonal directions of a five-dimensional lattice. If one projects onto the three-dimensional lattice spanned by the polygon-plane and one of the two directions, one either finds an $SU(2)$ or an $SU(3)$ top, each introducing resolution divisors $E_{0/1} = \{e_{0/1} =0\}$ and $F_{0/1/2} = \{f_{0/1/2}=0\}$. If the toric diagram is reflexive (which has to be checked for each base manifold), then toric geometry guarantees the smoothness of the fourfold $\hat{Y}_4$ that is cut out by the hypersurface equation inside the five-dimensional ambient space. A triangulation of the full polytope will in particular give rise to a triangulation of the $SU(2)$ and $SU(3)$ sub-top, so the corresponding full SR-ideal will -- as sub-ideals -- contain an SR-ideal of each the $SU(2)$ and the $SU(3)$ sub-model; in 
addition, there will be further generators that involve both $e_i$ and $f_j$. To use the results of sections \ref{sec:SU(2)-tops} and \ref{sec:SU(3)-tops}, we choose a triangulation of the full polytope that leads to an SR-ideal which as sub-ideals contains the SR-ideals we used when studying the corresponding $SU(2)$ and $SU(3)$ tops individually.

Because the $E_i$ and $F_j$ are fibred over different divisors of the base, intersections of the form
\begin{align}
 \int_{\hat{Y}_4} \, E_i \wedge F_j \wedge \pi^{-1} D_a \wedge \pi^{-1} D_b \, ,
\end{align}
with $D_{a/b}$ generic divisors of base, will yield zero. This just means that the roots of $SU(2)$ are uncharged under $SU(3)$ and vice versa -- as one would expect from a product structure $SU(3)\times SU(2)$ of the gauge group. Because the enhancement loci over $W_2$ away from $W_3$ 
are of the same form as in a model with only the $SU(2)$ singularity over $W_2$ (and similarly for the $SU(3)$ singularity over $W_3$ away from $W_2$), one will also find the same spectrum of matter charged only under $SU(2)$ or $SU(3)$. In addition one will find matter charged both under $SU(2)$ and $SU(3)$ at the enhancement loci $W_2 \cap W_3$. As it turns out, this matter transforms in the bifundamental representation $({\bf 3}, {\bf 2})$.

The $U(1)$ generators are now subject to the condition that the $SU(2)$ and $SU(3)$ roots are uncharged under them. However, since the $SU(2)$ and $SU(3)$ roots are mutually uncharged, this condition is met by setting 
\bea \label{ShiodaSu2Su3}
\omega_i^{SU(2) \times SU(3)} = \omega_i + \Sigma_j \, t_j \, E_j + \Sigma_j \, \tilde{t}_j \, F_j,
\eea
where $\omega_i$ are the generators of the form (\ref{eq:U(1)-generators}) and the correction terms $t_j$ and $\tilde t_j$ are the same as for the individual Shioda maps for the $SU(2)$ and $SU(3)$ tops.

Due to the equivalences amongst the $SU(2)$ and $SU(3)$ tops described in the previous sections, some of the combined $SU(3) \times SU(2) \times U(1)_1 \times U(1)_2$ models are also equivalent. In fact those models whose spectrum and Yukawa couplings can be mapped onto each other with the $U(1)$ transformation $U(1)'_1 = -U(1)_1, \, U(1)'_2 = U(1)_2 - U(1)_1$ are equivalent. One finds four inequivalent pairs of $SU(2) \times SU(3)$ top combinations, ${\rm I} \times {\rm A} \simeq {\rm II} \times {\rm C}$, ${\rm I} \times {\rm B} \simeq {\rm II} \times {\rm B}$, ${\rm I} \times {\rm C} \simeq {\rm II} \times {\rm A}$, ${\rm III} \times {\rm A} \simeq {\rm III} \times {\rm C}$, and the invariant model ${\rm III} \times {\rm B}$. A more detailed explanation based on the tops can be found in appendix \ref{app:tops}.

To summarise, in order to construct toric F-theory models with $SU(3) \times SU(2) \times U(1)_1 \times U(1)_2$ gauge symmetry we take an $SU(2)$ and an $SU(3)$ top of the form studied in the sections \ref{sec:SU(2)-tops} and \ref{sec:SU(3)-tops} and combine them into one new top. Some of the tops obtained in this way are equivalent. The previous sections directly give us the spectrum of $\mathbf{2}$- and $\mathbf{3}$-matter including their Yukawa couplings. What we need to compute is the bifundamental matter $(\mathbf{3},\mathbf{2})$ as well as all the Yukawa couplings it is involved in.

The result of this analysis is shown in table \ref{tab:SU(2)xSU(3)} for all five mutually inequivalent combinations of tops. In all cases the geometrically realised Yukawa couplings are in 1-to-1 correspondence with the set of gauge theoretically allowed couplings including the $U(1)_i$ selection rules.

\begin{table}[ht]
\begin{align*}
  \begin{array}{c|c|c}
    \text{top-combination} & (U(1)_1, U(1)_2)- & \text{additional gauge invariant Yukawas,} \\
    SU(2) \times SU(3) & \text{charge of } (\mathbf{3},\mathbf{2}) &  (\mathbf{3},\mathbf{2})- \ldots \\ \hline\hline \rule{0pt}{3ex}
   \mathrm{I} \times \mathrm{A} & (\frac{1}{6},-\frac{1}{3}) & \overline{\mathbf{3}}^\mathrm{A}_1-\mathbf{2}^\mathrm{I}_3, \, \overline{\mathbf{3}}^\mathrm{A}_2-\overline{\mathbf{2}}^\mathrm{I}_2, \, \overline{\mathbf{3}}^\mathrm{A}_3-\overline{\mathbf{2}}^\mathrm{I}_1, \, \overline{\mathbf{3}}^\mathrm{A}_4-\mathbf{2}^\mathrm{I}_2, \, \overline{\mathbf{3}}^\mathrm{A}_5-\overline{\mathbf{2}}^\mathrm{I}_3; \\ \rule{0pt}{3ex}
    & & (\mathbf{3},\mathbf{2})-\mathbf{3}^\mathrm{A}_3 \\ \hline \rule{0pt}{3ex}
    \mathrm{I} \times \mathrm{B} & (-\frac{1}{6},-\frac{1}{3}) & \overline{\mathbf{3}}^\mathrm{B}_1-\overline{\mathbf{2}}^\mathrm{I}_2, \, \overline{\mathbf{3}}^\mathrm{B}_2-\overline{\mathbf{2}}^\mathrm{I}_1, \, \overline{\mathbf{3}}^\mathrm{B}_3-\overline{\mathbf{2}}^\mathrm{I}_3, \, \overline{\mathbf{3}}^\mathrm{B}_4-\mathbf{2}^\mathrm{I}_2, \, \overline{\mathbf{3}}^\mathrm{B}_5-\mathbf{2}^\mathrm{I}_3; \\ \rule{0pt}{3ex}
    & & (\mathbf{3},\mathbf{2})-\mathbf{3}^\mathrm{B}_4 \\ \hline \rule{0pt}{3ex}
   \mathrm{I} \times \mathrm{C} & (-\frac{1}{6},0) & \overline{\mathbf{3}}^\mathrm{C}_1-\overline{\mathbf{2}}^\mathrm{I}_2, \, \overline{\mathbf{3}}^\mathrm{C}_2-\mathbf{2}^\mathrm{I}_1, \, \overline{\mathbf{3}}^\mathrm{C}_3-\mathbf{2}^\mathrm{I}_2, \, \overline{\mathbf{3}}^\mathrm{C}_4-\overline{\mathbf{2}}^\mathrm{I}_3, \, \overline{\mathbf{3}}^\mathrm{C}_5-\mathbf{2}^\mathrm{I}_3; \\ \rule{0pt}{3ex}
    & & (\mathbf{3},\mathbf{2})-\mathbf{3}^\mathrm{C}_5 \\ \hline \rule{0pt}{3ex}
    \mathrm{III} \times \mathrm{A} & (-\frac{1}{3},-\frac{1}{3}) & \overline{\mathbf{3}}^\mathrm{A}_1-\mathbf{2}^\mathrm{III}_1, \, \overline{\mathbf{3}}^\mathrm{A}_2-\overline{\mathbf{2}}^\mathrm{III}_3, \, \overline{\mathbf{3}}^\mathrm{A}_3-\mathbf{2}^\mathrm{III}_3, \, \overline{\mathbf{3}}^\mathrm{A}_4-\mathbf{2}^\mathrm{III}_2; \\ \rule{0pt}{3ex}
    & & (\mathbf{3},\mathbf{2})-\mathbf{3}^\mathrm{A}_4 \\ \hline \rule{0pt}{3ex}
    \mathrm{III} \times \mathrm{B} & (-\frac{2}{3},-\frac{1}{3}) & \overline{\mathbf{3}}^\mathrm{B}_1-\overline{\mathbf{2}}^\mathrm{III}_3, \, \overline{\mathbf{3}}^\mathrm{B}_2-\mathbf{2}^\mathrm{III}_3, \, \overline{\mathbf{3}}^\mathrm{B}_4-\mathbf{2}^\mathrm{III}_2, \, \overline{\mathbf{3}}^\mathrm{B}_5-\mathbf{2}^\mathrm{III}_1; \\    \rule{0pt}{3ex}
&&(\mathbf{3},\mathbf{2})-\mathbf{3}^\mathrm{B} \,  \text{non-existent}
  \end{array}
\end{align*}
\caption{$U(1)$ charges of the bifundamental matter and additional Yukawa couplings involving at least one  $(\mathbf{3},\mathbf{2})$, as arising in the five inequivalent combinations of the $SU(2)$ and $SU(3)$ tops studied in the previous chapters.}
\label{tab:SU(2)xSU(3)}
\end{table}

We explain the first example in a bit more detail. Combining the $SU(2)$-I and the $SU(3)$-A top amounts to restricting the sections appearing in (\ref{eq:hypersurface-equation})
as 
\begin{align}\label{eq:I-A-coeffs}
  \begin{split}
    b_0 &= b_{0;1,1}\, e_0 \, f_0 ,\quad b_2 = b_{2;0,0}\,e_1 \, f_1\,f_2 ,\quad c_1 = c_{1;0,0}\,e_1 \, f_2 ,\quad c_2 = c_{2;0,1}\,f_0\,f_2 , \\
    d_0 &= d_{0;1,1}\,e_0 \, f_0\,f_1,\quad d_1 = d_{1;0,0}\,f_1, \quad d_2 = d_{2;1,1}\,e_0 \, f_0\,f_1^2.
  \end{split}
\end{align}
In table \ref{tab:IxA-divisor-classes} we list the divisor classes and the corresponding scaling relations among the fibre coordinates. The last part shows the lattice vectors of the top that describes the ambient space. The vectors $\underline{x}$ and $\underline{y}$ should be linearly independent, but otherwise unspecified for a generic base $\cal B$. For this top there exist 16 different triangulations. We choose a triangulation for which the SR-ideal is the union of the individual ideals (\ref{eq:SU(2)-I-SR-ideal}) and (\ref{eq:SU(3)-A-SR-ideal}) together with the element  $\{f_0 \, e_1\}$, i.e. it is generated by
\begin{align}\label{eq:SR-ideal-IxA}
\su \, \sv , \su \, \sw , \sw \, s_0 , \sv \, s_1 , s_0 \, s_1 , e_0 \, \sw , e_1 \, s_0 , e_1 \, \su , \, f_0\,\sw , f_0\,s_1 , f_1\,s_0 , f_1\,\sv , f_2\,s_0 , f_2\,s_1 , f_2\,\su , f_0\,e_1.
\end{align}

%\{ {\rm u v, u w, s_0 w, s_1 v, s_0 s_1, e_0 w, e_1 s_0,
%e_1 u, f_0 w, f_0 s_1, f_1 s_0, f_1 v, f_2 s_0, f_2 s_1, f_2 u, f_0 e_1

\begin{table}[ht]
	\begin{align*}
	\begin{array}{c || c c c c c | c c | c c c}
			& \su & \sv & \sw & s_0 & s_1 & e_0 & e_1 & f_0 & f_1 & f_2	\\ \hline \hline
		\left[W_2 \right] & \cdot & \cdot & \cdot & \cdot & \cdot & 1 & \cdot & \cdot & \cdot & \cdot \\
		\left[W_3 \right] & \cdot & \cdot & \cdot & \cdot & \cdot & \cdot & \cdot & 1 & \cdot & \cdot \\
		\alpha & \cdot & \cdot & 1 & \cdot & \cdot & \cdot & \cdot & \cdot & \cdot & \cdot \\
		\beta & \cdot & 1 & \cdot & \cdot & \cdot & \cdot & \cdot & \cdot & \cdot & \cdot \\ \hline
		U & 1 & 1 & 1 & \cdot & \cdot & \cdot & \cdot & \cdot & \cdot & \cdot \\
		S_0 & \cdot  & \cdot & 1 & 1 & \cdot & \cdot & \cdot & \cdot & \cdot & \cdot \\
		S_1 & \cdot & 1 & \cdot & \cdot & 1 & \cdot & \cdot & \cdot & \cdot & \cdot \\ \hline
		E_1 & \cdot & \cdot & -1 & \cdot & \cdot & -1 & 1 & \cdot & \cdot & \cdot \\ \hline
		F_1 & \cdot & 1 & \cdot & \cdot & \cdot & \cdot & \cdot & -1 & 1 & \cdot \\
		F_2 & \cdot & \cdot & -1 & \cdot & \cdot & \cdot & \cdot & -1  & \cdot & 1 \\ \hline \hline
		\multirow{3}{*}{\text{toric data}}	& -1 & 0 & 1 & -1 & 0 & 0 & 1 & 0 & 0 & 1 \\ 
			& 1 & -1 & 0 & 0 & 1 & 0 & 0 & 0 & 1 & 0 \\
			& \underline{0} & \underline{0} & \underline{0} & \underline{0} & \underline{0} & \underline{x} & \underline{x} & \underline{y} & \underline{y} & \underline{y}
	\end{array}
	\end{align*}
	\caption{Divisor classes and coordinates of the ambient space for top-combination $\mathrm{I} \times \mathrm{A}$.}
	\label{tab:IxA-divisor-classes}
\end{table}

Irrespective of the chosen triangulation,
%  the discriminant of the blow-down takes the form
% \bea
% 256  w_2^2  w_3^3      c_{2,0,1}   c_{1,0,0}   \Big(       c_{2,0,1}}   b1 b_{2,0,0} c_{1,0,0} - c_{1,0,0}^2 d_{1,0,0} - b_{2,0,0}^2  c_{2,0,1}  w_3) 
% (  b2^2
% - 4 c_{2,0,1}} d_{1,0,0}  )^2  
%  (b_{0,1,1}^2 d_{1,0,0}^2 + b_{0,1,1} (-B1 d_{0,1,1} d_{1,0,0} + B1^2 d_{2,1,1} - 2 c_{2,0,1}} d_{1,0,0} d_{2,1,1} w_3 ) +
% c_{2,0,1}} w_2 (d_{0,1,1}^2 d_{1,0,0} - B1 d_{0,1,1} d_{2,1,1} + c_{2,0,1}} d_{2,1,1}^2 w_2)     )
% b_{0,1,1} B1^3 c_{1,0,0} (B1 b_{2,0,0} - c_{1,0,0} d_{1,0,0})  (  -B1 d_{0,1,1} d_{1,0,0} + b_{0,1,1} d_{1,0,0}^2
% + B1^2 d_{2,1,1})  (B1 c_{2,0,1}} - b_{0,1,1} c_{1,0,0} w_2)    \Big)
% \eea
 one recovers from the discriminant the ${\bf 2}$- and ${\bf 3}$-matter curves of the two individual tops, (\ref{tab:SU(2)-I-matter}) and (\ref{tab:SU(3)-A-matter}), where of course the base sections $g_{m,k}$
defining the matter curves are  modified in agreement with  (\ref{eq:I-A-coeffs}).\footnote{In addition, the same type $III$ and $IV$ enhancement loci arise as before, which do not carry matter representations.}
For example, the representation ${\bf 2}^{\rm I}_1$ is located at the intersection $\{w_2\} \cap \{c_{2;0,1}\}$ (because in the blow-down the locus $\{w_2\} \cap \{c_2\}$ appearing in $(\ref{tab:SU(2)-I-matter})$ splits into $\{w_2\} \cap \{c_{2;0,1}\}$ and $\{w_2\} \cap \{w_3\}$, but the latter contributes to matter charged under both $SU(2)$ and $SU(3)$), 
and the curve hosting ${\bf 3}^{\rm A}_3$ is now given by $\{w_3\} \cap \{ b_{0;1,1} \, w_2 \, c_{1;0,0}- b_1 \, c_{2;0,1}\}$.

What is new is that along the curve $\{w_2\} \cap \{w_3\}$ the Kodaira type of the fibre enhances to split type $I_5$, corresponding to vanishing orders $(0,0,5)$ of $(f,g,\Delta)$ in the Weierstrass model. 
Indeed, the fibre components straightforwardly split to form the affine Dynkin diagram of $SU(5)$; more precisely the five fibre components are given by 
\begin{align}
\begin{split}
\mathbb{P}^1_{00} \equiv P_T|_{\{e_0\} \cap \{f_0\}} &= b_{2;0,0} \, f_1 \, f_2 \, {\rm u} + d_{1;0,0} \, f_1 \, s_0 \, {\rm u}^2
+ c_{1;0,0} \, f_2 {\rm v} + b_1 \, s_0 \, {\rm u} \, {\rm v}, \\
\mathbb{P}^1_{01} \equiv P_T|_{\{e_0\} \cap \{f_1\}} &= c_{2;0,1} \, f_0 \, f_2 + c_{1;0,0}\, e_1 \, f_2 \, s_1 + b_1 \, s_1 \, {\rm u}, \\
\mathbb{P}^1_{02} \equiv P_T|_{\{e_0\} \cap \{f_2\}} &= d_{1;0,0} \, f_1 + b_1 \, {\rm v}, \\
\mathbb{P}^1_{11} \equiv P_T|_{\{e_1\} \cap \{f_1\}} &= b_{0;1,1} \, e_0 + c_{2;0,1} \, f_2 \, {\rm w} + b_1 \, s_1 \, {\rm w}, \\
\mathbb{P}^1_{12} \equiv P_T|_{\{e_1\} \cap \{f_2\}} &= d_{2;1,1} \, e_0 \, f_1^2\, + d_{0;1,1} \, e_0 \, f_1 \, {\rm v} + b_{0;1,1} \, e_0 \, {\rm v}^2 + d_{1;0,0} \, f_1 \, {\rm w} + b_1 \, {\rm v} \, {\rm w},
\end{split}
\label{fiveP1}
\end{align}
where we used the SR-ideal to set as many coordinates to one as possible. Note that $e_1$ and $f_0$ do not intersect due to the SR-ideal relations so that the locus $P_T|_{\{e_1\} \cap \{f_0\}}$ is absent in (\ref{fiveP1}).

The fibre component $\mathbb{P}^1_{00} + \mathbb{P}^1_{02}$ is identified with the highest weight of the bifundamental representation $({\bf 3},{\bf 2})$; the $U(1)_1$ and $U(1)_2$ charges are found to be $(\frac{1}{6}, -\frac{1}{3})$ by computing the intersection product with the generators
\begin{align}\label{eq:I-A-U(1)-generators}
 \begin{split}
    \omega_1^{{\rm I} \times {\rm A}} &= S_1 - S_0 - \overline{\mathcal{K}} +    \frac{1}{2} E_1 +  \frac{2}{3} F_1 + \frac{1}{3} F_2 ,\\
    \omega_2^{{\rm I} \times {\rm A}} &= U - S_0 - \overline{\mathcal{K}} - [c_{1;0,0}] + \frac{2}{3} F_1 + \frac{1}{3} F_2.
  \end{split}
\end{align}

Finally, we have analysed  the intersection of $\{w_2\} \cap \{w_3\}$ with each of the ${\bf 3}$- and ${\bf 2}$-curves to identify extra fibre enhancements signalling Yukawa couplings involving the new $({\bf 3},{\bf 2})$-state. 
The fibres over the Yukawa points ${\bf (3,2)} - {\bf \bar 3} - {\bf 2}/{\bf \bar 2}$ are of split Kodaira type $I_6$, and the fibre components can be explicitly
\begin{wrapfigure}[13]{r}{.2\textwidth}
\vspace{-13pt}
\begin{center}
	\def\svgwidth{.9\hsize}
	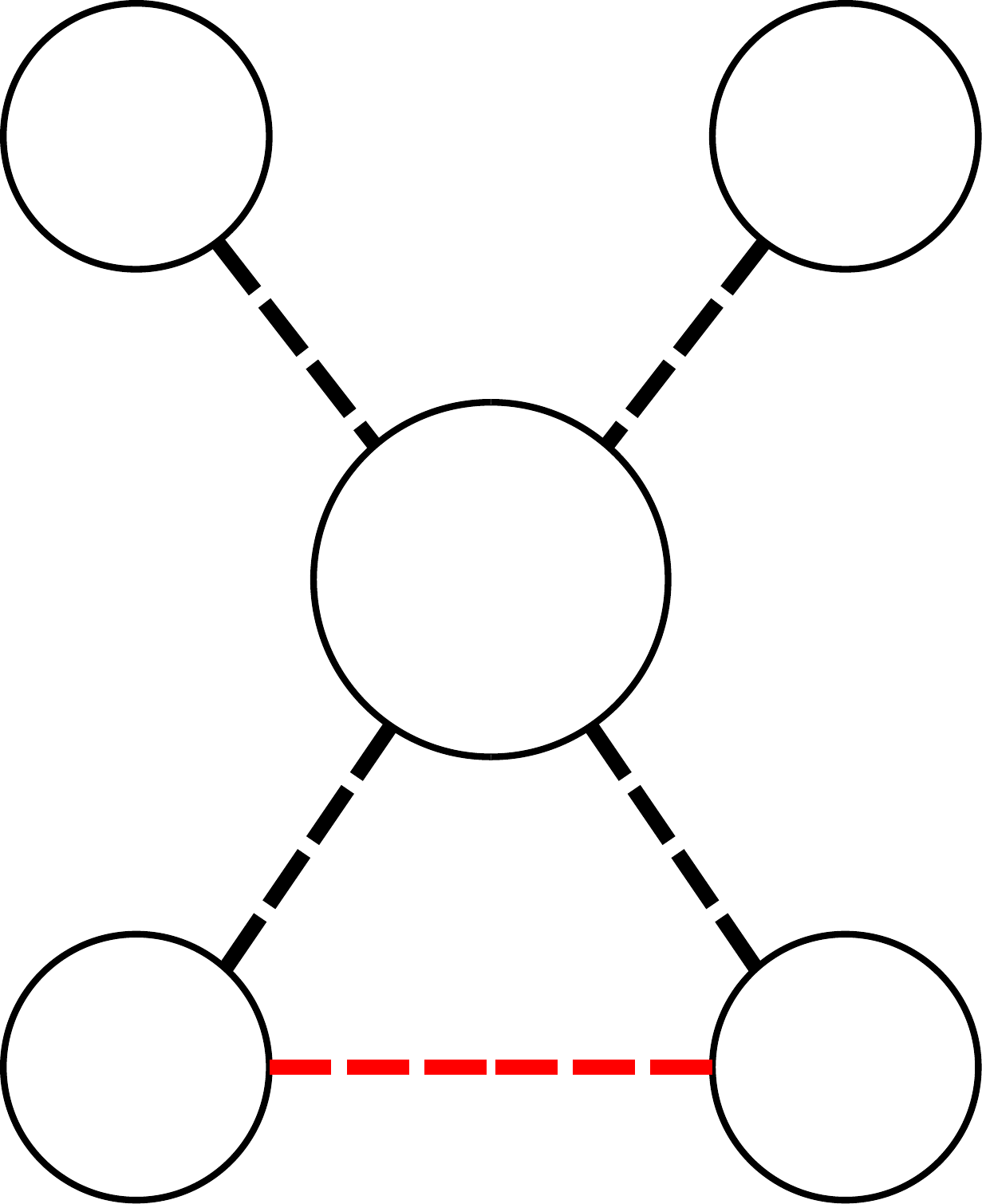
	\caption{Non-standard fibre structure at the $(\mathbf{3},\mathbf{2}) - (\mathbf{3},\mathbf{2}) -  \mathbf{3}$ Yukawa point.}
	\label{fig:non-standard}
\end{center}
\end{wrapfigure}
checked to form the affine Dynkin diagram of $SU(6)$.
The base points are found by intersecting $W_3$ and $W_2$ with the five ${\bf 3}$-curves and noting that the intersection points also lie on top of one of the ${\bf 2}$-curves.

%One can then convince oneself that precisely the fibre components associated with the three representations indicated in table (\ref{tab:I-A-Yukawas}) form a trivial combination of cycles, corresponding to the presence of a Yukawa coupling. 
A special role is played by the intersection locus $\{w_3\} \cap \{w_2 \} \cap \{ b_{0;1,1} \, w_2 \, c_{1;0,0}- b_1 \, c_{2;0,1}\}$, where the last term is the ${\bf 3}^{\rm A}_3$-curve.
Apart from the intersection $\{w_3\} \cap \{w_2 \} \cap \{c_{2;0,1}\}$, where the  $(\mathbf{3},\mathbf{2}) -    \overline{\mathbf{3}}^\mathrm{A}_3-\overline{\mathbf{2}}^\mathrm{I}_1 $ Yukawa coupling is localised,
there is also the intersection $\{w_3\} \cap \{w_2 \} \cap \{b_1\}$, which does not lie on any of the ${\bf 2}$-curves. The fibre components almost form an affine Dynkin diagram of $SO(8)$ (including the multiplicity $2$ for the interior node) except for an additional intersection point between two of four the exterior nodes which makes this fibre non-standard (cf.~figure \ref{fig:non-standard}).
The associated Yukawa coupling is  $(\mathbf{3},\mathbf{2}) - (\mathbf{3},\mathbf{2}) - \mathbf{3}^\mathrm{A}_3$.

\begin{table}[b]
\begin{align*}
  \begin{array}{c|c|c}
    \text{coupling} & \text{locus} = W_2 \cap W_3 \cap \ldots & \text{fibre type} \\ \hline\hline \rule{0pt}{3ex}
 (\mathbf{3},\mathbf{2}) - \overline{\mathbf{3}}^\mathrm{A}_1-\mathbf{2}^\mathrm{I}_3         & \{b_{0;1,1}\} & I_6   \\[.5ex] \hline \rule{0pt}{3ex}
 (\mathbf{3},\mathbf{2}) -   \overline{\mathbf{3}}^\mathrm{A}_2-\overline{\mathbf{2}}^\mathrm{I}_2       &  \{c_{1;0,0} \}  &   I_6 \\[.5ex] \hline\rule{0pt}{3ex}
  (\mathbf{3},\mathbf{2}) -    \overline{\mathbf{3}}^\mathrm{A}_3-\overline{\mathbf{2}}^\mathrm{I}_1    & \{c_{2;0,1}\} & I_6   \\[.5ex] \hline \rule{0pt}{3ex}
 (\mathbf{3},\mathbf{2}) - \overline{\mathbf{3}}^\mathrm{A}_4-{\mathbf{2}}^\mathrm{I}_2     & \{b_1 \, b_{2;0,0} - c_{1;0,0} \, d_{1;0,0} \}  & I_6      \\[.5ex] \hline \rule{0pt}{3ex}
 (\mathbf{3},\mathbf{2}) - \overline{\mathbf{3}}^\mathrm{A}_5-\overline{{\mathbf{2}}}^\mathrm{I}_3        & \{b_{0;1,1,} \, d_{1;0,0}^2 - b_1 \, d_{0;1,1} \, d_{1;0,0}  + b_1^2 \, d_{2;1,1} \}  &    I_6   \\[.5ex] \hline\hline \rule{0pt}{3ex}
 (\mathbf{3},\mathbf{2}) - (\mathbf{3},\mathbf{2}) -  {\mathbf{3}}^\mathrm{A}_3   & \{b_1\}  & {\rm non-standard}   \\[.5ex]
 %
%
 %    \mathbf{2}^\mathrm{I}_2 - \mathbf{2}^\mathrm{I}_2 - \overline{\mathbf{1}}^{(3)} & \{b_{2,0}\} \cap \{c_{1,0}\} & \mathbb{P}^1_0 \rightarrow \mathbb{P}^1_{0B} + \mathbb{P}^1_{0Cs_0} + \mathbb{P}^1_{0C'} \\[.5ex] \hline \rule{0pt}{3ex}
%
 %    \mathbf{2}^\mathrm{I}_3 - \mathbf{2}^\mathrm{I}_3 - \overline{\mathbf{1}}^{(2)} & \{b_{0,1}\,d_1 - c_2\,d_{2,1}\} \cap \{b_1\,d_{2,1} - d_{0,1}\,d_1\} & \mathbb{P}^1_1 \rightarrow \mathbb{P}^1_{1A} + \mathbb{P}^1_{1B'} + \mathbb{P}^1_{1B''}
  \end{array}
\end{align*}
\caption{Details on the additional Yukawas involving bifundamental matter in the top combination $\mathrm{I} \times \mathrm{A}$.}
\label{tab:I-A-Yukawas}
\end{table}

We have checked the above calculations for a specific fibration over the base ${\cal B} = \mathbb{P}^3$ with $H^{1,1}(\mathbb{P}^3) = \{ n \cdot H | n \in \mathbb{Z}\}$, where $H$ is the hyperplane class, and $\overline{\cal K}_{\cal B} = 4H$. For simplicity we take $w_2$ and $w_3$ to be two of the four homogeneous coordinates $(z_0 : z_1 : z_2 : z_3)$, e.g.~$w_2 = z_0$ and $w_3 = z_1$; then $W_2 = W_3 = H$. Recall that  $b_i,c_j,d_k$ must transform as sections of specific line bundles, see table~\ref{coeff}, where there is freedom left in choosing $\alpha$ and $\beta$. They are subject to further constraints as the restricted sections (\ref{eq:I-A-coeffs}) in the presence of the non-abelian symmetry must be effective classes. Over ${\cal B} = \mathbb{P}^3$, these constraints are met with $\alpha = 2H$, $\beta = H$, implying the classes of the restricted sections to be
\begin{align}\label{eq:explicit-base-classes}
\begin{split}
& [b_{0;1,1}] = 3 H, \quad [b_1] = 4 H , \quad [b_{2;0,0}] = 3 H, \quad [c_{1;0,0}] = 2 H, \\
& [c_{2;0,1}] = 2 H, \quad [d_{0;1,1}] = 4 H , \quad [d_{1;0,0}] = 5H , \quad [d_{2;1,1}] = 5 H \, .
\end{split}
\end{align}
From the choice of $\alpha$ and $\beta$ we have to impose the condition $2 \vec{\sw} + \vec{\sv} + \vec{e_0} + \vec{f_0} + \vec{z_2} + \vec{z_3} =0$, where $\vec{(\cdot)}$ is the $(\cdot)$-coordinate's lattice vector of the toric diagram of the full fibration (`toric data' in table \ref{tab:IxA-divisor-classes}). This condition is met by the toric ambient space $\hat{X}_5$, whose toric diagram has the lattice vectors shown in table \ref{tab:toric-table-IxA}. The resulting polytope is reflexive, guaranteeing that the fourfold cut out by the hypersurface equation inside this toric ambient space is smooth. The Euler characteristic of the fourfold is 1440.

\begin{table}[ht]
\begin{align*}
\begin{array}{c | c | c | c | c | c | c | c | c | c | c | c}
\vec{\su} & \vec{\sv} & \vec{\sw} & \vec{s_0} & \vec{s_1} & \vec{e_0} & \vec{f_0} & \vec{z_2} & \vec{z_3} & \vec{e_1} & \vec{f_1} & \vec{f_2} \\ [.3ex] \hline
-1 & 0 & 1 & -1 & 0 & 0 & 0 & -2 & 0 & 1 & 0 & 1 \\
1 & -1 & 0 & 0 & 1 & 0 & 0 & 1 & 0 & 0 & 1 & 0 \\
0 & 0 & 0 & 0 & 0 & 1 & 0 & 0 & -1 & 1 & 0 & 0\\
0 & 0 & 0 & 0 & 0 & 0 & 1 & 0 & -1 & 0 & 1 & 1\\
0 & 0 & 0 & 0 & 0 & 0 & 0 & 1 & -1 & 0 & 0 & 0
\end{array}
\end{align*}
\caption{Toric diagram for the ambiente space $\hat{X}_5$ of the model $ \mathrm{I} \times \mathrm{A}$ over the base ${\cal B} = \mathbb{P}^3$.}
\label{tab:toric-table-IxA}
\end{table}

The above analysis can be repeated for the remaining four inequivalent combinations of tops and leads to the couplings listed in (\ref{tab:SU(2)xSU(3)}). Note that for the top combinations ${\rm III} \times {\rm A}$ and ${\rm III} \times {\rm B}$ no gauge invariant coupling ${\bf (3,2)} - {\overline{\bf 3}}^\mathrm{A}_5 - {\bf 2}$ resp.~${\bf (3,2)} - {\overline{\bf 3}}^\mathrm{B}_3 - {\bf 2}$ exists. In both cases, the intersection point of the ${\bf (3,2)}$- and the corresponding ${\overline{\bf 3}}$-curve lies on the curve of non-split $I_3$-enhancement described after (\ref{eq:SU(2)-III-splitting-quadratic-curve-2}) in appendix \ref{SU2DetailsApp}.
Were it not for a monodromy along that curve, an additional ${\bf 2}$-representation would arise, which in fact would have the correct quantum numbers to couple as in ${\bf (3,2)} - {\overline{\bf 3}}^\mathrm{A}_5 - {\bf 2}$ (or ${\bf (3,2)} - {\overline{\bf 3}}^\mathrm{B}_3 - {\bf 2}$). Correspondingly,
 the fibre over the triple intersection of these curves does enhance to form an $I_6$ Kodaira fibre, but due to the described monodromy no physical Yukawa couplings result as the {\bf 2}-state in question is projected out. 

More drastically, the top combination ${\rm III} \times {\rm B}$ exhibits a non-Kodaira enhancement in the fibre over $\{w_3\} \cap \{w_2 \} \cap \{b_1\}$.\footnote{In the other four top combinations this point corresponds to the $({\bf 3},{\bf 2})-({\bf 3},{\bf 2})-{\bf 3}$ coupling. For ${\rm III} \times {\rm B}$, however, there is no suitable $\bf 3$ state.}
At this set of points, several matter curves coincide and the vanishing orders of $(f,g,\Delta)$ in the Weierstrass model take the values $(4,6,12)$. For such high enhancement no flat crepant resolution can be found. To see how this manifest itself, consider the triangulation of the ${\rm III} \times {\rm B}$ top leading to an SR-ideal generated by (\ref{eq:SU(2)-III-SR-ideal}) (for the $SU(2)$ part), (\ref{eq:SU(3)-B-SR-ideal}) (for the $SU(3)$ part) and the new elements $ \{ s_1 \, e_0 \, f_1 , e_1 \, f_0 \, f_2 \} $: For this SR-ideal the fibration becomes non-flat over $\{w_3\} \cap \{w_2 \} \cap \{b_1\}$ as the hypersurface equation $P_T$ becomes trivial for $b_1 = e_0 = f_1 = 0$.
Therefore, in order for the top combination ${\rm III} \times {\rm B}$ to give rise to a well-defined F-theory compactification, the point set  $\{w_3\} \cap \{w_2 \} \cap \{b_1\}$ must be empty.
Since $b_1$ is universally in the class $\overline{\cal K}$, these points cannot be turned off by a suitable choice of classes $\alpha$ and $\beta$ appearing in table (\ref{coeff}).
While it is not excluded that $W_2$ and $W_3$ can be found such that no intersection points  $\{w_3\} \cap \{w_2 \} \cap \{b_1\}$ arise, we do currently not have an example of this type.

%% file: non-standard-fibre.pdf_tex
%% Creator: Inkscape inkscape 0.48.2, www.inkscape.org
%% Accompanies image file 'non-standard-fibre.pdf' (pdf, eps, ps)
\begingroup%
  \makeatletter%
  \providecommand\color[2][]{%
    \errmessage{(Inkscape) Color is used for the text in Inkscape, but the package 'color.sty' is not loaded}%
    \renewcommand\color[2][]{}%
  }%
  \providecommand\transparent[1]{%
    \errmessage{(Inkscape) Transparency is used (non-zero) for the text in Inkscape, but the package 'transparent.sty' is not loaded}%
    \renewcommand\transparent[1]{}%
  }%
  \providecommand\rotatebox[2]{#2}%
  \ifx\svgwidth\undefined%
    \setlength{\unitlength}{354.40952148bp}%
    \ifx\svgscale\undefined%
      \relax%
    \else%
      \setlength{\unitlength}{\unitlength * \real{\svgscale}}%
    \fi%
  \else%
    \setlength{\unitlength}{\svgwidth}%
  \fi%
  \global\let\svgwidth\undefined%
  \global\let\svgscale\undefined%
  \makeatother%
  \begin{picture}(1,1.22570883)%
    \put(0,0){\includegraphics[width=\unitlength]{non-standard-fibre.pdf}}%
    \put(0.11286377,1.04963321){\color[rgb]{0,0,0}\makebox(0,0)[lb]{\smash{1}}}%
    \put(0.83519203,1.04963321){\color[rgb]{0,0,0}\makebox(0,0)[lb]{\smash{1}}}%
    \put(0.47402792,0.598178){\color[rgb]{0,0,0}\makebox(0,0)[lb]{\smash{2}}}%
    \put(0.11286377,0.11286373){\color[rgb]{0,0,0}\makebox(0,0)[lb]{\smash{1}}}%
    \put(0.83519203,0.11286373){\color[rgb]{0,0,0}\makebox(0,0)[lb]{\smash{1}}}%
  \end{picture}%
\endgroup%

%% file: section6_F-theory_MSSM_v3.tex
\section{ Standard Model Embeddings}\label{sec_SM}

% The next step towards constructing a realisation of the Standard Model is to identify the two $U(1)$ factors from the models (\ref{tab:SU(2)xSU(3)}) with linear combinations of $U(1)_Y$ and one further $U(1)$ symmetry. The latter $U(1)$ serve phenomenological purposes as to implement selection rules and must in the end be absent in the massless spectrum. For our further analyses we look at models where the extra symmetry is a Peccei-Quinn-type symmetry $U(1)_{PQ}$\TODO{Referenz!}, which has the phenomenologically desirable feature that it forbids dimension-5 proton decay. Under such a symmetry, the Higgs-up and -down doublets carry different charges.
%, so their states must be localised on different curves. Conveniently, all our models (\ref{tab:SU(2)xSU(3)}) precisely have three $\mathbf{2}$-curves, predestined to support the lepton doublet $L$ and the two Higgs-doublets $H_u / H_d$.

\subsection{Criteria for Standard Model Embeddings }\label{subsec_spectrum_search}

The toric fibrations with gauge group $SU(3) \times SU(2) \times U(1)_1 \times U(1)_2$ constructed in the previous section are the starting point of our search for F-theory vacua with  Standard Model gauge group and matter.  Our discussion will be phrased in the framework of the ${\cal N}=1$ Minimal Supersymmetric Standard Model (MSSM), potentially extended by further singlets, with the understanding that supersymmetry is broken at an priori unknown energy below the compactification scale in agreement with current lower collider bounds.
     To fix our conventions we recall the MSSM spectrum plus right-handed neutrinos $\nu^c_R$ (taking all fields to be chiral ${\cal N}=1$ superfields) in table \ref{tab:SM-spectrum}. 
 We also allow for a generation of the 
  $\mu$-term in the Higgs sector via the VEV of an MSSM singlet ${\bf 1}_\mu$, as studied extensively in the literature in the framework of the NMSSM (see e.g. \cite{Ellwanger:2009dp} and references therein).\footnote{A detailed and systematic analysis of such singlet extensions of the MSSM in perturbative Type II intersecting brane quivers has been performed in \cite{Cvetic:2010dz}. }

\begin{table}[ht]
\begin{align*}
  \begin{array}{cc|c|c}
    \text{matter} & & \text{representation} & \text{hypercharge} \\ \hline \hline \rule{0pt}{3ex}
    \text{left-handed quarks} & Q & (\mathbf{3},\mathbf{2}) & \hphantom{-}\frac{1}{6} \\ \rule{0pt}{3ex}
    \text{right-handed up-quarks} & u^c_R & (\overline{\mathbf{3}},\mathbf{1}) \equiv \overline{\mathbf{3}}_u & -\frac{2}{3} \\ \rule{0pt}{3ex}
    \text{right-handed down-quarks} & d^c_R & (\overline{\mathbf{3}},\mathbf{1}) \equiv \overline{\mathbf{3}}_d & \hphantom{-}\frac{1}{3} \\ \rule{0pt}{3ex}
    \text{Higgs-up} & H_u & (\mathbf{1},\mathbf{2}) \equiv \mathbf{2}_u & \hphantom{-}\frac{1}{2} \\ \rule{0pt}{3ex}
    \text{Higgs-down} & H_d & (\mathbf{1},\mathbf{2}) \equiv \mathbf{2}_d & -\frac{1}{2} \\ \rule{0pt}{3ex}
    \text{left-handed leptons} & L & (\mathbf{1},\mathbf{2}) \equiv \mathbf{2}_L & -\frac{1}{2} \\ \rule{0pt}{3ex}
    \text{right-handed electrons} & e^c_R & (\mathbf{1},\mathbf{1}) \equiv \mathbf{1}_e & \hphantom{-}1 \\ \rule{0pt}{3ex}
    \text{right-handed neutrinos} & \nu^c_R & (\mathbf{1},\mathbf{1}) \equiv \mathbf{1}_\nu & \hphantom{-}0 \\ \rule{0pt}{3ex}
    \text{$\mu$-singlet} &&   (\mathbf{1},\mathbf{1}) \equiv \mathbf{1}_\mu & \hphantom{-}0
  \end{array}
\end{align*}
\caption{Matter spectrum of the MSSM.}
\label{tab:SM-spectrum}
\end{table}

At the level of renormalisable couplings, the superpotential of the singlet-extended MSSM takes the form
\bea \label{W1W2}
W  &=& W_1 + W_2 + W_{\rm singlet}, \\
W_1 & =&   Y_u \,  Q\,H_u\,u^c_R + Y_d \, Q\,H_d\,d^c_R + Y_e \, L \, H_d \, e_R^c + Y_{\nu}   \, L\,H_u\,\nu^c_R +  \mu \, H_u \, H_d, \\
W_2 & = &    \alpha  \, Q \, L \, d_R^c   + \beta   \, u_R^c \, d_R^c \, d_R^c  + \gamma \, L \, L \, e_R^c + \kappa \, L \, H_u,  \label{eq:couplings_W2}\\
W_{\rm singlet} &=&   \delta_{3,0} \,  {\bf 1_{\mu}}  \, {\bf 1_{\mu}}  \, {\bf 1_{\mu}}  + \delta_{2,1} \,  {\bf 1_{\mu}}  \, {\bf 1_{\mu}}  \, \nu_R^c  +   \delta_{1,2} \, {\bf 1_{\mu}} \,   \nu_R^c   \,   \nu_R^c  +  \delta_{0,3}  \, \nu_R^c \, \nu_R^c    \,  \nu_R^c, \label{eq:couplings_W_singlet}
\eea
where we are suppressing family indices.
Here, $W_1$ contains the Yukawa couplings that give rise to the masses for the up-quarks, down-quarks and the charged leptons as well as potential Dirac masses for the right-handed neutrinos. We also include here the $\mu$-term for the Higgs sector, with the understanding that this term might  originate from a Yukawa coupling $Y_{\mu} \, {\bf 1}_{\mu} H_u \, H_d $ if the scalar in the superfield ${\bf 1}_{\mu}$ acquires a non-trivial VEV.
For completeness we have furthermore listed possible dimension-four singlet couplings 
$W_{\rm singlet} $. If $\langle {\bf 1}_{\mu} \rangle \neq 0$ the third term in $W_{\rm singlet} $ effectively contributes to the Majorana mass term for the right-handed neutrinos, while the first term would induce an F-term in the vacuum and is therefore of interest in the context of supersymmetry breaking. We have not listed potential tadpole and holomorphic mass terms involving the singlets, which are also allowed by the MSSM gauge group.

%\bea
%\delta_{3,0} {\bf 1_{\mu}}  \, {\bf 1_{\mu}}  \, {\bf 1_{\mu}}  + \delta_{2,1} {\bf 1_{\mu}}  \, {\bf 1_{\mu}}  \, \nu_R^c  +   \delta_{1,2} {\bf 1_{\mu}} \,   \nu_R^c   \,   \nu_R^c +  \delta_{0,3}  \, \nu_R^c \, \nu_R^c    \,  \nu_R^c
%\eea

The couplings in $W_2$ each violate R-parity $(-1)^{2S + 3 (B-L)}$ with $S$ the spin and $B$, $L$ baryon and lepton number. The second term does in addition not conserve baryon number, while the remaining terms are lepton-number violating.
In particular, some combinations of terms within $W_2$ lead to rapid proton decay and are therefore severely constrained \cite{Allanach:1999ic,Nath:2006ut}.
Proton decay due to dimension-four operators requires both baryon and lepton-number violating contributions. The most severe constraints arise from tree-level induced proton decay, which is generated only if
both $\alpha$ and $\beta$ are non-zero simultaneously. 
However, the precise bounds on the couplings depend, amongst other things, on the scale of supersymmetry breaking. In models with intermediate or high-scale supersymmetry breaking some of the constraints on $W_2$ are considerably relaxed compared to TeV-scale supersymmetric scenarios. For more details of the extremely rich phenomenology of R-partiy violating couplings we refer in addition to \cite{Allanach:1999ic,Allanach:2003eb,Nath:2006ut} and references therein.

% Proton decay due to dimension-4 operators requires both baryon and lepton-number violating contributions. The most severe constraints arise from tree-level induced proton decay, which is induced only if
% both $\alpha$ and $\beta$ is non-zero simultaneously. The dependence of the concrete bounds on the supersymmetry breaking scale can be found in \cite{Allanach:1999ic,Nath:2006ut}.

%A detailed discussion of the phenomenology of R-partiy violating couplings is beyond the scope of the present article and left for future work.
% To be as open-minded as possible, in our search for MSSM realizations we insist that not both couplings  $\alpha$ and $\beta$ are present at the same time, but we stress that for a high supersymmetry breaking scale this might be too restrictive and a more quantitative analysis of the couplings is required in order to decide if a model with non-vanishing $\alpha \times \beta$ is phenomenologically viable.

At mass dimension five, the MSSM allows for the following baryon or lepton number violating operators \cite{Allanach:2003eb},
\begin{align}
 \begin{split}\label{W3K}
W_3 &= \lambda_1 \, Q \, Q \, Q \, L + \lambda_2 \, u_R^c \,  u_R^c \, d_R^c \, e_R^c + \lambda_3 \, Q \, Q \, Q \, H_d + \lambda_4 \, Q \, u_R^c \, e_R^c \, H_d \\
& + \lambda_5 \, L \, L \, H_u \, H_u +  \lambda_6 \, L \, H_d \, H_u \, H_u,
\end{split} \\
K &\supset \lambda _7 \, u_R^c \, (d_R^c)^* \, e_R^c + \lambda_8 \, H_u^* \, H_d \, e_R^c + \lambda_9 \, Q \, u_R^c \, L^* + \lambda_{10} \, Q \, Q \, (d_R^c)^*. \label{eq:couplings_K}
\end{align}
Additional dimension-five terms are possible which involve the singlets $\nu_R^c$ and ${\bf 1}_\mu$. In particular, any of the dimension-four operators present in (\ref{W1W2}) can in principle be dressed with such a singlet.
We do not list these couplings explicitly here.

Our attitude towards lepton and baryon number violating  couplings is as follows:
 In order to fully explore the parameter space of possible Standard Models within our framework we do not insist on TeV scale supersymmetry a priori, but rather allow for the possibility of intermediate scale supersymmetry breaking. While the viability of such a scenario will ultimately be determined experimentally, a higher supersymmetry breaking scale is in fact a natural option in direct Standard Model constructions. After all, the exact unification of the gauge couplings at a scale around $10^{16}$ GeV, which is one of the predictions of the TeV scale MSSM, is not immediate if the gauge groups $SU(3)$, $SU(2)$ and $U(1)_Y$ are constructed independently.
 More importantly perhaps, intermediate scale supersymmetry is well-motivated by a $126$ GeV Higgs 
as studied in string theoretic frameworks recently in \cite{Hebecker:2012qp,Ibanez:2012zg,Hebecker:2013lha,Ibanez:2013gf,Hebecker:2014uaa} (see also \cite{Chatzistavrakidis:2012bb,Hall:2013eko,Ibanez:2014zsa,Hall:2014vga} and references therein for other recent examples in the literature motivating an intermediate supersymmetry breaking scale).
 Keeping an open mind towards the supersymmetry breaking scale, we do therefore not require absence of all dimension-four and -five lepton and baryon number violating  couplings in our search criterion for Standard Model configurations, but will only list which of these couplings are present. A more detailed study of the associated phenomenology, taking into account the details of supersymmetry breaking, is left for future explorations. Having said that, in many cases the $U(1)$ selection rules do prevent potentially dangerous such operators as we will see explicitly. 

For each of the five combinations of tops with gauge group $SU(3) \times SU(2) \times U(1)_1 \times U(1)_2$ a plethora of possibilities arises for identifying the massless representations with the MSSM fields.
This identification will in particular determine which linear combination of $U(1)_1$ and $U(1)_2$ corresponds to hypercharge $U(1)_Y$. The orthogonal combination is then massless in absence of gauge fluxes and will remain as a perturbative selection rule after gauge fluxes induce a St\"uckelberg mass for the associated gauge potential. At the same time it must be ensured that the gauge fluxes do not render hypercharge massive, see section \ref{sec_fluxes}. 
In the sequel we classify the possible identifications along the following lines:
\begin{itemize}
\item
Since the fibrations under consideration contain only one type of $({\bf 3},{\bf 2})$-curve, all three generations of left-handed quark fields $Q$ must reside on this single $({\bf 3},{\bf 2})$-curve. The  $U(1)_1 \times U(1)_2$ charges of $Q$ for the five possible tops are listed in table \ref{tab:SU(2)xSU(3)}. 
\item
In a second step we identify the fields $(H_u,H_d)$ with two of the ${\bf 2}_i$-representations or their conjugate representations ${\bf \overline 2}_i$, $i=1,2,3$. 
A definite assignment of $(H_u,H_d)$ together with the $U(1)$ charges of $Q$ determines $U(1)_Y$ as a linear combination
\bea \label{hyper-ass}
U(1)_Y = a\,U(1)_1 + b\,U(1)_2, \quad a,b \in \mathbb{R}.
\eea
We then identify the different possible choices of ${\bf 2}_i$- or ${\bf \overline 2}_i$-states for the left-handed leptons $L$ based on their hypercharge.
The same value of $(a,b)$ and identification of $(H_u,H_d)$ may be compatible with more than one choice for $L$.
In this case, different generations of leptons $L$ may reside on different matter  curves and will then be distinguished by their charge under the linear combination of $U(1)_1$ and $U(1)_2$ orthogonal to $U(1)_Y$. 

\item 
For the specific values of $(a,b)$ in (\ref{hyper-ass}) we next check which of the six singlets ${\bf 1}^{(k)}$, $k=1, \ldots,6$ (and their conjugates) have the correct hypercharge to be identified with the fields $\nu^c_R$ and $e_R^c$,
and similarly which of the ${\bf \overline 3}_j$-representations for $j=1, \ldots, 5$ have the correct hypercharge to be identified with $u_R^c$ and $d_R^c$. 
If there is no possible assignment of $e_R^c$, $u_R^c$ or $d_R^c$ we discard this choice of hypercharge. However, to be as general as possible, we do allow for configurations with no right-handed neutrinos $\nu^c_R$.

\item
There are now two types of right-handed leptons and quarks: If the Yukawa couplings $W_1$ in (\ref{W1W2}) are indeed among the geometrically realised couplings as analysed in the previous sections, the fields acquire a perturbative mass term upon electro-weak symmetry breaking. Those generations of MSSM matter for which this is the case will therefore be called `heavy'.\footnote{\label{footcaveat}Note that if two or more families are localised on the same matter curve the rank of the perturbative Yukawa coupling matrix is non-maximal, at least if there exists only one Yukawa coupling point, as studied in the F-theory GUT literature \cite{Heckman:2008qa,Font:2009gq,Cecotti:2009zf,Conlon:2009qq}. In this case some of the `heavy' fields do not receive a perturbative mass after all. Non-perturbative effects can solve this rank-one problem \cite{Marchesano:2009rz,Aparicio:2011jx,Font:2012wq,Font:2013ida}.}
Otherwise, the Yukawas, which are now forbidden by the extra $U(1)$ selection rules, must be generated either by non-perturbative effects \cite{Blumenhagen:2006xt,Ibanez:2006da,Haack:2006cy,Florea:2006si,Blumenhagen:2009qh}, here by M5-brane instantons\footnote{The generation of charged operators via D3/M5-branes in F-theory along the lines of \cite{Blumenhagen:2006xt,Ibanez:2006da,Haack:2006cy,Florea:2006si,Blumenhagen:2009qh} has been studied recently in \cite{Marsano:2008py,Marsano:2008py,Blumenhagen:2010ja,Donagi:2010pd,Cvetic:2011gp,Marsano:2011nn,Grimm:2011dj,Kerstan:2012cy}, and related aspects of such instantons in F-theory appear in \cite{Cvetic:2009ah,Cvetic:2010rq,Bianchi:2011qh,Bianchi:2012kt,Cvetic:2012ts,Martucci:2014ema}.}, or via higher non-renormalisable couplings involving one or more extra singlet states as these acquire a VEV. 
In both cases the mass terms will generically be suppressed and the corresponding fields will be called `light'. Again it is understood that different generations can be distributed over the various matter curves.
In particular, if only one of the generations enjoys a perturbative coupling, this could serve as a realisation of the observed mass hierarchies in the MSSM.\footnote{The generation of such mass hierarchies in perturbative Type II MSSM quivers has been studied systematically in \cite{Ibanez:2008my,Cvetic:2009yh,Cvetic:2009ez}.} 
Note that while the generation of masses for the `light' generations by M5-instantons depends on the specific geometry of the base ${\cal B}$, the mechanism involving singlet fields could be analysed already at this general level by checking for the existence of singlets with appropriate $U(1)_i$-charges to form a dimension-5 coupling of the required type. We leave such a more advanced analysis for further studies.
\item
The $U(1)_Y$ charges together with the spectrum of perturbative couplings also provide candidates for $\mu$-singlets ${\bf 1}_\mu$ with a Yukawa coupling  ${\bf 1}_{\mu} H_u \, H_d $, which we list. As anticipated, if $\langle {\bf 1}_{\mu} 
\rangle \neq 0$ this will induce a $\mu$-term in the Higgs sector. In absence of such a VEV the $\mu$-term can in principle be generated via M5-instantons. 
Note that sometimes the same type of singlets can also have several interpretations. We furthermore list which other couplings in $W_{\rm singlet}$ are allowed. Since there is only one type of ${\bf 1}_\mu$, the term ${\bf 1}_\mu^3$  term is always forbidden perturbatively, but for $\nu_R^c$ a cubic coupling for the neutrino involving families distributed over different curves can exist. Note that tadpole terms in the superpotential, linear in the singlets, can only be generated non-perturbatively. Holomorphic quadratic terms involving either different families of $\nu_R^c$ or one $\nu_R^c$ and ${\bf 1}_\mu$ are allowed by gauge invariance for vector-like pairs of such fields, even though we do not list this explicitly. Determining their presence amounts to computing the vector-like spectrum of massless states. Otherwise quadratic singlet terms, especially Majorana mass terms for $\nu_R^c$, are only generated non-perturbatively \cite{Blumenhagen:2006xt,Ibanez:2006da} or as effective couplings from the cubic 
interactions with $\langle {\bf 1}_{\mu} \rangle \neq 0$.

\item Based on the various assignments of fields we list the perturbative $R$-parity violating dimension-four couplings $W_2$ in  (\ref{W1W2}) which are allowed in view of the structure of geometrically realised Yukawa couplings. 
More precisely, we list for which of the possible choice of (`heavy' or `light') right-handed quark and lepton fields a coupling of type $\alpha$, $\beta$, $\gamma$ is realised. The coupling $\kappa$ is allowed by the $U(1)$ section rules whenever $L$ and $H_u$ reside on the same matter curve and are thus vector-like with respect to all gauge symmetries. Such terms correspond to effective mass couplings and their presence can  be read off from the precise vector-like spectrum of the compactification, which can be computed in F-theory once the gauge background is specified \cite{Bies:2014sra}.

\item Finally we check for which matter identifications the potentially dangerous dimension-five couplings (\ref{W3K}) are perturbatively allowed, based on the $U(1)_i$ charges of the involved fields.
Note that in principle, this does not necessarily imply that the couplings are actually non-zero; to check this one would have to analyse in more detail how precisely the non-renormalisable couplings arise by exchange of heavy intermediate states. Depending on the details of the setup the resulting couplings can be negligibly small. This is left for a more in-depth analysis, and we take the results based purely on $U(1)_i$ selection rules merely as a first indication. 
Let us also note that some of the non-perturbative effects required to induce the Yukawa couplings for the `light' generations may at the same time induce other baryon or lepton-number violating or other undesirable operators \cite{Ibanez:2008my,Kiritsis:2009sf,Cvetic:2009yh,Cvetic:2009ez}.  We do not check for this possibility here.  
Furthermore we reiterate that the constraints on both dimension-four and -five couplings are relaxed in scenarios with intermediate or even high-scale supersymmetry breaking. 
\end{itemize}

The results of this scan over possible Standard Model-like embeddings is presented in appendix \ref{appsec:huge_table}. 
In configurations where the matter states can be localised at different curves, we do not list all possible combinations separately. In particular our analysis so far does not make any statements about whether it is possible to realise precisely the Standard Model context by inclusion of fluxes.
Irrespective of our relaxed attitude towards baryon and lepton number violation, a number of configurations exists in which all dangerous dimension-four and in particular the dimensions-five operators $\lambda_1$ and $\lambda_2$ in (\ref{W3K}) are absent as a result of the $U(1)$ selection rules. 
 
\subsection{A Specific Example}

As an example consider model number 5 in the top combination ${\rm I} \times {\rm A}$ listed in table \ref{tab:match_IxA} with $U(1)_Y = U(1)_1$. 
In perhaps the simplest scenario, the $H_u$, $H_d$ and all families of  left-handed leptons $L$ are realised as the states ${\bf \overline 2}^{\rm I}_1$, ${\bf \overline 2}^{\rm I}_2$ and  ${\bf \overline 2}^{\rm I}_3$ respectively. 
In particular,  $U(1)_2$ therefore distinguishes these states. Perturbative lepton masses arise if we identify $\nu_R^c  = { \bf \overline 1}^{(6)} $ and $e_R^c  = { \bf  1}^{(1)} $. For the choice $(u_R^c, d_R^c)  = ({\bf \overline 3}^{\rm A}_4, {\bf \overline 3}^{\rm A}_3)$ the quark masses are also realised perturbatively with the caveat noted in footnote \ref{footcaveat}.
For  the described assignment of matter all R-parity violating dimension-four couplings are perturbatively forbidden, as are the potentially problematic dimension-five couplings  $\lambda_1 \, Q \, Q \, Q \, L$  and $ \lambda_2 \, u_R^c \,  u_R^c \, d_R^c \, e_R^c $.

However, more complicated assignments are possible. For instance, if one or more families of leptons $L$ are instead identified with the state ${\bf \overline 2}_2$, then in this family the right-handed neutrino, which could be any of the states ${\bf 1}^{(5)}$, ${\bf 1}^{(6)}$ or their conjugates, does not have a perturbative Dirac mass. In this case, the Dirac mass would have to be generated directly by non-perturbative effects as proposed in \cite{Cvetic:2008hi}, naturally explaining the smallness of the neutrino masses via the non-perturbative suppression. If both types of matter identifications are combined for different families, a lepton-number violating dimension-four term ${\bf 2}_3 \,  {\bf 2}_3 \, {\bf 1}_2$ arises, where ${\bf 1}^{(2)}$ is now the `massive' $e_R^c$ which couples perturbatively  to the $L$-family ${\bf \overline 2}_2$. This coupling is innocuous for the proton as no baryon-lepton number violating terms are created.

%% file: section7_F-theory_Fluxes_v3.tex
\section{Fluxes and Chiral Spectrum} \label{sec_fluxes}

A chiral charged matter spectrum, ideally with three generations of MSSM matter and no chiral exotics, requires the introduction of suitable gauge fluxes.
In F/M-theory, gauge fluxes are described by 4-form fluxes $G_4 \in H^{2,2}(\hat Y_4)$ subject to a number of consistency conditions.
Apart from obeying transversality  \cite{oai:arXiv.org:hep-th/9908088},
\bea
\int_{\hat Y_4} G_4 \wedge Z \wedge \pi^{-1} D_a = \int_{\hat Y_4} G_4  \wedge \pi^{-1} D_a  \wedge \pi^{-1} D_b = 0  \qquad\quad \forall D_a, D_b \in H^{1,1}(\cal B)
\eea
with $Z$ the zero-section, and the quantisation condition \cite{oai:arXiv.org:hep-th/9609122,oai:arXiv.org:1011.6388,oai:arXiv.org:1203.4542}
\bea \label{quantisation}
G_4 + \frac{1}{2} c_2({\hat Y_4}) \in H^4(\hat Y_4, \mathbb Z),
\eea
the gauge flux $G_4$ must not break the $SU(3)$ and $SU(2)$ gauge group,
\bea
\int_{\hat Y_4} G_4 \wedge E_i \wedge \pi^{-1} D_a =0 \qquad \quad  \forall D_a  \in H^{1,1}(\cal B),
\eea
satisfy the D-term supersymmetry conditions for both $U(1)$ gauge groups with generating 2-forms $\omega_i$ \cite{Grimm:2010ks,Grimm:2011tb},
\bea
\int_{\hat Y_4} G_4 \wedge \omega_i \wedge J, \qquad i=1,2
\eea
with K\"ahler form $J$ inside the K\"ahler cone, as well as the D3-tadpole cancellation condition
$
\int_{\hat Y_4} G_4 \wedge G_4 + N_{D_3} = \frac{\chi(\hat Y_4)}{24}$.
In presence of such fluxes, the chiral index of matter in representation $R$ localised on matter curve $C_R \subset {\cal B}$ is given by the topological intersection number   \cite{oai:arXiv.org:0802.2969, oai:arXiv.org:0904.1218, Braun:2011zm,oai:arXiv.org:1108.1794,Krause:2011xj, oai:arXiv.org:1111.1232,oai:arXiv.org:1202.3138,oai:arXiv.org:1203.6662}      
\bea
\chi(R) = \int_{{\cal C}_R} G_4 .
\eea
Here ${\cal C}_R$ denotes the matter surface given by the fibration over $C_R$ of the linear combination of fibre $\mathbb P^1$s associated with the highest weight of representation $R$. 
Eventually we will need to know not only the chiral index, but the exact massless vector-like spectrum.
A proposal for the computation of the charged localised vector-like matter based on a description of the 3-form potential underlying $G_4$ in terms of Chow groups has been given recently in \cite{Bies:2014sra}.

An important restriction arises from the requirement that the MSSM hypercharge 
\bea
U(1)_Y = a \, U(1)_1 + b \, U(1)_2
\eea
must not receive a St\"uckelberg mass. This is guaranteed precisely if 
\bea
\int_{\hat Y_4} G_4 \wedge \omega_Y \wedge \pi^{-1} D_a = 0 \qquad \quad \forall D_a \in H^{1,1}({\cal B}), \quad {\rm where} \quad \omega_Y = a \, \omega_1 + b \, \omega_2
\eea
%where 
%\bea
%\omega_Y = a \, \omega_1 + b \, \omega_2
%\eea
is the hypercharge generator defined in terms of the generators of the two $U(1)_i$, c.f. (\ref{ShiodaSu2Su3}). This condition ensures that no $U(1)_Y$-D-term is induced by the flux, which is equivalent to stating that the fluxes do not lead to a $U(1)_Y$-dependent gauging of the axions as would be the case if the $U(1)_Y$ boson received a St\"uckelberg mass. 

To explicitly analyse these conditions, recall that the group $H^{2,2}(\hat Y_4)$ splits into a vertical part $H^{2,2}_{\rm vert}(\hat Y_4)$ and a horizontal part $H^{2,2}_{\rm hor}(\hat Y_4)$, the first of which is generated by products of elements in $H^{1,1}(\hat Y_4)$ \cite{Greene:1993vm,oai:arXiv.org:1203.6662}. Elements in $H^{2,2}_{\rm vert}(\hat Y_4)$ are comparatively straightforward to describe, and can in principle be classified completely, see e.g.~\cite{oai:arXiv.org:1202.3138} in the context of so-called $U(1)$-restricted $SU(N)\times U(1)$ models (for $N \leq 5$) and \cite{Cvetic:2013uta,Cvetic:2013jta} for $SU(5) \times U(1)_1 \times U(1)_2$ fibrations. While we leave a more systematic analysis of the possible vertical fluxes for the nine combinations of $SU(3) \times SU(2) \times U(1)_1 \times U(1)_2$ for future work, we here exemplify the general procedure.

First, the gauge fluxes associated with the $U(1)_1$ and $U(1)_2$ gauge symmetries are guaranteed to satisfy the transversality condition as an immediate consequence of the properties of the Shioda map that leads to the definition of the $U(1)_i$ generators $\omega_i$. In our situation therefore
\bea
G^{(i)}_4 = \pi^{-1} F_i \wedge \omega_i, \qquad F_i  \in H^{1,1}({\cal B}),    \qquad\quad i=1,2
\eea 
represent viable gauge fluxes subject to the remaining conditions listed above.
%In addition, other elements of $H^{2,2}_{\rm vert}(\hat Y_4)$ exist which satisfy transversality.
One more independent vertical flux exists already for the $U(1)_1 \times U(1)_2$-fibration without further non-abelian gauge group \cite{Cvetic:2013uta,Borchmann:2013hta,Cvetic:2013jta}. 
We will make use of the representation of this extra flux as the 4-form dual to one of the two singlet matter surfaces associated with the singlet ${\bf 1}^{(3)}$ appearing in table \ref{tab:singlets-charges}. More precisely, the extra flux can be described, in the notation of \cite{Borchmann:2013hta}, as
\bea \label{G4gamma}
G_4^{\gamma} = [\gamma] = \pi^{-1} [b_2] \wedge  \pi^{-1}[c_1]  - [b_2 \cap c_1 \cap s_0],
\eea
where the matter surface $\gamma$ is given as a complete intersection inside the ambient 5-fold of the fibration $\hat Y_4$,
\bea
 \gamma = b_2 \cap c_1 \cap \tilde P, \qquad\quad {P_T}|_{b_2 = c_1=0} = s_0 \tilde P.
\eea
In the second expression in (\ref{G4gamma}) it is used that the fibre over the curve $\{b_2\} \cap \{c_1\}$ splits into two $\mathbb P^1$s.
Additional vertical fluxes may exist in presence of extra non-abelian gauge groups. In fact, each of the 4-cycles associated with the charged matter surfaces defines a transverse cycle and thus defines a valid type of gauge flux subject to the remaining constraints. However, more work is required to check which of these lead to fluxes independent of the fluxes $G_4^{(i)}$ and $G_4^\gamma$, see \cite{Cvetic:2013uta,Cvetic:2013jta} for this analysis with extra non-abelian gauge group $SU(5)$. %We leave a systematic analysis for future work and merely exemplify the use of fluxes in our context for the $U(1)$-fluxes $G_4^{i}$ as well as $G_4^\gamma$. 

Evaluating the described constraints depends on the details of the tops under consideration.
Let us consider the combination of tops ${\rm I} \times {\rm A}$.
We first need to evaluate the constraints on the fluxes for $U(1)_Y$ to remain massless.
With the help in particular of the intersection numbers (2.18) - (2.22) in \cite{Borchmann:2013hta} and using the explicit the form  (\ref{eq:I-A-U(1)-generators}) of the $U(1)_i$ generators, one computes
\bea
\int_{\hat Y_4} \omega_1 \wedge \omega_1 \wedge \pi^{-1} D_a \wedge  \pi^{-1} D_b &=&  \int_{\cal B} (-2 \overline{\cal K} + \frac{1}{2} W_2 + \frac{2}{3} W_3)  \wedge  D_a \wedge  D_b, \\
\int_{\hat Y_4} \omega_2 \wedge \omega_2  \wedge \pi^{-1} D_a \wedge  \pi^{-1} D_b &=& \int_{\cal B} (-2 \overline{\cal K} - 2 [c_{1;0,0}] + \frac{2}{3} W_3)  \wedge  D_a \wedge  D_b, \\
\int_{\hat Y_4} \omega_1 \wedge \omega_2 \wedge \pi^{-1} D_a \wedge  \pi^{-1} D_b &=&  \int_{\cal B} (- \overline{\cal K} + [c_{2;0,1}] - [c_{1;0,0}] + \frac{2}{3} W_3)  \wedge  D_a \wedge  D_b,
\eea
where $D_a, D_b \in H^{1,1}({\cal B})$ and $W_2 = \{w_2 \}$ and $W_3=\{w_3 \}$ represent the $SU(2)$ and $SU(3)$ divisors on the base. The base sections $c_{1;0,0}$ and $c_{2;0,1}$ are defined in (\ref{eq:I-A-coeffs}).
% From this
% \bea
% \int_{\hat Y_4}(F_1 \wedge w_1 + F_2 \wedge w_2) \wedge w_Y \wedge J = xxx
% \eea

To compute the intersection numbers $\int G_4^\gamma \wedge \omega_i \wedge \pi^{-1}D_a$, we follow the procedure in  \cite{Borchmann:2013hta}  and exploit that the 4-cycle $\gamma$ dual to the flux $G_4^\gamma$ represents the weight of the singlet state ${\bf \overline 1}^{(3)}$ with charges $(-1,-2)$. With  $\int G_4^\gamma \wedge \omega_i \wedge \pi^{-1}D_a =  \int_\gamma \omega_i  \wedge \pi^{-1}D_a$  therefore
\bea
\int G_4^\gamma \wedge \omega_1 \wedge  \pi^{-1} D_a &=& (-1) \int_{\cal B} [b_{2;0,0}] \wedge [c_{1;0,0}] \wedge D_a, \\
\int G_4^\gamma \wedge \omega_2 \wedge  \pi^{-1} D_a &=& (-2) \int_{\cal B} [b_{2;0,0}] \wedge [c_{1;0,0}] \wedge D_a,
\eea 
because integration of the $U(1)_i$ generators $\omega_i$ over the highest-weight $\mathbb P^1$ in the fibre gives the $U(1)_i$ charge.
Altogether, if we only switch on flux $G_4 = G_4^{(1)} + G_4^{(2)} + \alpha \, G^\gamma_4$, the parameters $F_i \in H^{1,1}(\cal B)$ and $\alpha$ are constrained by masslessness of $U(1)_Y$ such that for all $D_a \in H^{1,1}({\cal B})$
\begin{align}
\begin{split}
\int_{\cal B} \, \Big[ & F_1 \wedge \Big(a (-2 \overline{\cal K} + \frac{1}{2} W_2 + \frac{2}{3} W_3)  +  b (-2 \overline{\cal K} - 2 [c_{1;0,0}] + \frac{2}{3} W_3)  \Big) - \alpha (a + 2b) \, [b_{2;0,0}] \wedge [c_{1;0,0}] + \\
& F_2 \wedge \Big(a  (-2 \overline{\cal K} - 2 [c_{1;0,0}] + \frac{2}{3} W_3) + b (- \overline{\cal K} + [c_{2;0,1}] - [c_{1;0,0}] + \frac{2}{3} W_3)   \Big)  \Big] \wedge D_a = 0 .
\end{split}
\end{align}
For a given base ${\cal B}$ one then expands all forms into a basis of $H^{1,1}_{\mathbb Z}(\cal B)$ and makes an ansatz for the $U(1)_i$ fluxes $F_i$ with suitably quantised coefficients (and same for $\alpha$)
in agreement with the flux quantisation condition (\ref{quantisation}).

The chiral index of the charged matter in representation $R$ with respect to the $U(1)_i$ fluxes is simply given by 
\bea
\int_{{\cal C}_R} G_4^{(i)} = q_i \int_{C_R} F_i,
\eea
where $q_i$ denotes the $U(1)_i$ charge.
To compute the index $\int_{{\cal C}_R} G_4^\gamma$ we analyse the geometric intersection of ${\cal C}_R$ with the 4-cycle ${\gamma}$.
For top ${\rm I} \times {\rm A}$ the only intersections occur with the surfaces associated with the states ${\bf 2}_2$ as well as  ${\bf \overline 3}_2$ and ${\bf \overline 3}_4$. The intersection product of $\gamma$ with the various matter surfaces is implicitly contained in tables \ref{tab:SU(2)-I-Yukawas} and \ref{tab:SU(3)-A-Yukawas}, which contain the Yukawa couplings with the singlet ${\bf 1}^{(3)}$.
More precisely, explicit analysis of the intersection shows that the topological intersection number of 4-cycle $\gamma$ with the fibration of $\mathbb P^1_{0B}$ over the ${\bf 2}_2$-curve  - see table \ref{tab:SU(2)-I-matter} - is given by 
$ - \int_{\cal B} [b_{2;0,0}] \wedge [c_{1;0,0}] \wedge W_2$. Since $\mathbb P^1_{0B}$ is the highest weight of the representation ${\bf \overline 2}_2$, this equals the chiral index for this state.
Similarly for $\mathbb P^1_{0u}$ over the ${\bf 3}_2$-curve listed in table \ref{tab:SU(3)-A-matter} one gets  $- \int_{\cal B}  [b_{2;0,0}] \wedge [c_{1;0,0}] \ \wedge  W_3$. 
Altogether thus $G_4^\gamma$ contributes 
\bea
\chi^\gamma({\bf \overline 2}_2) =  - \int_{\cal B} [b_{2;0,0}] \wedge [c_{1;0,0}] \wedge W_2, \qquad 
\chi^\gamma({\bf \overline 3}_2) =  \int_{\cal B}  [b_{2;0,0}] \wedge [c_{1;0,0}] \ \wedge  W_3 = - \chi^\gamma({\bf \overline 3}_4).
\eea

This merely exemplifies the use of fluxes and the constraints that have to be met in constructing vacua with a realistic particle spectrum. An explicit analysis of the set of possible fluxes is clearly beyond the scope of this work and left for future investigations.

%% file: section8_Conclusions_v3.tex
\section{Summary and Outlook}\label{sec_Conclusions}

F-theory is a unifying framework for the description of Type IIB compactifications with 7-branes which extends to an intrinsically non-perturbative regime. 
While recent phenomenological studies of F-theory have exploited its non-perturbative nature in the context of GUT models of particle physics, 
it is an equally exciting question to what extent direct, non-GUT realisations of the Standard Model within F-theory go beyond the known possibilities of perturbative models.
In view of the very different structure of Yukawa couplings in perturbative and non-perturbative brane vacua, F-theory is expected to encompass realisations of the Standard Model which cannot be studied in purely perturbative approaches. 

In this work we have taken some first steps towards a direct embedding of the Standard Model gauge group and matter fields into F-theory.
To this end we have constructed elliptic fibrations for F-theory compactifications with gauge group $SU(3) \times SU(2) \times U(1)_1 \times U(1)_2$. 
The fibrations considered arise as specialisations of the class of ${\rm Bl}_2 \mathbb P^3$-fibrations constructed in \cite{Borchmann:2013jwa,Cvetic:2013nia,Cvetic:2013uta,Borchmann:2013hta,Cvetic:2013jta} (see also \cite{Klemm:1996hh} for earlier work) with gauge group $U(1)_1 \times U(1)_2$. We have focused on the class of toric singularity 
enhancements leading to extra gauge group $SU(3) \times SU(2)$ along two in principle unrelated divisors $W_3$ and $W_2$. In this sense our construction differs from the earlier approaches to F-theory Standard Models reported in \cite{Choi:2013hua,Choi:2010su,Choi:2010nf}, which geometrically deform an underlying $SU(5)$ theory to the Standard Model. The structure of singularities in the class of toric $SU(3) \times SU(2)\times U(1)_1 \times U(1)_2$ models and their resolution is described by the combination of the $3 \times 3$ possible tops with gauge group $SU(3) \times SU(2)$ over polygon 5 in the classification of \cite{Bouchard:2003bu}; of the nine combinations only five are mutually inequivalent.
For generic choice of $SU(3)$ and $SU(2)$ divisors on the base, the spectrum of charged matter representations consists, in absence of fluxes, of a state $({\bf 3},{\bf  2})$, three types of $({\bf 1},{\bf 2})$-states and five types of $({\bf 3},{\bf 1})$-states plus conjugates, whose $U(1)_i$ charges we have computed. We have analysed the perturbative Yukawa interactions among these states including the sector of $U(1)_i$ charged singlets  \cite{Borchmann:2013jwa,Cvetic:2013nia,Cvetic:2013uta,Borchmann:2013hta,Cvetic:2013jta}. 
This analysis is independent of the specific base space ${\cal B}$ of the fibration. Given a concrete base space which allows for all the sections defining the fibration  one can then construct explicit elliptically fibred Calabi-Yau fourfolds, whose smoothness can be checked torically. We have exemplified this for a toy model over ${\cal B} = \mathbb P^3$; more complicated explicit realisations, ideally with rigid divisors for the non-abelian gauge groups along the lines of \cite{Blumenhagen:2009yv,Grimm:2009yu,Chen:2010ts,Knapp:2011wk}, are left for future work.

Based on these investigations, we have classified the possible identifications of the MSSM matter fields with the various representations present in each of the five inequivalent types of fibrations.
We allow for an extension of the MSSM by right-handed neutrinos and possibly by an extra singlet whose non-vanishing VEV generates a $\mu$-term in the Higgs sector.
One linear combination of  $U(1)_1$ and $U(1)_2$ describes the MSSM hypercharge, which must remain massless upon inclusion of gauge fluxes, while the orthogonal linear combination acts as a $U(1)$ selection and must acquire a flux-induced St\"uckelberg mass.
For suitable matter assignments in our list of possibilities, this extra $U(1)$ selection rule indeed forbids all dimension-four R-parity violating and the most dangerous dimension-five lepton- and baryon-number violating  effective couplings. As stressed and investigated in \cite{Ibanez:2008my,Kiritsis:2009sf,Cvetic:2009yh,Cvetic:2009ez}, such couplings can be introduced non-perturbatively if some of the required Yukawas are generated by instantons. 
A study of this effect is left for future work. In fact, in our classification of possible Standard Model identifications we  
do a priori not insist on absence of all lepton- and baryon-number violating dimension-four and -five couplings even at the perturbative level, but merely list which of these couplings are geometrically realised. While most of them would certainly have to be excluded for a TeV-scale supersymmetry breaking scale -- see \cite{Cvetic:2009yh,Cvetic:2009ez,Anastasopoulos:2012zu} for this approach in the context of perturbative MSSM quivers and \cite{Cvetic:2010dz} in the context of MSSM quivers with singlet extensions -- the constraints on some of the couplings are more relaxed in intermediate or high-scale supersymmetry breaking scenarios. 
%Depending on the specifics of supersymmetry breaking the existence of  such couplings may be tolerated to a certain extent.  
Based on our classification of potential Standard Model identifications, an interesting task for future work will be an in-depth analysis of the effects of such couplings depending on the spectrum of superpartners.

In view of our original motivation as stated at the beginning of this section, it will furthermore be interesting to investigate which of the configurations admit a well-defined weak-coupling limit. This will in turn identify those potential Standard Model realisations which truly go beyond perturbative models, and it will be illuminating to distill their characteristic physical properties in contradistinction to perturbative D-brane vacua.

Another important question to study will be the constraints on the moduli of the compactification for the running of the gauge couplings to be consistent with their observed value at the weak scale. 
The gauge couplings depend on the K\"ahler moduli of the compactification, which therefore must obey certain relations in order to reproduce the approximate unification of gauge couplings at a higher scale, the detailed form of which depends of course on the precise  spectrum of intermediate states. Such relations have been studied in the perturbative Type IIA framework in \cite{Blumenhagen:2003jy}. The need for the moduli to obey such relations might seem ad hoc to GUT model builders; on the other hand even in GUT realisations of string theory important corrections, in the example of F-theory due to hypercharge flux or Kaluza-Klein states \cite{Donagi:2008kj,Blumenhagen:2008aw,Conlon:2009qa,Mayrhofer:2013ara,Hebecker:2014uaa}, render unification less non-trivial  than one might think.

The construction of phenomenologically viable F-theory vacua requires the inclusion of gauge fluxes in such a way as  to reproduce the chiral spectrum of the Standard Model.
As we have discussed, the fluxes must be chosen such as to respect masslessness of the specific linear combination of $U(1)_1$ and $U(1)_2$ corresponding to hypercharge.
In future work we plan to systematically analyse large classes of consistent gauge fluxes in order to determine which of the potential Standard Model identifications are actually compatible with three families of MSSM spectrum and absence of chiral exotics. Indeed, consistency of the compactification including the gauge flux data is known to restrict the allowed spectrum, sometimes even beyond a purely 4-dimensional field theory analysis as studied in the perturbative framework in \cite{Cvetic:2011iq} (see also \cite{Halverson:2013ska}). 
Apart from the inclusion of suitable fluxes, there are various modifications of the geometry conceivable in order to avoid unwanted exotic states:
Some of the matter curves are inexistent for suitable choices of classes $\alpha$ and $\beta$ which determine the fibration as summarised in table \ref{coeff}. It will be interesting to analyse in detail which special choices are possible such as to `switch off' a maximal number of exotic matter curves without affecting the Standard Model sector. Furthermore, the spectrum of massless states changes in the presence of vacuum expectation values for some of the singlets. This corresponds either to a recombination process or to a gluing \cite{Donagi:2011jy,Donagi:2011dv} of branes and can turn out instrumental in concrete model building.

Finally we stress that the analysis of the massless spectrum  has to go beyond the chiral index and include also a computation of the  the vector-like spectrum of states. 
This implies that the $C_3$-gauge field background must be specified more accurately than merely in terms of the gauge flux $G_4$. In \cite{Bies:2014sra} it was shown how to specify the gauge data via  rational equivalence classes of 4-cycles, and  a natural candidate was proposed for the cohomology groups counting the exact vector-like matter spectrum individually. 
We look forward to applying this technology in our search for realistic F-theory non-GUTs.

\subsubsection*{Acknowledgements}

\noindent We thank Philipp Arras, Arthur Hebecker, Luis Ibanez, Christoph Mayrhofer, Dave Morrison, Eran Palti, Hernan Piragua and Oskar Till for important discussions.
This work was supported in part by Deutsche Forschungsgeimeinschaft under TR 33 `The Dark Universe' and by Studienstiftung des Deutschen Volkes.

%% file: section_appendix_v3.tex
\newpage

\begin{appendix}

\section[Details on the Toric Diagrams]{Details on the Toric Diagrams}\label{app:tops}

In this appendix we present the toric diagrams of the $SU(2)$ and $SU(3)$ tops. Special emphasis will be put on the symmetries which identify some of the models as equivalent pairs. The relationship between toric geometry and the geometry of Calabi-Yau hypersurfaces is described in e.g.~\cite{Bouchard:2003bu}.

\begin{figure}[ht]
\begin{center}
	\def\svgwidth{.98\hsize}
	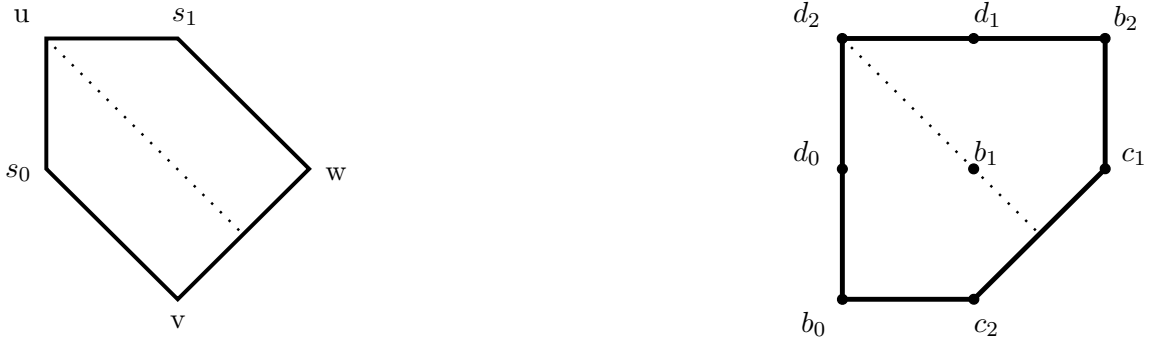
\end{center}
\caption{Polygon 5 (in the classification of \cite{Bouchard:2003bu}) describing the fibre ambient space ${\rm Bl}_2 {\mathbb P}^2$; every lattice point of the dual polygon (right) gives an individual term of the hypersurface equation. The reflection symmetry along the dotted diagonal is manifest.}
\label{fig:base_polygon}
\end{figure}

The ambient space ${\rm Bl}_2 {\mathbb P}^2$ has the toric diagram depicted on the left in figure \ref{fig:base_polygon}. The lattice points of  the dual diagram on the right correspond to the terms of the hypersurface equation (\ref{eq:hypersurface-equation}). Clearly the diagrams share a common reflection symmetry along the dotted diagonal axis.

The symmetry exchanges fibre coordinates $s_0 \leftrightarrow s_1, \, \sv \leftrightarrow \sw$ and coefficients $b_0 \leftrightarrow b_2, \, c_1 \leftrightarrow c_2 , \, d_0 \leftrightarrow d_1$ of the hypersurface equation. Consequently, the $U(1)$ generators (\ref{eq:U(1)-generators}) are also transformed, namely as
\begin{align}\label{eq:omega-transformation}
\begin{split}
	\omega_1 &= S_1 - S_0 - \overline{\cal K} \longrightarrow S_0 - S_1 - \overline{\cal K} = - \omega_1 + 2\overline{\cal K}\, , \\
	\omega_2 &= U - S_0 - \overline{\cal K} - [c_1] \longrightarrow U -S_1 - \overline{\cal K} - [c_2] = \omega_2 - \omega_1 - \overline{\cal K}+[c_1]-[c_2] \, .
\end{split}
\end{align}
These forms do not satisfy the verticality condition, i.e.~they are not  in the image of the Shioda map. However they only differ from such by the pullback of divisors of the base. Since such pullbacks never contribute to the $U(1)$ charges of any states, the $U(1)$ charges indeed transform as
\begin{align}
\begin{split}
	U(1)'_1 &= - U(1)_1 \, , \\
	U(1)'_2 &= U(1)_2 - U(1)_1 \, .
\end{split}
\end{align}
The symmetry also exchanges the singlets (as the coefficients $b_i, c_j, d_k$ are exchanged), namely ${\bf 1}^{(1)} \leftrightarrow \overline{\bf 1}^{(3)}, {\bf 1}^{(2)} \leftrightarrow \overline{\bf 1}^{(4)}$, while ${\bf 1}^{(5)}$ and ${\bf 1}^{(6)}$ are invariant.

\begin{figure}[ht]
\begin{center}
	\def\svgwidth{.98\hsize}
	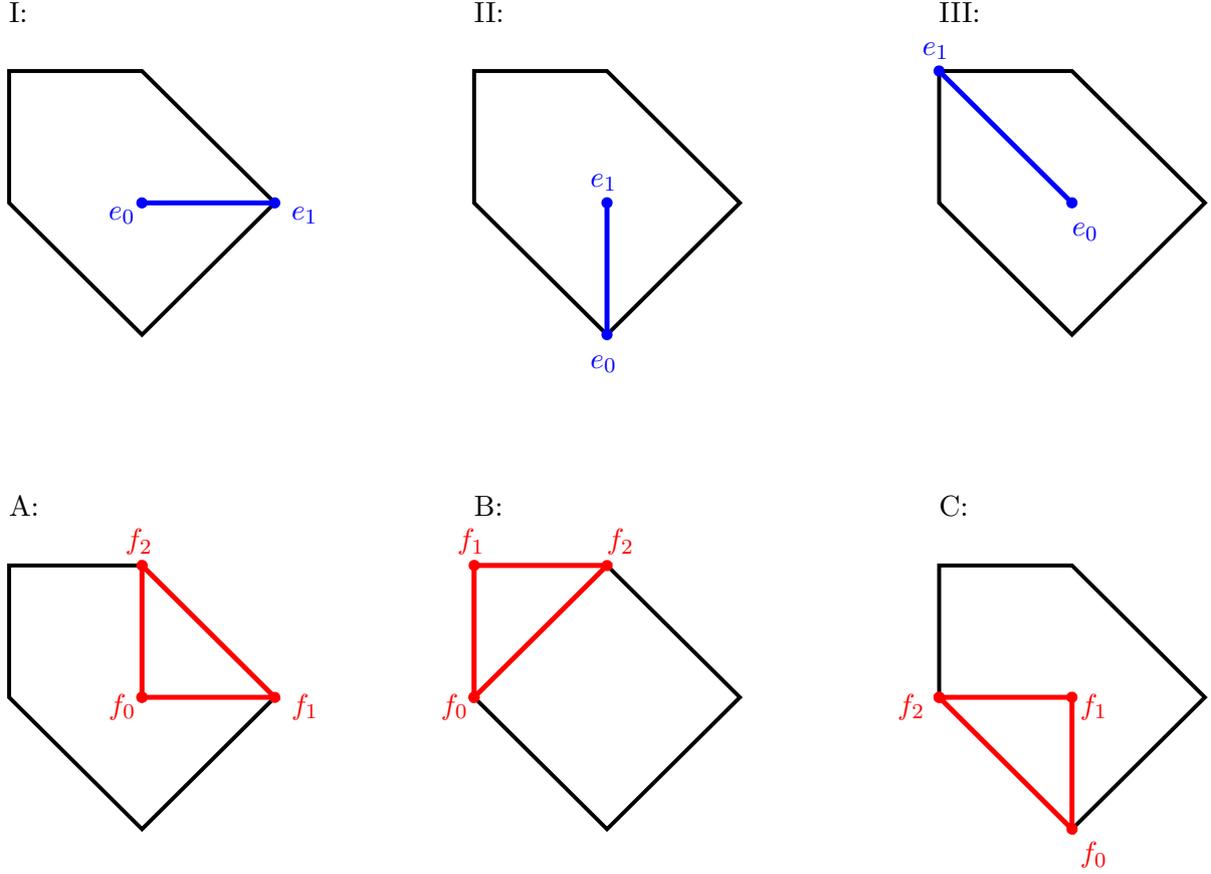
\end{center}
\caption{Possible $SU(2)$ (upper) and $SU(3)$ (lower) tops. The coloured lines and vertices are the `top layer' of the three-dimensional toric diagram, projected down onto the layer containing the base polygon representing the fibre ambient space.}
\label{fig:tops}
\end{figure}
The same symmetry relates the $SU(2)$ and $SU(3)$ tops as well as combinations of those. As shown in figure \ref{fig:tops}, the $SU(2)$-I and -II tops are precisely matched onto each other, however only if one also exchanges the resolution coordinates $e_0 \leftrightarrow e_1$. The $U(1)$ generators on both sides can be similarly matched. The first generator transforms as $\omega^{\rm I}_1 \longrightarrow S^{\rm II}_0 - S^{\rm II}_1 - \overline{\cal K} + \frac{1}{2} E^{\rm II}_0 = -(S^{\rm II}_1 - S^{\rm II}_0 - \overline{\cal K} + \frac{1}{2} E^{\rm II}_1) + 2\overline{\cal K} + \frac{1}{2} \pi^* W_2 = - \omega^{\rm II}_1 + 2\overline{\cal K} + \frac{1}{2} \pi^* W_2$, where the first equality exploits the relation $E_0 + E_1 = \pi^* W_2$ for the resolution divisors of an $SU(2)$ singularity. For the second generator one now needs to take into account that $c^{\rm I}_1 = c^{\rm I}_{1,0} \, e^{\rm I}_1$ is mapped onto $c^{\rm II}_2 = c^{\rm II}_{2,1} \, e^{\rm II}_0$, from which one can easily verify the transformation $\omega^{\rm I}_2 \longrightarrow \omega^{\rm II}_2 - \omega^{\
\rm II}_1 + [c^{\rm II}_1] - [c^{\rm II}_{2,1}] - \overline{\cal K}$. As mentioned after (\ref{eq:U1-charge-transformation}), the spectrum of $SU(2)$-charged states is also exchanged as ${\bf 2}^{\rm I}_i \leftrightarrow \overline{\bf 2}^{\rm II}_i, i=1,2,3$. Obviously the $SU(2)$-III top is invariant under the reflection symmetry. The spectrum and $U(1)$ charges transform as stated in subsection \ref{subsec:SU(2)-tops-short}. Similarly one can see that the $SU(3)$-A and -C tops are equivalent to each other, while the -B top is invariant under reflection. Analogous calculations as above show that the $U(1)$ generators transform accordingly.

\begin{figure}[ht]
\begin{center}
	\def\svgwidth{.98\hsize}
	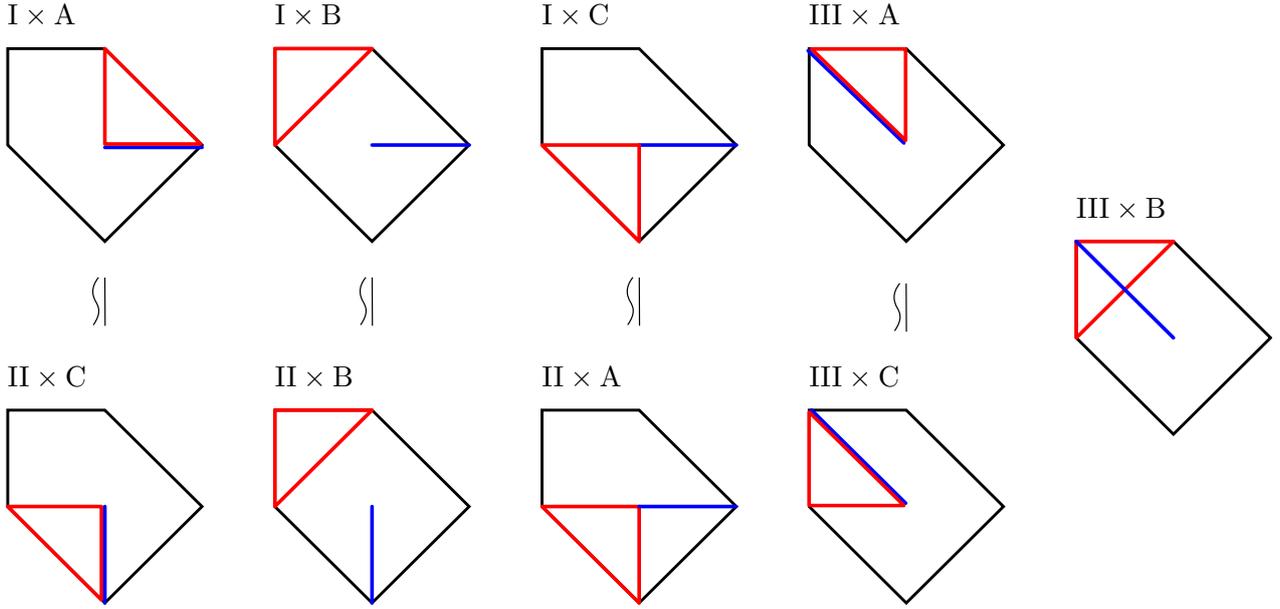
\end{center}
\caption{The combination of $SU(2)$ and $SU(3)$ tops gives rise to five inequivalent models. Their toric diagram lies in a four-dimensional lattice, where the `top layers' corresponding to $SU(2)$ and $SU(3)$ resolution divisors extend into two linearly independent directions that do not lie in the plane spanned by the base polygon. For this figure we have projected the tops down into said plane. The four pairs of tops that are equivalent are related to each other by reflection along the diagonal in the plane of the base polygon.}
\label{fig:combined_tops}
\end{figure}

When we combine the $SU(2)$ and $SU(3)$ tops, we see that among the nine possibilities there are in fact five inequivalent models with $SU(2) \times SU(3)$ gauge group. The redundancy comes again from reflecting along the symmetry axis of the base polygon, which identifies four pairs of models to be equivalent (cf.~figure \ref{fig:combined_tops}). The combination ${\rm III} \times {\rm B}$ is invariant under the reflection transformation. The equivalence of the pairs of models can also be checked in a similar fashion as above, by inspecting the transformation of the $U(1)$ generators and the matter states.

\section[Details on \texorpdfstring{\boldmath $SU(2)$}{SU(2)}-II and -III Tops]{Details on \texorpdfstring{\boldmath $SU(2)$}{SU(2)}-II and -III Tops} \label{SU2DetailsApp}

In this part we provide more details on the matter and Yukawa couplings of the $SU(2)$-II and -III top which have not been discussed at length in the corpus of this paper.

\subsection[\texorpdfstring{$SU(2)$}{SU(2)}-II Top]{\texorpdfstring{{\boldmath $SU(2)$}}{SU(2)}-II Top} \label{SU2App1}

The top corresponds to restricting the hypersurface coefficients as
\begin{align}\label{eq:SU(2)-II-coeffs}
  b_0 = b_{0,1} \, e_0, \quad b_2 = b_{2,0} \, e_1, \quad c_2 = c_{2,1} \, e_0 , \quad d_1 = d_{1,0} \, e_1 , \quad d_2 = d_{2,0} \, e_1. 
\end{align}
There are two possible SR-ideals, of which we choose
\begin{align}\label{eq:SU(2)-II-SR-ideal}
  \su \, \sv, \su \, \sw , \sw \, s_0, \sv \, s_1 , s_0 \, s_1 , e_0 \, s_1 , e_0 \, \su , e_1 \, \sv .
\end{align}
The scaling relations and divisor classes for this top are as follows:
\begin{align}\label{tab:SU(2)-II-divisor-classes}
  \begin{array}{c|cccccc|c}
    \hphantom{U} & \su & \sv & \sw & s_0 & s_1 & e_1 & e_0 \\ \hline
    \mathrm{U} & 1 & 1 & 1 & \cdot & \cdot & \cdot & \cdot \\
    S_0 & \cdot & \cdot & 1 & 1& \cdot & \cdot & \cdot \\
    S_1 & \cdot & 1 & \cdot & \cdot & 1 & \cdot & \cdot \\
    E_1 & \cdot & 1 & \cdot & \cdot & \cdot & 1 & -1
  \end{array}
\end{align}
The $U(1)$ generators, normalised such that the $SU(2)$ root remains uncharged, are given by
\begin{align}\label{eq:SU(2)-II-U(1)-generators}
  \begin{split}
    \omega_1^\text{II} &= S_1 - S_0 - \overline{\mathcal{K}} + \frac{1}{2} E_1, \\
    \omega_2^\text{II} &= U - S_0 - \overline{\mathcal{K}} - [c_1] + \frac{1}{2} E_1.
  \end{split}
\end{align}

The discriminant locus now takes the form
\bea \label{Dis2II}
\Delta \simeq w_2^2 \Big(  c_{1}  \,  ( c_{2,1}^2 \, d_0 - b_{0,1} \, b_1 \, c_{2,1} + b_{0,1}^2 \, c_1 ) \,  \ell_3   \,  (b_1^2 - 4 c_1 d_0)^2 \, + {\cal O}(w_2) \,   \Big).  
\eea

% \subsubsection*{Matter Curves}
The fibres over the intersection of the first three factors inside the big bracket in (\ref{Dis2II})  with $\{w_2\}$ are indeed of split Kodaira type $I_3$.
A similar analysis as for the  $SU(2)$-I top confirms matter in the $\mathbf{2}$ representation (together with their charge conjugates $\overline{\mathbf{2}}$) on the matter curves displayed in table \ref{tab:SU(2)-II-matter}.
\begin{table}[ht]
\begin{align*}
  \begin{array}{c|c|c|c}
    \text{matter} & \text{locus} = W_2 \cap \ldots & \text{splitting of fibre components} & U(1)-\text{charges} \\ \hline \rule{0pt}{3ex}
    \mathbf{2}^{\mathrm{II}}_1 & \{c_{1}\} & \mathbb{P}^1_1 \rightarrow \mathbb{P}^1_{1s_0}\!+ \mathbb{P}^1_{1A} & (\frac{1}{2}, \frac{3}{2}) \\[.5ex]\hline \rule{0pt}{3ex}
    \mathbf{2}^{\mathrm{II}}_2 & \{ c_{2,1}^2 \, d_0 - b_{0,1} \, b_1 \, c_{2,1} + b_{0,1}^2 \, c_1 \} & \mathbb{P}^1_1 \rightarrow \mathbb{P}^1_{1B} + \mathbb{P}^1_{1C} & (\frac{1}{2},-\frac{1}{2}) \\[.5ex]\hline \rule{0pt}{3ex}
    & \{\ell_3\} := \{ b_{2,0}^2\,d_0^2 & & \\
    \mathbf{2}^\mathrm{II}_3 & + b_{2,0}\,( b_1^2\,d_{2,0} - 2\,c_1\,d_0\,d_{2,0} - b_1\,d_0\,d_{1,0}) & \mathbb{P}^1_0 \rightarrow \mathbb{P}^1_{0A} + \mathbb{P}^1_{0B} & (\frac{1}{2}, \frac{1}{2}) \\
    & + c_1 (d_0\,d_{1,0}^2 + d_{2,0}(c_1\,d_{2,0} - b_1\,d_{1,0} ) ) \} & &
  \end{array}
\end{align*}
\caption{Matter states in the $SU(2)$-II top.}
\label{tab:SU(2)-II-matter}
\end{table}

The splitting of the fibre components over the first curve and the resulting enhancement of the intersection structure to that of an affine $SU(3)$ diagram is straightforward to see. For the second curve, we factorise 
\bea\label{eq:SU(2)-II-quadratic-curve}
 c_{2,1}^2 \, d_0 - b_{0,1} \, b_1 \, c_{2,1} + b_{0,1}^2 \, c_1 =\frac{1}{d_0}  \,     {\mathcal{C}_+}  \, {\mathcal{C}_-} \quad {\rm with} \quad  {\mathcal{C}_\pm} =  c_{2,1}\,d_0 - b_{0,1} \left( \frac{b_1}{2} \pm \sqrt{ \frac{b_1^2}{4} - c_1\,d_0} \right), 
\eea 
which splits the curve into two parts $W_2 \cap \{C_\pm=0\}$ that are connected at a branch cut. Similar to the results (\ref{eq:SU(2)-I-splitting-quadratic-curve}) of the first $SU(2)$ top, we find that over these two parts the fibre of the divisor $E_1$ splits into two components, $\mathbb{P}^1_{1B}$ and $\mathbb{P}^1_{1C}$, that can be extended over the whole curve without being interchanged by any monodromy. The intersection structure is again that of the affine $SU(3)$ diagram.
Analogously, the third curve can be written as $ W_2 \cap \ell_3$  with 
\begin{align}\label{eq:SU(2)-II-complicated-quadratic-curve}
\ell_3 = 1/d_0^2 \, \mathcal{D}_+ \, \mathcal{D}_-, \qquad \quad             \mathcal{D}_\pm = b_{2,0}\,d_0^2 - \left[ c_1\,d_0\,d_{2,0} + (d_0\,d_{1,0} - b_1\,d_{2,0}) \left( \frac{b_1}{2} \pm \sqrt{ \frac{b_1^2}{4} - c_1\,d_0 } \right)\!\right] .
\end{align}
A similar calculation as (\ref{eq:SU(2)-I-splitting-complicated-quadratic-curve}) shows that the divisor $E_0$ splits into $\mathbb{P}^1_{0A} + \mathbb{P}^1_{0B}$, with both components well-defined over the whole curve. As expected one finds the intersection structure to be an affine $SU(3)$ diagram.

Note that apart from the three matter curves discussed above, the vanishing order of the discriminant increases from $2$ to $3$ also along the curve $\{ w_2 \} \cap \{b_1^2 - 4 c_1 d_0  \}$; however, the fibre is of Kodaira type $III$ since the Weierstrass sections $f$ and  $g$
vanish to order $1$ and $2$ respectively. Thus no extra  charged matter representations arise here, in agreement with the formalism of \cite{Grassi:2011hq}.

The Yukawa couplings involving $SU(2)$ matter are summarised in table \ref{tab:SU(2)-II-Yukawas}. The fibre structure enhancement for each Yukawa point can be read off from the last column in an analogous fashion as with the first $SU(2)$ top (cf.~table \ref{tab:SU(2)-I-Yukawas}). We find the affine $SU(4)$ diagram as the intersection structure over all Yukawa points. Note that for the second and third pair of Yukawa points, the same split products of each pair arrange themselves into different intersection \textit{patterns}, realising either $\mathbf{2}_i-\mathbf{2}_j-\mathbf{1}/\overline{\mathbf{1}}$ or $\mathbf{2}_i-\overline{\mathbf{2}}_j-\mathbf{1}/\overline{\mathbf{1}}$ couplings.

\begin{table}[ht]
\begin{align*}
  \begin{array}{c|c|c}
    \text{coupling} & \text{locus} = W_2 \cap \ldots & \text{splitting of fibre components} \\ \hline\hline \rule{0pt}{3ex}
    \mathbf{2}_1^{\mathrm{II}} - \mathbf{2}^{\mathrm{II}}_2 - \overline{\mathbf{1}}^{(4)} & \{c_1\} \cap \{c_{2,1}\,d_0 - b_{0,1}\,b_1 \} & \mathbb{P}^1_1 \rightarrow \mathbb{P}^1_{1s_0 C} \! + \mathbb{P}^1_{1AB} + \mathbb{P}^1_{1AC} \\[.5ex] \hline \rule{0pt}{3ex}
    \mathbf{2}^{\mathrm{II}}_1 - \overline{\mathbf{2}}^{\mathrm{II}}_2 - \overline{\mathbf{1}}^{(5)} & \{c_1\} \cap \{c_{2,1}\} & \mathbb{P}^1_1 \rightarrow \mathbb{P}^1_{1s_0 B} \! + \mathbb{P}^1_{1AB'} + \mathbb{P}^1_{1AC'} \\[.5ex] \hline\hline \rule{0pt}{3ex}
    \mathbf{2}^{\mathrm{II}}_1 - \mathbf{2}^{\mathrm{II}}_3 - \overline{\mathbf{1}}^{(3)} & \{c_1\} \cap \{b_{2,0}\} & \mathbb{P}^1_0 \rightarrow \mathbb{P}^1_{0A} + \mathbb{P}^1_{0B}, \, \mathbb{P}^1_1 \rightarrow \mathbb{P}^1_{1s_0}\!+ \mathbb{P}^1_{1A}  \\[.5ex] \hline \rule{0pt}{3ex}
    \mathbf{2}^{\mathrm{II}}_1 - \overline{\mathbf{2}}^{\mathrm{II}}_3 - \overline{\mathbf{1}}^{(6)} & \{c_1\} \cap \{b_{2,0}\,d_0^2 + b_1\,(d_0\,d_{1,0} - b_1\,d_{2,0}) \} & \mathbb{P}^1_0 \rightarrow \mathbb{P}^1_{0A} + \mathbb{P}^1_{0B}, \, \mathbb{P}^1_1 \rightarrow \mathbb{P}^1_{1s_0}\!+ \mathbb{P}^1_{1A}  \\[.5ex] \hline\hline \rule{0pt}{3ex}
    \mathbf{2}^{\mathrm{II}}_2 - \mathbf{2}^{\mathrm{II}}_3 - \overline{\mathbf{1}}^{(2)} & \left( \{\mathcal{C}_+\} \cap \{\mathcal{D}_+\} \right) \cup \left( \{\mathcal{C}_-\} \cap \{\mathcal{D}_-\} \right) & \mathbb{P}^1_0 \rightarrow \mathbb{P}^1_{0A} + \mathbb{P}^1_{0B}, \, \mathbb{P}^1_1 \rightarrow \mathbb{P}^1_{1B} + \mathbb{P}^1_{1C}  \\[.5ex] \hline \rule{0pt}{3ex}
    \mathbf{2}^{\mathrm{II}}_2 - \overline{\mathbf{2}}^{\mathrm{II}}_3 - \mathbf{1}^{(6)} & \left( \{\mathcal{C}_+\} \cap \{\mathcal{D}_-\} \right) \cup \left( \{\mathcal{C}_-\} \cap \{\mathcal{D}_+\} \right) & \mathbb{P}^1_0 \rightarrow \mathbb{P}^1_{0A} + \mathbb{P}^1_{0B}, \, \mathbb{P}^1_1 \rightarrow \mathbb{P}^1_{1B} + \mathbb{P}^1_{1C}  \\[.5ex] \hline\hline \rule{0pt}{3ex}
    \mathbf{2}^{\mathrm{II}}_2 - \mathbf{2}^{\mathrm{II}}_2 - \overline{\mathbf{1}}^{(1)} & \{b_{0,1}\} \cap \{c_{2,1}\} & \mathbb{P}^1_1 \rightarrow \mathbb{P}^1_{1B} \! + \mathbb{P}^1_{1s_1 C} + \mathbb{P}^1_{1C'} \\[.5ex] \hline \rule{0pt}{3ex}
    \mathbf{2}^{\mathrm{II}}_3 - \mathbf{2}^{\mathrm{II}}_3 - \overline{\mathbf{1}}^{(4)} & \{b_1\,d_{2,0} - d_0\,d_{1,0}\} \cap \{c_1\,d_{2,0} - b_{2,0}\,d_0\} & \mathbb{P}^1_0 \rightarrow \mathbb{P}^1_{0A} \! + \mathbb{P}^1_{0B'} + \mathbb{P}^1_{0B''}
  \end{array}
\end{align*}
\caption{Yukawa couplings in the $SU(2)$-II top.}
\label{tab:SU(2)-II-Yukawas}
\end{table}

\subsection[\texorpdfstring{{$SU(2)$}}{SU(2)}-III Top]{\texorpdfstring{{\boldmath $SU(2)$}}{SU(2)}-III Top}\label{SU2App2}

This top restricts the hypersurface coefficients as 
\begin{align}\label{eq:SU(2)-III-coeffs}
  c_1 = c_{1,1} \, e_0, \quad c_2 = c_{2,1} \, e_0, \quad d_0 = d_{0,0} \, e_1 , \quad d_1 = d_{1,0} \, e_1 , \quad d_2 = d_{2,0} \, e_1^2. 
\end{align}
For this top there are four possible SR-ideals, of which we choose
\begin{align}\label{eq:SU(2)-III-SR-ideal}
  \su \, \sv, \su \, \sw , \sw \, s_0, \sv \, s_1 , s_0 \, s_1 , e_0 \, \su , e_0 \, s_0 , e_1 \, \sw .
\end{align}
The scaling relations and divisor classes for this top are
\begin{align}\label{tab:SU(2)-III-divisor-classes}
  \begin{array}{c|cccccc|c}
    \hphantom{U} & \su & \sv & \sw & s_0 & s_1 & e_1 & e_0 \\ \hline
    \mathrm{U} & 1 & 1 & 1 & \cdot & \cdot & \cdot & \cdot \\
    S_0 & \cdot & \cdot & 1 & 1& \cdot & \cdot & \cdot \\
    S_1 & \cdot & 1 & \cdot & \cdot & 1 & \cdot & \cdot \\
    E_1 & \cdot & 1 & 1 & \cdot & \cdot & 1 & -1
  \end{array}
\end{align}
The $SU(2)$ root in this top is uncharged under the generators (\ref{eq:U(1)-generators}) so that no correction term is needed,
\begin{align}\label{eq:SU(2)-III-U(1)-generators}
  \begin{split}
    \omega_1^\text{III} &= S_1 - S_0 - \overline{\mathcal{K}}, \\
    \omega_2^\text{III} &= U - S_0 - \overline{\mathcal{K}} - [c_{1,1}].
  \end{split}
\end{align}
This time the discriminant of the singular blow-down takes the form
\bea
\Delta  &=& w_2^2 \Big(  b_0 \, b_2 \,  (b_0 \, c_{1,1}^2 - b_1\,c_{1,1}\,c_{2,1} + b_2\,c_{2,1}^2)  (b_1^2 - 4 b_0 b_2)^2 \,   \nonumber \\
&  &  \phantom{w_2^2 \Big( }   ( -b_2 d_0^2 + b_1 d_{0,0} d_{1,0} - b_0 d_{1,0}^2  - b_1^2 d_{2,0} + 4 b_0 b_2 d_{2,0})  + {\cal O}(w_2) \Big). \label{DisSu2III}
\eea

Over the intersection of  $\{w_2\}$ with the first three factors in the bracket in (\ref{DisSu2III}), the fibre type enhances to  split Kodaira type $I_3$. This gives rise to matter states in the $\mathbf{2}$ representation (together with their charge conjugate $\overline{\mathbf{2}}$ states) summarised in tabe \ref{tab:SU(2)-III-matter}. 

By setting $b_0 / b_2=0$ in the hypersurface equation, it is again straightforward to see the splitting process and the enhancement of the intersection structure to an affine $SU(3)$ diagram over the first/second $\mathbf{2}$-curve.
The quadratic equation defining the third curve can be again factorised analogously to (\ref{eq:SU(2)-I-quadratic-curve}). However, because of the Yukawa points that are present in this top (see below), we need two different factorisations,
\bea
 b_0 \, c_{1,1}^2 - b_1\,c_{1,1}\,c_{2,1} + b_2\,c_{2,1}^2    =  \frac{1}{b_0} \, {\cal C}_+ \, {\cal C}_- =    \frac{1}{b_2} \, {\cal D}_+ \, {\cal D}_-   
 \eea
 with
 \bea \label{eq:SU(2)-III-quadratic-curve-2}
  {\cal C}_\pm   =  c_{1,1}\,b_0 - c_{2,1} \left( \frac{b_1}{2} \pm  \sqrt{ \frac{b_1^2}{4} - b_0\,b_2 } \right), \qquad 
   {\cal D}_{\pm} =   c_{2,1}\,b_2 - c_{1,1}\left( \frac{b_1}{2} \mp \sqrt{ \frac{b_1^2}{4} - b_0\,b_2 } \right). 
  \eea

%     \notag \\[1ex]
%
%  = \, & \frac{1}{b_0} \underbrace{\left(c_{1,1}\,b_0 - c_{2,1} \left( \frac{b_1}{2} + \sqrt{ \frac{b_1^2}{4} - b_0\,b_2 } \right) \right)}_{\mathcal{C}} \underbrace{\left( c_{1,1}\,b_0 - c_{2,1} \left( \frac{b_1}{2} - \sqrt{ \frac{b_1^2}{4} - b_0\,b_2 } \right) \right)}_{\mathcal{D}} \label{eq:SU(2)-III-quadratic-curve-1} \\[-.6ex]
%
%  = \, & \frac{1}{b_2} \overbrace{\left(c_{2,1}\,b_2 - c_{1,1}\left( \frac{b_1}{2} - \sqrt{ \frac{b_1^2}{4} - b_0\,b_2 } \right) \right)} \overbrace{\left(c_{2,1}\,b_2 - c_{1,1}\left( \frac{b_1}{2} + \sqrt{ \frac{b_1^2}{4} - % b_0\,b_2 } \right) \right)}\label{eq:SU(2)-III-quadratic-curve-2}
% \end{align}
The two factorisations describe the same splittings of the curve into two parts which are connected at the branch cut of the square root. 
%To get from (\ref{eq:SU(2)-III-quadratic-curve-1}) to (\ref{eq:SU(2)-III-quadratic-curve-2}), simply multiply the first factor ($\mathcal{C}$) with $1=(b_1/2 - \sqrt{\ldots})/(b_1/2 - \sqrt{\ldots})$ and the second %factor ($\mathcal{D}$) with $(b_1/2 + \sqrt{\ldots})/(b_1/2 + \sqrt{\ldots})$. 
With these two factorisations, one can analyse the splitting of the fibre when either $b_0$ or $b_2$ is non-zero.\footnote{On a generic base of complex dimension 3, $b_0$ and $b_2$ cannot both vanish on the codimension 2 curve.} When $b_0 \neq 0$, we can solve $\mathcal{C}_\pm$ for $c_{1,1}$ and plug the result into the hypersurface equation; if $b_2 \neq 0$, we solve ${\cal D}_\pm$ for $c_{2,1}$. Doing so, we find no splitting for $\mathbb{P}^1_0$, but $\mathbb{P}^1_1$ 
splits as follows:
\begin{align}
  & P_T (e_1=0 , \, \mathcal{C}_\pm/\mathcal{D}_\pm=0) \notag \\
  \overset{b_0 \neq 0}{=} \, & \frac{1}{b_0} \underbrace{\left[\!\left(\frac{b_1}{2} \pm \sqrt{ \frac{b_1^2}{4} - b_0\,b_2} \right)\!s_1 + b_0\,s_0\,\sv \right]}_{\mathbb{P}^1_{1B}} \! \underbrace{\left[\!\left( \frac{b_1}{2} \mp \sqrt{ \frac{b_1^2}{4} - b_0\,b_2} \right) \! s_1\,\su + c_{2,1}\,e_0\,\sv + b_0\,s_0\,\su\,\sv \right]}_{\mathbb{P}^1_{1C}} \label{eq:SU(2)-III-splitting-quadratic-curve-1} \\[-.6ex]
  \overset{b_2 \neq 0}{=} \, & \frac{1}{b_2} \overbrace{\left[\!b_2\,s_1 + \left( \frac{b_1}{2} \mp \sqrt{ \frac{b_1^2}{4} - b_0\,b_2} \right)\!s_0\,\sv \right]} \! \overbrace{\left[\! b_2\,s_1\,\su + c_{1,1}\,e_0\,\sv + \left( \frac{b_1}{2} \pm \sqrt{ \frac{b_1^2}{4} - b_0\,b_2} \right)\!s_0\,\su\,\sv \right]} \label{eq:SU(2)-III-splitting-quadratic-curve-2}
\end{align}
% The notation is analogous to (\ref{eq:SU(2)-I-splitting-quadratic-curve}): Over $\mathcal{C}=0$, only the signs upstairs in both lines apply, while over $\mathcal{D}=0$ one has to take the signs downstairs. To switch from (\ref{eq:SU(2)-III-splitting-quadratic-curve-1}) to (\ref{eq:SU(2)-III-splitting-quadratic-curve-2}), one multiplies the factor defining $\mathbb{P}^1_{1B}$ by $(b_1/2 \mp \sqrt{\ldots})/(b_1/2 \mp \sqrt{\ldots})$ and the factor defining $\mathbb{P}^1_{1C}$ by $(b_1/2 \pm \sqrt{\ldots})/(b_1/2 \pm \sqrt{\ldots})$. %; the sign depends on whether $\mathcal{C}=0$ (signs upstairs) or $\mathcal{D}=0$ (signs downstairs).
% \subsubsection*{Yukawa Points}

The fibre over $\{w_2\} \cap  \{ b_1^2 - 4 b_0 b_2 \}$ is of Kodaira type $III$ and thus no massless matter arises. 
Interestingly, over the remaining locus  $\{w_2\} \cap    \{ -b_2 d_0^2 + b_1 d_{0,0} d_{1,0} - b_0 d_{1,0}^2  - b_1^2 d_{2,0} + 4 b_0 b_2 d_{2,0} \} $,  $(f,g,\Delta)$ vanish to order $(0,0,3)$, but  the fibre is of {\it non-split } Kodaira type $I_3$. This can be read off from the specifics of the Weierstrass sections $f$ and $g$ following Tate's algorithm. Moreover, an explicit analysis of the resolved fibre confirms that it locally factors into three $\mathbb P^1$s, two of which are however exchanged by a monodromy 
along the curve in the base.  Since the corresponding singularity type is merely ${\rm Sp}(1)$ (as opposed to $SU(3)$) no massless matter arises here. Note that this conclusion is not in contradiction with the results of \cite{Grassi:2011hq}, especially table 9, which would naively indicate fundamental matter along this curve. However, the analysis of \cite{Grassi:2011hq} holds on Calabi-Yau 3-folds and therefore does not account for potential monodromies along the matter loci. 

The possible Yukawa couplings are summarised in table \ref{tab:SU(2)-III-Yukawas}. The splitting process over the first type of Yukawa points is straightforward to see when one evaluates the hypersurface equation on the locus.

\begin{table}[t]
\begin{align*}
  \begin{array}{c|c|c|c}
    \text{matter} & \text{locus} = W_2 \cap \ldots & \text{splitting of fibre components} & U(1)-\text{charges} \\ \hline \rule{0pt}{3ex}
    \mathbf{2}^{\mathrm{III}}_1 & \{b_0\} & \mathbb{P}^1_0 \rightarrow \mathbb{P}^1_{0s_1}\!+ \mathbb{P}^1_{0A} & (1,0) \\ \rule{0pt}{3ex}
    \mathbf{2}^{\mathrm{III}}_2 & \{b_2 \} & \mathbb{P}^1_1 \rightarrow \mathbb{P}^1_{1\sv} + \mathbb{P}^1_{1A} & (1,1)\\ \rule{0pt}{3ex}
    \mathbf{2}^{\mathrm{III}}_3 & \{b_0 \, c_{1,1}^2 - b_1\,c_{1,1}\,c_{2,1} + b_2\,c_{2,1}^2 \} & \mathbb{P}^1_1 \rightarrow \mathbb{P}^1_{1B} + \mathbb{P}^1_{1C} & (0,1)
  \end{array}
\end{align*}
\caption{Matter states in the $SU(2)$-III top.}
\label{tab:SU(2)-III-matter}
\end{table}

\begin{table}[ht]
\begin{align*}
  \begin{array}{c|c|c}
    \text{coupling} & \text{locus} = W_2 \cap \ldots & \text{splitting of fibre components} \\ \hline \rule{0pt}{3ex}
    \mathbf{2}^{\mathrm{III}}_1 - \overline{\mathbf{2}}^{\mathrm{III}}_2 - \mathbf{1}^{(6)} & \{b_0\} \cap \{b_2\} & \mathbb{P}^1_0 \rightarrow \mathbb{P}^1_{0s_1} \! + \mathbb{P}^1_{0A}, \, \mathbb{P}^1_1 \rightarrow \mathbb{P}^1_{1\sv} + \mathbb{P}^1_{1A} \\[.5ex] \hline \rule{0pt}{3ex}
    \mathbf{2}^{\mathrm{III}}_1 - \mathbf{2}^{\mathrm{III}}_3 - \overline{\mathbf{1}}^{(4)} & \{b_0\} \cap \{c_{2,1}\,b_2 - c_{1,1}\,b_1\} & \mathbb{P}^1_0 \rightarrow \mathbb{P}^1_{0s_1} \! + \mathbb{P}^1_{0A}, \, \mathbb{P}^1_1 \rightarrow \mathbb{P}^1_{1B} + \mathbb{P}^1_{1C} \\[.5ex] \hline \rule{0pt}{3ex}
    \mathbf{2}^{\mathrm{III}}_1 - \overline{\mathbf{2}}^{\mathrm{III}}_3 - \overline{\mathbf{1}}^{(1)} & \{b_0\} \cap \{c_{2,1}\} & \mathbb{P}^1_0 \rightarrow \mathbb{P}^1_{0s_1} \! + \mathbb{P}^1_{0A}, \, \mathbb{P}^1_1 \rightarrow \mathbb{P}^1_{1B'} + \mathbb{P}^1_{1C} \\ \hline \hline \rule{0pt}{3ex}
    \mathbf{2}^{\mathrm{III}}_2 - \mathbf{2}^{\mathrm{III}}_3 - \overline{\mathbf{1}}^{(3)} & \{b_2\} \cap \{c_{1,1}\} & \mathbb{P}^1_1 \rightarrow \mathbb{P}^1_{1\sv B} + \mathbb{P}^1_{1AB} + \mathbb{P}^1_{1AC} \\ \hline \rule{0pt}{3ex}
    \mathbf{2}^{\mathrm{III}}_2 - \overline{\mathbf{2}}^{\mathrm{III}}_3 - \overline{\mathbf{1}}^{(2)} & \{b_2\} \cap \{c_{1,1}\,b_0 - c_{2,1}\,b_1 \} & \mathbb{P}^1_1 \rightarrow \mathbb{P}^1_{1\sv C} + \mathbb{P}^1_{1AB'} + \mathbb{P}^1_{1AC'} \\ \hline \rule{0pt}{3ex}
    \mathbf{2}^{\mathrm{III}}_3 - \mathbf{2}^{\mathrm{III}}_3 - \overline{\mathbf{1}}^{(5)} & \{c_{1,1}\} \cap \{c_{2,1}\} & \mathbb{P}^1_1 \rightarrow \mathbb{P}^1_{1B} + \mathbb{P}^1_{1\su C} + \mathbb{P}^1_{1C'}
  \end{array}
\end{align*}
\caption{Yukawa couplings in the $SU(2)$-III top.}
\label{tab:SU(2)-III-Yukawas}
\end{table}

The second and third groups of couplings arise over the intersection of the $\mathbf{2}^\mathrm{III}_3$-curve with $b_0=0$ and hence require the factorisation  (\ref{eq:SU(2)-III-splitting-quadratic-curve-2}). The second Yukawa point lies over $\mathcal{D}_-=0$, corresponding to the downstairs signs in (\ref{eq:SU(2)-III-splitting-quadratic-curve-2}); the third point lies over $\mathcal{D}_+=0$, corresponding to upstairs signs. In both cases, we see that there is no further splitting of $\mathbb{P}^1_{1B}$ and $\mathbb{P}^1_{1C}$ when we set $b_0=0$. Rather, $\mathbb{P}^1_0$ splits, coming from the already present enhancement over the $\mathbf{2}^\mathrm{III}_1$ curve.

The fourth and fifth Yukawa point are the intersection points of $\mathbf{2}^\mathrm{III}_3$ and $b_2=0$, hence we make use of the factorisation (\ref{eq:SU(2)-III-splitting-quadratic-curve-1}). The fourth/fifth point lies on $\mathcal{C}_-/\mathcal{C}_+ =0$, correspondingly we take the downstairs/upstairs signs in (\ref{eq:SU(2)-III-splitting-quadratic-curve-1}). Setting $b_2=0$, we see that for the fourth coupling, $\mathbb{P}^1_{1B}$ splits off a factor $\sv$, while $\mathbb{P}^1_{1C}$ remains irreducible; for the fifth coupling, it is $\mathbb{P}^1_{1C}$ that splits.

Finally, over the last Yukawa point, which is a self-intersection point, neither $b_0$ nor $b_2$ are 0, and so both factorisations (\ref{eq:SU(2)-III-splitting-quadratic-curve-1}) and (\ref{eq:SU(2)-III-splitting-quadratic-curve-2}) should give the same splitting process in the fibre. Indeed, setting $c_{2,1}=0$ in (\ref{eq:SU(2)-III-splitting-quadratic-curve-1}) and $c_{1,1}=0$ in (\ref{eq:SU(2)-III-splitting-quadratic-curve-2}) shows that $\mathbb{P}^1_{1B}$ remains irreducible while $\mathbb{P}^1_{1C}$ splits off a factor $\su$.

Over all Yukawa points we find that the intersection structure of the $\mathbb{P}^1$ components is the affine $SU(4)$ diagram.

\section[Details on \texorpdfstring{{\boldmath $SU(3)$}}{SU(3)}-B and -C Tops]{Details on \texorpdfstring{{\boldmath $SU(3)$}}{SU(3)}-B and -C Tops} \label{app-SU3}

Here we go through the remaining $SU(3)$ tops in more detail.

\subsection[\texorpdfstring{{$SU(3)$}}{SU(3)}-B Top]{\texorpdfstring{{\boldmath $SU(3)$}}{SU(3)}-B Top} \label{app-SU3-B}

The $SU(3)$-B top leads to the restrictions of the following coefficients
\begin{align}\label{eq:SU(3)-B-coeffs}
  \begin{split}
    b_0 &= b_{0,2}\,f_0^2\,f_1 ,\quad b_2 = b_{2,0}\,f_1\,f_2^2 ,\quad c_1 = c_{1,0}\,f_2 ,\quad c_2 = c_{2,1}\,f_0 , \\
    d_0 &= d_{0,1}\,f_0\,f_1 ,\quad d_1 = d_{1,0}\,f_1\,f_2 ,\quad d_2 = d_{2,0}\,f_1 ,
  \end{split}
\end{align}
while $b_1$ remain unrestricted. There are eight different triangulations. For definiteness, we choose the one leading to the following SR-ideal:
\begin{align}\label{eq:SU(3)-B-SR-ideal}
  \su\,\sv , \su\,\sw , \sw\,s_0 , \sv\,s_1 , s_0\,s_1 , f_0\,\su , f_0\,s_1 , f_1\,\sv , f_1\,\sw , f_2\,\su , f_2\,s_0 , f_2\,\sv .
\end{align}

The coordinates and their corresponding divisor classes are summarised in the following table:
\begin{align}\label{tab:SU(3)-B-divisor-classes}
  \begin{array}{c|ccccccc|c}
    \hphantom{U} & \su & \sv & \sw & s_0 & s_1 & f_1 & f_2 & f_0 \\ \hline
    \mathrm{U} & 1 & 1 & 1 & \cdot & \cdot & \cdot & \cdot & \cdot \\
    S_0 & \cdot & \cdot & 1 & 1& \cdot & \cdot & \cdot & \cdot \\
    S_1 & \cdot & 1 & \cdot & \cdot & 1 & \cdot & \cdot & \cdot \\
    F_1 & \cdot & 1 & \cdot & \cdot & \cdot & 1 & \cdot & -1 \\
    F_2 & \cdot & 1 & -1 & \cdot & \cdot & \cdot & 1 & -1
  \end{array}
\end{align}

For $SU(3)$ roots to have zero $U(1)$ charge, the generators (\ref{eq:U(1)-generators}) receive the following correction:
\begin{align}\label{eq:SU(3)-B-U(1)-generators}
  \begin{split}
    \omega_1^\text{B} &= S_1 - S_0 - \overline{\mathcal{K}} + \frac{1}{3} F_1 + \frac{2}{3} F_2 ,\\
    \omega_2^\text{B} &= U - S_0 - \overline{\mathcal{K}} - [c_{1,0}] + \frac{2}{3} F_1 + \frac{1}{3} F_2.
  \end{split}
\end{align}

We find codimension 2 enhancement with $\mathbf{3}$ and $\overline{\mathbf{3}}$ matter over loci and with charges as presented in table \ref{tab:SU(3)-B-matter}. Over the curve $\{ w_3 \} \cap \{ b_1 \}$ the fibre type changes to Kodaira type $IV$, but such fibres do not give rise to additional charged matter.

The gauge invariant Yukawa coupling appearing are listed in table \ref{tab:SU(3)-B-Yukawas}.

\begin{table}[h!]
\begin{align*}
  \begin{array}{c|c|c|c}
    \text{matter} & \text{locus} = W_3 \cap \ldots & \text{splitting of fibre components} & U(1)-\text{charges} \\ \hline \rule{0pt}{3ex}
    \mathbf{3}^{\mathrm{B}}_1 & \{c_{1,0}\} & \mathbb{P}^1_1 \rightarrow \mathbb{P}^1_{1s_0} \!+ \mathbb{P}^1_{1A} & (-\frac{2}{3}, -\frac{4}{3}) \\ \rule{0pt}{3ex}
    \mathbf{3}^{\mathrm{B}}_2 & \{c_{2,1}\} & \mathbb{P}^1_1 \rightarrow \mathbb{P}^1_{1s_1} \!+ \mathbb{P}^1_{1B} & (-\frac{2}{3}, \frac{2}{3}) \\ \rule{0pt}{3ex}
    \mathbf{3}^{\mathrm{B}}_3 & \{d_{2,0}\} & \mathbb{P}^1_0 \rightarrow \mathbb{P}^1_{0\sw} + \mathbb{P}^1_{0A} & (-\frac{2}{3}, -\frac{1}{3}) \\ \rule{0pt}{3ex}
    \mathbf{3}^{\mathrm{B}}_4 & \{b_1^2\,b_{2,0} - b_1\,c_{1,0}\,d_{1,0} + c_{1,0}^2\,d_{2,0}\} & \mathbb{P}^1_0 \rightarrow \mathbb{P}^1_{0B} + \mathbb{P}^1_{0C} & (\frac{1}{3}, \frac{2}{3}) \\ \rule{0pt}{3ex}
    \mathbf{3}^{\mathrm{B}}_5 & \{b_{0,2}\,b_1^2 + c_{2,1}^2\,d_{2.0} - b_1\,c_{2,1}\,d_{0,1}\} & \mathbb{P}^1_2 \rightarrow \mathbb{P}^1_{2A} + \mathbb{P}^1_{2B} & (\frac{1}{3},-\frac{1}{3})
  \end{array}
\end{align*}
\caption{Matter states in the $SU(3)$-B top.}
\label{tab:SU(3)-B-matter}
\end{table}

\begin{table}[h!]
\begin{align*}
  \begin{array}{c|c|c}
    \text{coupling} & \text{locus}=W_3 \cap \ldots & \text{splitting of fibre components} \\ \hline \rule{0pt}{3ex}    
    \mathbf{3}^{\mathrm{B}}_1 - \overline{\mathbf{3}}^{\mathrm{B}}_2 - \mathbf{1}^{(5)} & \{c_{1,0}\} \cap \{c_{2,1}\} & \mathbb{P}^1_1 \rightarrow \mathbb{P}^1_{1s_0 B} + \mathbb{P}^1_{1 s_1 A} + \mathbb{P}^1_{1AB} \\ \hline \rule{0pt}{3ex}
    \mathbf{3}^{\mathrm{B}}_1 - \overline{\mathbf{3}}^{\mathrm{B}}_3 - \mathbf{1}^{(6)} & \{c_{1,0}\} \cap \{d_{2,0}\} & \mathbb{P}^1_0 \rightarrow \mathbb{P}^1_{0 \sw} + \mathbb{P}^1_{0A} , \, \mathbb{P}^1_1 \rightarrow \mathbb{P}^1_{1s_0} \!+ \mathbb{P}^1_{1A} \\ \hline \rule{0pt}{3ex}
    \mathbf{3}^{\mathrm{B}}_1 - \overline{\mathbf{3}}^{\mathrm{B}}_4 - \mathbf{1}^{(3)} & \{c_{1,0}\} \cap \{b_{2,0}\} & \mathbb{P}^1_0 \rightarrow \mathbb{P}^1_{0B} + \mathbb{P}^1_{0C}, \, \mathbb{P}^1_1 \rightarrow \mathbb{P}^1_{1 s_0} \!+ \mathbb{P}^1_{1A} \\ \hline \rule{0pt}{3ex}
    \mathbf{3}^{\mathrm{B}}_1 - \overline{\mathbf{3}}^{\mathrm{B}}_5 - \mathbf{1}^{(4)} & \{c_{1,0}\} \cap \left(\mathbf{3}^\mathrm{B}_5\right) & \mathbb{P}^1_1 \rightarrow \mathbb{P}^1_{1s_0} \!+ \mathbb{P}^1_{1A} , \, \mathbb{P}^1_2 \rightarrow \mathbb{P}^1_{2A} + \mathbb{P}^1_{2B} \\ \hline \rule{0pt}{3ex}
    \mathbf{3}^{\mathrm{B}}_2 - \overline{\mathbf{3}}^{\mathrm{B}}_3 - \overline{\mathbf{1}}^{(6)} & \{c_{2,1}\} \cap \{d_{2,0}\} & \mathbb{P}^1_0 + \mathbb{P}^1_{0\sw} + \mathbb{P}^1_{0A} , \, \mathbb{P}^1_1 \rightarrow \mathbb{P}^1_{1 s_1} \!+ \mathbb{P}^1_{1B} \\ \hline \rule{0pt}{3ex}
    \mathbf{3}^{\mathrm{B}}_2 - \overline{\mathbf{3}}^{\mathrm{B}}_4 - \mathbf{1}^{(2)} & \{c_{2,1}\} \cap \left( \mathbf{3}^\mathrm{B}_4 \right) &  \mathbb{P}^1_0 \rightarrow \mathbb{P}^1_{0B} + \mathbb{P}^1_{0C} , \, \mathbb{P}^1_1 \rightarrow \mathbb{P}^1_{1s_1} \! + \mathbb{P}^1_{1B} \\ \hline \rule{0pt}{3ex}
    \mathbf{3}^{\mathrm{B}}_2 - \overline{\mathbf{3}}^{\mathrm{B}}_5 - \mathbf{1}^{(1)} & \{b_{0,2}\} \cap \{c_{2,1}\} & \mathbb{P}^1_1 \rightarrow \mathbb{P}^1_{1s_1} \! + \mathbb{P}^1_{1B} , \, \mathbb{P}^1_2 \rightarrow \mathbb{P}^1_{2A} + \mathbb{P}^1_{2B} \\ \hline \rule{0pt}{3ex}
    \mathbf{3}^{\mathrm{B}}_3 - \overline{\mathbf{3}}^{\mathrm{B}}_4 - \mathbf{1}^{(4)} & \{d_{2,0}\} \cap \{b_1\,b_{2,0} - c_{1,0}\,d_{1,0}\} & \mathbb{P}^1_0 \rightarrow \mathbb{P}^1_{0\sw C} + \mathbb{P}^1_{0 AC} + \mathbb{P}^1_{0AB} \\ \hline \rule{0pt}{3ex}
    \mathbf{3}^{\mathrm{B}}_3 - \overline{\mathbf{3}}^{\mathrm{B}}_5 - \mathbf{1}^{(2)} & \{d_{2,0}\} \cap \{b_{0,2}\,b_1 - c_{2,1}\,d_{0,1}\} & \mathbb{P}^1_0 + \mathbb{P}^1_{0\sw} + \mathbb{P}^1_{0A} , \, \mathbb{P}^1_2 \rightarrow \mathbb{P}^1_{2A} + \mathbb{P}^1_{2B} \\ \hline \rule{0pt}{3ex}
    \mathbf{3}^{\mathrm{B}}_4 - \overline{\mathbf{3}}^{\mathrm{B}}_5 - \overline{\mathbf{1}}^{(6)} & \left(\mathbf{3}^\mathrm{B}_4 \right) \cap \left( \mathbf{3}^\mathrm{B}_5 \right) \setminus (\{d_{2,0}\} \cap \{b_1\}) & \mathbb{P}^1_0 \rightarrow \mathbb{P}^1_{0B} + \mathbb{P}^1_{0C} , \, \mathbb{P}^1_2 \rightarrow \mathbb{P}^1_{2A} + \mathbb{P}^1_{2B} \\ \hline \rule{0pt}{3ex}
    \mathbf{3}^{\mathrm{B}}_3 - \mathbf{3}^{\mathrm{B}}_4 - \mathbf{3}^{\mathrm{B}}_5 & \{d_{2,0}\} \cap \{b_1\} & \mathbb{P}^1_0 \rightarrow \mathbb{P}^1_{0\sw B} + \mathbb{P}^1_{0AB'} + \mathbb{P}^1_{0AC'} , \, \mathbb{P}^1_2 \rightarrow \mathbb{P}^1_{2A} + \mathbb{P}^1_{2B} \\ \rule{0pt}{3ex}
    & & \mathbb{P}^1_{0AB'} = \mathbb{P}^1_{2A} \\ \hline \rule{0pt}{3ex}
    \mathbf{3}^{\mathrm{B}}_1 - \mathbf{3}^{\mathrm{B}}_4 - \mathbf{3}^{\mathrm{B}}_4 & \{c_{1,0}\} \cap \{b_1\} & \mathbb{P}^1_0 \rightarrow \mathbb{P}^1_{0B} + \mathbb{P}^1_{0C'} + \mathbb{P}^1_{0C1} , \, \mathbb{P}^1_1 \rightarrow \mathbb{P}^1_{1 s_0} \! + \mathbb{P}^1_{1A} \\ \rule{0pt}{3ex}
    & & \mathbb{P}^1_{0C1} = \mathbb{P}^1_{1A} \\ \hline \rule{0pt}{3ex}
    \mathbf{3}^{\mathrm{B}}_2 - \mathbf{3}^{\mathrm{B}}_5 - \mathbf{3}^{\mathrm{B}}_5 & \{c_{2,1}\} \cap \{b_1\} & \mathbb{P}^1_1 \rightarrow \mathbb{P}^1_{1 s_1} \!+ \mathbb{P}^1_{1B'}, \, \mathbb{P}^1_2 \rightarrow \mathbb{P}^1_{2A} + \mathbb{P}^1_{2B'} + \mathbb{P}^1_{2B1} \\ \rule{0pt}{3ex}
    & & \mathbb{P}^1_{1B} = \mathbb{P}^1_{2B1}
  \end{array}
\end{align*}
\caption{Yukawa couplings in the $SU(3)$-B top.}
\label{tab:SU(3)-B-Yukawas}
\end{table}

\subsection[\texorpdfstring{{$SU(3)$}}{SU(3)}-C Top]{\texorpdfstring{{\boldmath $SU(3)$}}{SU(3)}-C Top} \label{app-SU3-C}

The third top leads to the restrictions of the following coefficients
\begin{align}\label{eq:SU(3)-C-coeffs}
  \begin{split}
    b_0 &= b_{0,1}\,f_0\,f_2 ,\quad b_2 = b_{2,0}\,f_1 ,\quad c_1 = c_{1,1}\,f_0\,f_1 ,\quad c_2 = c_{2,1}\,f_0 , \\
    d_0 &= d_{0,0}\,f_2 ,\quad d_1 = d_{1,0}\,f_1\,f_2 ,\quad d_2 = d_{2,0}\,f_1\,f_2^2 ,
  \end{split}
\end{align}
while $b_1$ remain unrestricted. The top allows 4 different triangulations. For definiteness, we choose the one leading to the following SR-ideal:
\begin{align}\label{eq:SU(3)-C-SR-ideal}
  \su\,\sv , \su\,\sw , \sw\,s_0 , \sv\,s_1 , s_0\,s_1 , f_0\,\su , f_0\,s_0 , f_0\,s_1 , f_1\,s_0 , f_1\,\sv , f_2\,\sw , f_2\,s_1 .
\end{align}

The coordinates and their corresponding divisor classes are summarised in the following table:
\begin{align}\label{tab:SU(3)-C-divisor-classes}
  \begin{array}{c|ccccccc|c}
    \hphantom{U} & \su & \sv & \sw & s_0 & s_1 & f_1 & f_2 & f_0 \\ \hline
    \mathrm{U} & 1 & 1 & 1 & \cdot & \cdot & \cdot & \cdot & \cdot \\
    S_0 & \cdot & \cdot & 1 & 1& \cdot & \cdot & \cdot & \cdot \\
    S_1 & \cdot & 1 & \cdot & \cdot & 1 & \cdot & \cdot & \cdot \\
    F_1 & \cdot & 1 & \cdot & \cdot & \cdot & 1 & \cdot & -1 \\
    F_2 & \cdot & 1 & 1 & \cdot & \cdot & \cdot & 1 & -1
  \end{array}
\end{align}

For $SU(3)$ roots to have zero $U(1)$ charge, the generators (\ref{eq:U(1)-generators}) receive the following correction:
\begin{align}\label{eq:SU(3)-C-U(1)-generators}
  \begin{split}
    \omega_1^\text{C} &= S_1 - S_0 - \overline{\mathcal{K}} + \frac{1}{3} F_1 - \frac{1}{3} F_2 ,\\
    \omega_2^\text{C} &= U - S_0 - \overline{\mathcal{K}} - [c_{1,1}] .
  \end{split}
\end{align}

The Kodaira type of the fibre enhances from $I_3$ to $I_4$ (split) over the codimension-2 loci displayed in table \ref{tab:SU(3)-C-matter}, which therefore give rise to $\mathbf{3}$ and $\overline{\mathbf{3}}$ matter. In addition, over the curve $\{ w_3 \} \cap \{ b_1 \}$ the fibre type changes to Kodaira type $IV$, but no matter representation arises over this locus.

The gauge invariant Yukawa coupling appearing are summarised in table \ref{tab:SU(3)-C-Yukawas}.

\begin{table}[ht]
\begin{align*}
  \begin{array}{c|c|c|c}
    \text{matter} & \text{locus} = W_3 \cap \ldots & \text{splitting of fibre components} & U(1)-\text{charges} \\ \hline \rule{0pt}{3ex}
    \mathbf{3}^{\mathrm{C}}_1 & \{b_{2,0}\} & \mathbb{P}^1_2 \rightarrow \mathbb{P}^1_{2\sv} + \mathbb{P}^1_{2A} & (-\frac{2}{3}, -1) \\ \rule{0pt}{3ex}
    \mathbf{3}^{\mathrm{C}}_2 & \{c_{2,1}\} & \mathbb{P}^1_1 \rightarrow \mathbb{P}^1_{1\su} + \mathbb{P}^1_{1A} & (\frac{1}{3}, -1) \\ \rule{0pt}{3ex}
    \mathbf{3}^{\mathrm{C}}_3 & \{b_1\,c_{1,1} - b_{2,0}\,c_{2,1}\} & \mathbb{P}^1_2 \rightarrow \mathbb{P}^1_{2B} + \mathbb{P}^1_{2C} & (\frac{1}{3}, 1) \\ \rule{0pt}{3ex}
    \mathbf{3}^{\mathrm{C}}_4 & \{b_{0,1}\,b_1 - c_{2,1}\,d_{0,0}\} & \mathbb{P}^1_1 \rightarrow \mathbb{P}^1_{1B} + \mathbb{P}^1_{1C} & (-\frac{2}{3}, 0) \\ \rule{0pt}{3ex}
    \mathbf{3}^{\mathrm{C}}_5 & \{b_{2,0}\,d_{0,0}^2 + b_1^2\,d_{2,0} - b_1\,d_{0,0}\,d_{1,0}\} & \mathbb{P}^1_0 \rightarrow \mathbb{P}^1_{0A} + \mathbb{P}^1_{0B} & (\frac{1}{3},0)
  \end{array}
\end{align*}
\caption{Matter states in the $SU(3)$-C top.}
\label{tab:SU(3)-C-matter}
\end{table}

\begin{table}[ht]
\begin{align*}
  \begin{array}{c|c|c}
    \text{coupling} & \text{locus}= W_3 \cap \ldots & \text{splitting of fibre components} \\ \hline \rule{0pt}{3ex}    
    \mathbf{3}^{\mathrm{C}}_1 - \overline{\mathbf{3}}^{\mathrm{C}}_2 - \mathbf{1}^{(2)} & \{b_{2,0}\} \cap \{c_{2,1}\} & \mathbb{P}^1_1 \rightarrow \mathbb{P}^1_{1\su} + \mathbb{P}^1_{1A} , \, \mathbb{P}^1_2 \rightarrow \mathbb{P}^1_{2\sv} + \mathbb{P}^1_{2A} \\ \hline \rule{0pt}{3ex}
    \mathbf{3}^{\mathrm{C}}_1 - \overline{\mathbf{3}}^{\mathrm{C}}_3 - \mathbf{1}^{(3)} & \{b_{2,0}\} \cap \{c_{1,1}\} & \mathbb{P}^1_2 \rightarrow \mathbb{P}^1_{2\sv C} + \mathbb{P}^1_{2AB} + \mathbb{P}^1_{2C'} \\ \hline \rule{0pt}{3ex}
    \mathbf{3}^{\mathrm{C}}_1 - \overline{\mathbf{3}}^{\mathrm{C}}_4 - \mathbf{1}^{(6)} & \{b_{2,0}\} \cap \{b_{0,1}\,b_1 - c_{2,1}\,d_{0,0}\} & \mathbb{P}^1_1 \rightarrow \mathbb{P}^1_{1B} + \mathbb{P}^1_{1C} , \, \mathbb{P}^1_2 \rightarrow \mathbb{P}^1_{2\sv} + \mathbb{P}^1_{2A} \\ \hline \rule{0pt}{3ex}
    \mathbf{3}^{\mathrm{C}}_1 - \overline{\mathbf{3}}^{\mathrm{C}}_5 - \mathbf{1}^{(4)} & \{b_{2,0}\} \cap \{b_1\,d_{2,0} - d_{0,0}\,d_{1,0}\} & \mathbb{P}^1_0 \rightarrow \mathbb{P}^1_{0A} + \mathbb{P}^1_{0B} , \, \mathbb{P}^1_2 \rightarrow \mathbb{P}^1_{2\sv} + \mathbb{P}^1_{2A} \\ \hline \rule{0pt}{3ex}
    \mathbf{3}^{\mathrm{C}}_2 - \overline{\mathbf{3}}^{\mathrm{C}}_3 - \mathbf{1}^{(5)} & \{c_{2,1}\} \cap \{c_{1,1}\} & \mathbb{P}^1_1 \rightarrow \mathbb{P}^1_{1\su} + \mathbb{P}^1_{1A} , \, \mathbb{P}^1_2 \rightarrow \mathbb{P}^1_{2B} + \mathbb{P}^1_{2C}  \\ \hline \rule{0pt}{3ex}
    \mathbf{3}^{\mathrm{C}}_2 - \overline{\mathbf{3}}^{\mathrm{C}}_4 - \overline{\mathbf{1}}^{(1)} & \{c_{2,1}\} \cap \{b_{0,1}\} & \mathbb{P}^1_1 \rightarrow \mathbb{P}^1_{1\su B} + \mathbb{P}^1_{1B'} + \mathbb{P}^1_{1AC} \\ \hline \rule{0pt}{3ex}
    \mathbf{3}^{\mathrm{C}}_2 - \overline{\mathbf{3}}^{\mathrm{C}}_5 - \mathbf{1}^{(6)} & \{c_{2,1}\} \cap \left( \mathbf{3}^\mathrm{C}_5 \right) & \mathbb{P}^1_0 \rightarrow \mathbb{P}^1_{0A} + \mathbb{P}^1_{0B} , \, \mathbb{P}^1_1 \rightarrow \mathbb{P}^1_{1\su} + \mathbb{P}^1_{1A}\\ \hline \rule{0pt}{3ex}
    \mathbf{3}^{\mathrm{C}}_3 - \overline{\mathbf{3}}^{\mathrm{C}}_4 - \overline{\mathbf{1}}^{(4)} & \left( \mathbf{3}^\mathrm{C}_3 \right) \cap \left( \mathbf{3}^\mathrm{C}_4 \right) \setminus (\{c_{2,1}\} \cap \{b_1\}) & \mathbb{P}^1_1 \rightarrow \mathbb{P}^1_{1B} + \mathbb{P}^1_{1C} , \, \mathbb{P}^1_2 \rightarrow \mathbb{P}^1_{2B} + \mathbb{P}^1_{2C} \\ \hline \rule{0pt}{3ex}
    \mathbf{3}^{\mathrm{C}}_3 - \overline{\mathbf{3}}^{\mathrm{C}}_5 - \overline{\mathbf{1}}^{(6)} & \left( \mathbf{3}^\mathrm{C}_3 \right) \cap \left( \mathbf{3}^\mathrm{C}_5 \right) \setminus ( \{b_{2,0}\} \cap \{b_1\} ) & \mathbb{P}^1_0 \rightarrow \mathbb{P}^1_{0A} + \mathbb{P}^1_{0B} , \, \mathbb{P}^1_2 \rightarrow \mathbb{P}^1_{2B} + \mathbb{P}^1_{2C} \\ \hline \rule{0pt}{3ex}
    \mathbf{3}^{\mathrm{C}}_4 - \overline{\mathbf{3}}^{\mathrm{C}}_5 - \mathbf{1}^{(2)} & \left( \mathbf{3}^\mathrm{C}_4 \right) \cap \left( \mathbf{3}^\mathrm{C}_5 \right) \setminus ( \{d_{0,0}\} \cap \{b_1\} ) & \mathbb{P}^1_0 \rightarrow \mathbb{P}^1_{0A} + \mathbb{P}^1_{0B} , \, \mathbb{P}^1_1 \rightarrow \mathbb{P}^1_{1B} + \mathbb{P}^1_{1C} \\ \hline \rule{0pt}{3ex}
    \mathbf{3}^{\mathrm{C}}_1 - \mathbf{3}^{\mathrm{C}}_3 - \mathbf{3}^{\mathrm{C}}_5 & \{b_{2,0}\} \cap \{b_1\} & \mathbb{P}^1_0 \rightarrow \mathbb{P}^1_{0A} + \mathbb{P}^1_{0B}, \, \mathbb{P}^1_2 \rightarrow \mathbb{P}^1_{2\sv B} + \mathbb{P}^1_{2 B'} + \mathbb{P}^1_{2AC} \\ \rule{0pt}{3ex}
    & & \mathbb{P}^1_{0A} = \mathbb{P}^1_{2B'} \\ \hline \rule{0pt}{3ex}
    \mathbf{3}^{\mathrm{C}}_2 - \mathbf{3}^{\mathrm{C}}_3 - \mathbf{3}^{\mathrm{C}}_4 & \{c_{2,1}\} \cap \{b_1\} & \mathbb{P}^1_1 \rightarrow \mathbb{P}^1_{1 \su C} + \mathbb{P}^1_{1C'} + \mathbb{P}^1_{1AB} , \, \mathbb{P}^1_2 \rightarrow \mathbb{P}^1_{2B} + \mathbb{P}^1_{2C} \\ \rule{0pt}{3ex} 
    & & \mathbb{P}^1_{1C'} = \mathbb{P}^1_{2C} \\ \hline \rule{0pt}{3ex}
    \mathbf{3}^{\mathrm{C}}_4 - \mathbf{3}^{\mathrm{C}}_5 - \mathbf{3}^{\mathrm{C}}_5 & \{d_{0,0}\} \cap \{b_1\} & \mathbb{P}^1_0 \rightarrow \mathbb{P}^1_{0B} + \mathbb{P}^1_{0C'} + \mathbb{P}^1_{0C1} , \, \mathbb{P}^1_1 \rightarrow \mathbb{P}^1_{1B} + \mathbb{P}^1_{1C} \\ \rule{0pt}{3ex}
    & & \mathbb{P}^1_{0C'} = \mathbb{P}^1_{1C}
  \end{array}
\end{align*}
\caption{Yukawa couplings in the $SU(3)$-C top.}
\label{tab:SU(3)-C-Yukawas}
\end{table}

\section{Matching the MSSM-Spectrum}\label{appsec:huge_table}

With the search criteria and algorithm presented in section \ref{subsec_spectrum_search}, we find, for each of the toric $SU(3) \times SU(2) \times U(1)_1 \times U(1)_2$ models described in section \ref{sec_3211},  a significant number of possibilities to match the geometric spectrum with matter states of the MSSM including right-handed neutrinos und $\mu$-singlets. The results are listed in the left column of the tables below, in the notation introduced in section \ref{subsec_spectrum_search}. In the right column, we have listed which of the baryon and lepton number violating couplings (cf.~(\ref{eq:couplings_W2}) to (\ref{eq:couplings_K})) are allowed by the $U(1)$ selection rules, although -- for space-saving reasons -- in a slightly altered order. Furthermore we do not explicitly write down the states associated with $H_u$, $H_d$ and $Q$ in each coupling that appears, as these states are fixed for each possible match. For example the $\alpha$-term comes from a coupling $Q\,L\,d^c_R$, where there can be, 
depending on the matching, several different states for $L$ and $d^c_R$, while $Q$ is given by the unique $({\bf 3}, {\bf 2})$-state. In our table we list such an existing coupling as $({\bf 2} , \overline{\bf 3})$, where the $\bf 2$-state is the lepton $L$ and $\overline{\bf 3}$ the down-quark, i.e.~the states appear in the same order as in the corresponding term in equations (\ref{eq:couplings_W2}) -- (\ref{eq:couplings_K}). When there is no $Q$ and $H$ involved in a coupling we give all the states involved, again in the order as they appear in the corresponding term; e.g.~for the $\beta$-term $u_R^c \, u_R^c \, d_R^c$ the corresponding entry in the table looks like $\overline{\bf 3}_i \, \overline{\bf 3}_j \, \overline{\bf 3}_k$, with the first up-quark involved being the state $\overline{\bf 3}_i$, the second one being $\overline{\bf 3}_j$, and the down-quark being $\overline{\bf 3}_k$. Note that we have summarised all possible terms of $W_{\text{singlet}}$ (\ref{eq:couplings_W_singlet}) in the entry $\delta$. Another special entry is the $\lambda_3$-term in (\ref{W3K}) of the form $Q\,Q\,Q\,H_d$; since there is no ambiguity in this term from the matching of the states, we simply list whether the coupling is allowed by the selection rules ($\checkmark$) or not ($-$).

We need to point out one case where the search algorithm does not completely fix the identification of the states with the MSSM fields. This happens when the choice of the Higgs states $H_{u/d}$ together with the charges of $Q = ({\bf 3},{\bf 2})$ does not completely fix the coefficients $a$ and $b$ of the hypercharge in terms of $U(1)_{1/2}$. In fact this is the case whenever there is a linear combination of $U(1)_{1/2}$ under which $Q$ and $H_{u/d}$ are all uncharged, which is the orthogonal linear combination to $U(1)_Y$. In such a case there might be some other states that are also uncharged under this particular $U(1)$ charge combination and can be identified with some Standard Model states. As these states are uncharged under the orthogonal $U(1)$, they are not subject to any selection rules, so that the dimension-four and -five operators in (\ref{eq:couplings_W2}) -- (\ref{eq:couplings_K}) will be present if all states involved are present. In this case there may be more possibilities to match the 
spectrum, which we do not work out explicitly here. For every top combination (except for ${\rm III} \times {\rm B}$, where there is no such case) we have listed the corresponding case at the end of the tables below. In the ${\rm I} \times {\rm A}$-model for example, this happens when $H_u = {\bf 2}^{\rm I}_1$, $H_d = \overline{\bf 2}^{\rm I}_1$, and any assignment $U(1)_Y = (2b + 1) \, U(1)_1 + b \, U(1)_2$ for arbitrary $b$ gives the correct hypercharge for the Higgs and the left-handed quarks, because they are uncharged under the linear combination $2 \, U(1)_1 + U(1)_2$. To match states charged under this $U(1)$ one needs to specify the value of $b$. It would be interesting to see if, after Higgsing the particular linear combination of $U(1)_{1/2}$ under which $H_{u/d}$ and $Q$ are uncharged, the geometric spectrum can be embedded into the most general F-theory compactification with only one abelian factor \cite{Morrison:2012ei}.

Finally, recall from section
\ref{sec_3211} that the top-combination $ {\rm III} \times {\rm B}$ generically suffers from a non-Kodaira point, which, at least as far as our current understanding of F-theory is concerned, must be absent in order for the fibration to describe a well-defined vacuum.

\newpage

\footnotesize

\input{outIxA.txt}
\input{outIxB.txt}
\input{outIxC.txt}
\newpage
\input{outIIIxA.txt}
\newpage
\input{outIIIxB.txt}

\end{appendix}

%% file: base_polygon.pdf_tex
%% Creator: Inkscape inkscape 0.48.2, www.inkscape.org
%% Accompanies image file 'base_polygon.pdf' (pdf, eps, ps)
\begingroup%
  \makeatletter%
  \providecommand\color[2][]{%
    \errmessage{(Inkscape) Color is used for the text in Inkscape, but the package 'color.sty' is not loaded}%
    \renewcommand\color[2][]{}%
  }%
  \providecommand\transparent[1]{%
    \errmessage{(Inkscape) Transparency is used (non-zero) for the text in Inkscape, but the package 'transparent.sty' is not loaded}%
    \renewcommand\transparent[1]{}%
  }%
  \providecommand\rotatebox[2]{#2}%
  \ifx\svgwidth\undefined%
    \setlength{\unitlength}{616.7109375bp}%
    \ifx\svgscale\undefined%
      \relax%
    \else%
      \setlength{\unitlength}{\unitlength * \real{\svgscale}}%
    \fi%
  \else%
    \setlength{\unitlength}{\svgwidth}%
  \fi%
  \global\let\svgwidth\undefined%
  \global\let\svgscale\undefined%
  \makeatother%
  \begin{picture}(1,0.27620366)%
    \put(0,0){\includegraphics[width=\unitlength]{base_polygon.pdf}}%
    \put(0.05019556,0.25518185){\color[rgb]{0,0,0}\makebox(0,0)[lb]{\smash{$\su$}}}%
    \put(0.17532968,0.25649224){\color[rgb]{0,0,0}\makebox(0,0)[lb]{\smash{$s_1$}}}%
    \put(0.29666435,0.12939256){\color[rgb]{0,0,0}\makebox(0,0)[lb]{\smash{$\sw$}}}%
    \put(0.17395342,0.01192313){\color[rgb]{0,0,0}\makebox(0,0)[lb]{\smash{$\sv$}}}%
    \put(0.04370954,0.13194746){\color[rgb]{0,0,0}\makebox(0,0)[lb]{\smash{$s_0$}}}%
    \put(0.66590658,0.25518185){\color[rgb]{0,0,0}\makebox(0,0)[lb]{\smash{$d_2$}}}%
    \put(0.91886139,0.25240213){\color[rgb]{0,0,0}\makebox(0,0)[lb]{\smash{$b_2$}}}%
    \put(0.80859904,0.25518185){\color[rgb]{0,0,0}\makebox(0,0)[lb]{\smash{$d_1$}}}%
    \put(0.92534742,0.1449195){\color[rgb]{0,0,0}\makebox(0,0)[lb]{\smash{$c_1$}}}%
    \put(0.80859904,0.00871306){\color[rgb]{0,0,0}\makebox(0,0)[lb]{\smash{$c_2$}}}%
    \put(0.6723926,0.00871306){\color[rgb]{0,0,0}\makebox(0,0)[lb]{\smash{$b_0$}}}%
    \put(0.66590658,0.1449195){\color[rgb]{0,0,0}\makebox(0,0)[lb]{\smash{$d_0$}}}%
    \put(0.80859904,0.1449195){\color[rgb]{0,0,0}\makebox(0,0)[lb]{\smash{$b_1$}}}%
  \end{picture}%
\endgroup%

%% file: tops.pdf_tex
%% Creator: Inkscape inkscape 0.48.3.1, www.inkscape.org
%% Accompanies image file 'tops.pdf' (pdf, eps, ps)
\begingroup%
  \makeatletter%
  \providecommand\color[2][]{%
    \errmessage{(Inkscape) Color is used for the text in Inkscape, but the package 'color.sty' is not loaded}%
    \renewcommand\color[2][]{}%
  }%
  \providecommand\transparent[1]{%
    \errmessage{(Inkscape) Transparency is used (non-zero) for the text in Inkscape, but the package 'transparent.sty' is not loaded}%
    \renewcommand\transparent[1]{}%
  }%
  \providecommand\rotatebox[2]{#2}%
  \ifx\svgwidth\undefined%
    \setlength{\unitlength}{609.92001953bp}%
    \ifx\svgscale\undefined%
      \relax%
    \else%
      \setlength{\unitlength}{\unitlength * \real{\svgscale}}%
    \fi%
  \else%
    \setlength{\unitlength}{\svgwidth}%
  \fi%
  \global\let\svgwidth\undefined%
  \global\let\svgscale\undefined%
  \makeatother%
  \begin{picture}(1,0.76390347)%
    \put(0,0){\includegraphics[width=\unitlength]{tops.pdf}}%
    \put(0.02780692,0.67051411){\color[rgb]{0,0,0}\makebox(0,0)[lb]{\smash{I:}}}%
    \put(0.39506819,0.67051411){\color[rgb]{0,0,0}\makebox(0,0)[lb]{\smash{II:}}}%
    \put(0.76232946,0.67051411){\color[rgb]{0,0,0}\makebox(0,0)[lb]{\smash{III:}}}%
    \put(0.10650577,0.51311642){\color[rgb]{0,0,0}\makebox(0,0)[lb]{\smash{$\color{blue}{e_0}$}}}%
    \put(0.25078698,0.51311642){\color[rgb]{0,0,0}\makebox(0,0)[lb]{\smash{$\color{blue}{e_1}$}}}%
    \put(0.48688351,0.39506815){\color[rgb]{0,0,0}\makebox(0,0)[lb]{\smash{$\color{blue}{e_0}$}}}%
    \put(0.48688351,0.53934937){\color[rgb]{0,0,0}\makebox(0,0)[lb]{\smash{$\color{blue}{e_1}$}}}%
    \put(0.86726125,0.49999995){\color[rgb]{0,0,0}\makebox(0,0)[lb]{\smash{$\color{blue}{e_0}$}}}%
    \put(0.74921299,0.64428116){\color[rgb]{0,0,0}\makebox(0,0)[lb]{\smash{$\color{blue}{e_1}$}}}%
    \put(0.02780693,0.2770199){\color[rgb]{0,0,0}\makebox(0,0)[lb]{\smash{A:}}}%
    \put(0.11962224,0.25078696){\color[rgb]{0,0,0}\makebox(0,0)[lb]{\smash{$\color{red}{f_2}$}}}%
    \put(0.25078698,0.11962222){\color[rgb]{0,0,0}\makebox(0,0)[lb]{\smash{$\color{red}{f_1}$}}}%
    \put(0.10650576,0.11962222){\color[rgb]{0,0,0}\makebox(0,0)[lb]{\smash{$\color{red}{f_0}$}}}%
    \put(0.39506821,0.2770199){\color[rgb]{0,0,0}\makebox(0,0)[lb]{\smash{B:}}}%
    \put(0.36883526,0.11962222){\color[rgb]{0,0,0}\makebox(0,0)[lb]{\smash{$\color{red}{f_0}$}}}%
    \put(0.5,0.25078696){\color[rgb]{0,0,0}\makebox(0,0)[lb]{\smash{$\color{red}{f_2}$}}}%
    \put(0.38195174,0.25078696){\color[rgb]{0,0,0}\makebox(0,0)[lb]{\smash{$\color{red}{f_1}$}}}%
    \put(0.76232948,0.2770199){\color[rgb]{0,0,0}\makebox(0,0)[lb]{\smash{C:}}}%
    \put(0.87381951,0.11962222){\color[rgb]{0,0,0}\makebox(0,0)[lb]{\smash{$\color{red}{f_1}$}}}%
    \put(0.87381951,0.00157391){\color[rgb]{0,0,0}\makebox(0,0)[lb]{\smash{$\color{red}{f_0}$}}}%
    \put(0.72953829,0.11962222){\color[rgb]{0,0,0}\makebox(0,0)[lb]{\smash{$\color{red}{f_2}$}}}%
  \end{picture}%
\endgroup%

%% file: combined_tops.pdf_tex
%% Creator: Inkscape inkscape 0.48.3.1, www.inkscape.org
%% Accompanies image file 'combined_tops.pdf' (pdf, eps, ps)
\begingroup%
  \makeatletter%
  \providecommand\color[2][]{%
    \errmessage{(Inkscape) Color is used for the text in Inkscape, but the package 'color.sty' is not loaded}%
    \renewcommand\color[2][]{}%
  }%
  \providecommand\transparent[1]{%
    \errmessage{(Inkscape) Transparency is used (non-zero) for the text in Inkscape, but the package 'transparent.sty' is not loaded}%
    \renewcommand\transparent[1]{}%
  }%
  \providecommand\rotatebox[2]{#2}%
  \ifx\svgwidth\undefined%
    \setlength{\unitlength}{834.15996094bp}%
    \ifx\svgscale\undefined%
      \relax%
    \else%
      \setlength{\unitlength}{\unitlength * \real{\svgscale}}%
    \fi%
  \else%
    \setlength{\unitlength}{\svgwidth}%
  \fi%
  \global\let\svgwidth\undefined%
  \global\let\svgscale\undefined%
  \makeatother%
  \begin{picture}(1,0.48824943)%
    \put(0,0){\includegraphics[width=\unitlength]{combined_tops.pdf}}%
    \put(0.8454014,0.30833414){\color[rgb]{0,0,0}\makebox(0,0)[lb]{\smash{${\rm III}\times{\rm B}$}}}%
    \put(0.00143857,0.46178193){\color[rgb]{0,0,0}\makebox(0,0)[lb]{\smash{${\rm I}\times{\rm A}$}}}%
    \put(0.21242928,0.46178193){\color[rgb]{0,0,0}\makebox(0,0)[lb]{\smash{${\rm I}\times{\rm B}$}}}%
    \put(0.42341999,0.46178193){\color[rgb]{0,0,0}\makebox(0,0)[lb]{\smash{${\rm I}\times{\rm C}$}}}%
    \put(0.63441069,0.46178193){\color[rgb]{0,0,0}\makebox(0,0)[lb]{\smash{${\rm III}\times{\rm A}$}}}%
    \put(0.00143857,0.17406733){\color[rgb]{0,0,0}\makebox(0,0)[lb]{\smash{${\rm II}\times{\rm C}$}}}%
    \put(0.21242928,0.17406733){\color[rgb]{0,0,0}\makebox(0,0)[lb]{\smash{${\rm II}\times{\rm B}$}}}%
    \put(0.42341999,0.17406733){\color[rgb]{0,0,0}\makebox(0,0)[lb]{\smash{${\rm II}\times{\rm A}$}}}%
    \put(0.63441069,0.17406733){\color[rgb]{0,0,0}\makebox(0,0)[lb]{\smash{${\rm III}\times{\rm C}$}}}%
  \end{picture}%
\endgroup%

%% file: outIxA.txt
\begin{center}
\begin{longtable}{| l | l |}
\caption[Possible matches for IxA]{Possible matches for $\mathrm{I} \times \mathrm{A}$ } \label{tab:match_IxA} \\*
\nobreakhline \textit{Matter spectrum} & \textit{Baryon- and Lepton number violation} \\* \nobreakhline
\endfirsthead
\multicolumn{2}{c}{{\bfseries \tablename\ \thetable{}, Possible matches for {\boldmath $\mathrm{I} \times \mathrm{A}$} -- continued from previous page}} \\
\hline
\textit{Matter spectrum} &
\textit{Baryon- and Lepton number violation} \\* \nobreakhline 
\endhead
\multicolumn{2}{|r|}{{Continued on next page}} \\ \hline
\endfoot
\endlastfoot
\hline
\begin{tabular}{@{} l }
	\textit{possibility no. 1} \rule{0pt}{3ex} \\ [1.5ex]
	$ a =0, b = -\frac{1}{2} ; \, (H_u,H_d) = ( \mathbf{2}^\mathrm{I}_{1} , \mathbf{2}^\mathrm{I}_{2} )$ \\[.3ex]
	heavy $(u_R^c , d_R^c)$: $(- \,  , \overline{\mathbf{3}}^\mathrm{A}_{4})$ \\ [.3ex]
	light $u_R^c$: $\overline{\mathbf{3}}^\mathrm{A}_{2}$; light $d_R^c$: $\overline{\mathbf{3}}^\mathrm{A}_{3}$ \\ [1ex]
	heavy generations of $(L, \nu_R^c, e_R^c)$: \\ [.3ex]$( \overline{\mathbf{2}}^\mathrm{I}_{1} , - \,  , \overline{\mathbf{1}}^\mathrm{(5)} )$, $( \mathbf{2}^\mathrm{I}_{2} , \overline{\mathbf{1}}^\mathrm{(2)} , \overline{\mathbf{1}}^\mathrm{(3)} )$\\ [1.5ex] light $\nu_R^c$: $\mathbf{1}^\mathrm{(2)}$; light $e_R^c$: $- \, $ \\ [.3ex]
	$\mathbf{1}_\mu$: $\overline{\mathbf{1}}^\mathrm{(2)}$\end{tabular}
	&
	\begin{tabular}{ @{} >{\raggedright\arraybackslash}p{.45\textwidth} @{}}
	$\alpha$: $ ( \overline{\mathbf{2}}^\mathrm{I}_{1},\overline{\mathbf{3}}^\mathrm{A}_{3} )$ , $( \mathbf{2}^\mathrm{I}_{2},\overline{\mathbf{3}}^\mathrm{A}_{4} )$ ;\rule{0pt}{3ex}  $\beta$: $ \overline{\mathbf{3}}^\mathrm{A}_{2}\overline{\mathbf{3}}^\mathrm{A}_{4}\overline{\mathbf{3}}^\mathrm{A}_{3}$; $\delta$: $ - \,$; \\ [.3ex]
	$\gamma$: $ \overline{\mathbf{2}}^\mathrm{I}_{1}\mathbf{2}^\mathrm{I}_{2}\overline{\mathbf{1}}^\mathrm{(5)}$, $\mathbf{2}^\mathrm{I}_{2}\mathbf{2}^\mathrm{I}_{2}\overline{\mathbf{1}}^\mathrm{(3)}$; \\ [.3ex]
	$\lambda_1$: $ \mathbf{2}^\mathrm{I}_{1} $; $\lambda_3$: $- \, $; $\lambda_6$: $ - \, $; $\lambda_8$: $\overline{\mathbf{1}}^\mathrm{(5)}$; $\lambda_{10}$:  $(\overline{\mathbf{3}}^\mathrm{A}_{3})^* $; \\ [.3ex]
	$\lambda_4$:  $( \overline{\mathbf{3}}^\mathrm{A}_{2} , \overline{\mathbf{1}}^\mathrm{(3)} )$; $\lambda_5$:  $( \overline{\mathbf{2}}^\mathrm{I}_{1} , \overline{\mathbf{2}}^\mathrm{I}_{1} )$; \\ [.3ex]
	$\lambda_9$:  $(\overline{\mathbf{3}}^\mathrm{A}_{2}, (\mathbf{2}^\mathrm{I}_{2})^*  )$; \\ [.3ex]
	$\lambda_7$:  $\overline{\mathbf{3}}^\mathrm{A}_{2} (\overline{\mathbf{3}}^\mathrm{A}_{4})^* \overline{\mathbf{1}}^\mathrm{(3)}$, $\overline{\mathbf{3}}^\mathrm{A}_{2}(\overline{\mathbf{3}}^\mathrm{A}_{3})^*\overline{\mathbf{1}}^\mathrm{(5)}$; \\ [.3ex]
	$\lambda_2$:  $\overline{\mathbf{3}}^\mathrm{A}_{2}\overline{\mathbf{3}}^\mathrm{A}_{2}\overline{\mathbf{3}}^\mathrm{A}_{4}\overline{\mathbf{1}}^\mathrm{(5)}$, $\overline{\mathbf{3}}^\mathrm{A}_{2}\overline{\mathbf{3}}^\mathrm{A}_{2}\overline{\mathbf{3}}^\mathrm{A}_{3}\overline{\mathbf{1}}^\mathrm{(3)}$
	\end{tabular}
	\\
\hline
\end{longtable}
\end{center}

%% file: outIxB.txt
\begin{center}
\begin{longtable}{| l | l |}
\caption[Possible matches for IxB]{Possible matches for $\mathrm{I} \times \mathrm{B}$ } \label{tab:match_IxB} \\*
\nobreakhline \textit{Matter spectrum} & \textit{Baryon- and Lepton number violation} \\* \nobreakhline
\endfirsthead
\multicolumn{2}{c}{{\bfseries \tablename\ \thetable{}, Possible matches for {\boldmath $\mathrm{I} \times \mathrm{B}$} -- continued from previous page}} \\
\hline
\textit{Matter spectrum} &
\textit{Baryon- and Lepton number violation} \\* \nobreakhline 
\endhead
\multicolumn{2}{|r|}{{Continued on next page}} \\ \hline
\endfoot
\endlastfoot
\hline
\begin{tabular}{@{} l }
	\textit{possibility no. 1} \rule{0pt}{3ex} \\ [1.5ex]
	$ a =0, b = -\frac{1}{2} ; \, (H_u,H_d) = ( \mathbf{2}^\mathrm{I}_{1} , \overline{\mathbf{2}}^\mathrm{I}_{1} )$ \\[.3ex]
	heavy $(u_R^c , d_R^c)$: $(- \,  , \overline{\mathbf{3}}^\mathrm{B}_{2})$ \\ [.3ex]
	light $u_R^c$: $\overline{\mathbf{3}}^\mathrm{B}_{1}$; light $d_R^c$: $\overline{\mathbf{3}}^\mathrm{B}_{4}$ \\ [1ex]
	heavy generations of $(L, \nu_R^c, e_R^c)$: \\ [.3ex]$( \overline{\mathbf{2}}^\mathrm{I}_{1} , - \,  , - \,  )$, $( \mathbf{2}^\mathrm{I}_{2} , \overline{\mathbf{1}}^\mathrm{(2)} , \overline{\mathbf{1}}^\mathrm{(5)} )$\\ [1.5ex] light $\nu_R^c$: $\mathbf{1}^\mathrm{(2)}$; light $e_R^c$: $\overline{\mathbf{1}}^\mathrm{(3)}$ \\ [.3ex]
	$\mathbf{1}_\mu$: $- \, $\end{tabular}
	&
	\begin{tabular}{ @{} >{\raggedright\arraybackslash}p{.45\textwidth} @{}}
	$\alpha$: $ ( \overline{\mathbf{2}}^\mathrm{I}_{1},\overline{\mathbf{3}}^\mathrm{B}_{2} )$ , $( \mathbf{2}^\mathrm{I}_{2},\overline{\mathbf{3}}^\mathrm{B}_{4} )$ ;\rule{0pt}{3ex}  $\beta$: $ \overline{\mathbf{3}}^\mathrm{B}_{1}\overline{\mathbf{3}}^\mathrm{B}_{4}\overline{\mathbf{3}}^\mathrm{B}_{4}$; $\delta$: $ - \,$; \\ [.3ex]
	$\gamma$: $ \overline{\mathbf{2}}^\mathrm{I}_{1}\mathbf{2}^\mathrm{I}_{2}\overline{\mathbf{1}}^\mathrm{(5)}$, $\mathbf{2}^\mathrm{I}_{2}\mathbf{2}^\mathrm{I}_{2}\overline{\mathbf{1}}^\mathrm{(3)}$; \\ [.3ex]
	$\lambda_1$: $ \overline{\mathbf{2}}^\mathrm{I}_{1} $; $\lambda_3$: $- \, $; $\lambda_6$: $\mathbf{2}^\mathrm{I}_{1}$; $\lambda_8$: $ - \, $; $\lambda_{10}$:  $(\overline{\mathbf{3}}^\mathrm{B}_{4})^* $; \\ [.3ex]
	$\lambda_4$:  $( \overline{\mathbf{3}}^\mathrm{B}_{1} , \overline{\mathbf{1}}^\mathrm{(5)} )$; $\lambda_5$:  $( \overline{\mathbf{2}}^\mathrm{I}_{1} , \overline{\mathbf{2}}^\mathrm{I}_{1} )$; \\ [.3ex]
	$\lambda_9$:  $(\overline{\mathbf{3}}^\mathrm{B}_{1}, (\mathbf{2}^\mathrm{I}_{2})^*  )$; \\ [.3ex]
	$\lambda_7$:  $\overline{\mathbf{3}}^\mathrm{B}_{1} (\overline{\mathbf{3}}^\mathrm{B}_{2})^* \overline{\mathbf{1}}^\mathrm{(5)}$, $\overline{\mathbf{3}}^\mathrm{B}_{1}(\overline{\mathbf{3}}^\mathrm{B}_{4})^*\overline{\mathbf{1}}^\mathrm{(3)}$; \\ [.3ex]
	$\lambda_2$:  $\overline{\mathbf{3}}^\mathrm{B}_{1}\overline{\mathbf{3}}^\mathrm{B}_{1}\overline{\mathbf{3}}^\mathrm{B}_{4}\overline{\mathbf{1}}^\mathrm{(3)}$
	\end{tabular}
	\\
\hline
\end{longtable}
\end{center}

%% file: outIxC.txt
\begin{center}
\begin{longtable}{| l | l |}
\caption[Possible matches for IxC]{Possible matches for $\mathrm{I} \times \mathrm{C}$ } \label{tab:match_IxC} \\*
\nobreakhline \textit{Matter spectrum} & \textit{Baryon- and Lepton number violation} \\* \nobreakhline
\endfirsthead
\multicolumn{2}{c}{{\bfseries \tablename\ \thetable{}, Possible matches for {\boldmath $\mathrm{I} \times \mathrm{C}$} -- continued from previous page}} \\
\hline
\textit{Matter spectrum} &
\textit{Baryon- and Lepton number violation} \\* \nobreakhline 
\endhead
\multicolumn{2}{|r|}{{Continued on next page}} \\ \hline
\endfoot
\endlastfoot
\hline
\begin{tabular}{@{} l }
	\textit{possibility no. 1} \rule{0pt}{3ex} \\ [1.5ex]
	$ a =-1, b = -1 ; \, (H_u,H_d) = ( \mathbf{2}^\mathrm{I}_{1} , \overline{\mathbf{2}}^\mathrm{I}_{1} )$ \\[.3ex]
	heavy $(u_R^c , d_R^c)$: $(\overline{\mathbf{3}}^\mathrm{C}_{2} , - \, )$ \\ [.3ex]
	light $u_R^c$: $\overline{\mathbf{3}}^\mathrm{C}_{4}$; light $d_R^c$: $\overline{\mathbf{3}}^\mathrm{C}_{5}$ \\ [1ex]
	heavy generations of $(L, \nu_R^c, e_R^c)$: \\ [.3ex]$( \overline{\mathbf{2}}^\mathrm{I}_{1} , - \,  , - \,  )$, $( \mathbf{2}^\mathrm{I}_{3} , \overline{\mathbf{1}}^\mathrm{(1)} , \overline{\mathbf{1}}^\mathrm{(6)} )$\\ [1.5ex] light $\nu_R^c$: $\mathbf{1}^\mathrm{(1)}$; light $e_R^c$: $\overline{\mathbf{1}}^\mathrm{(2)}$ \\ [.3ex]
	$\mathbf{1}_\mu$: $- \, $\end{tabular}
	&
	\begin{tabular}{ @{} >{\raggedright\arraybackslash}p{.45\textwidth} @{}}
	$\alpha$: $ ( \mathbf{2}^\mathrm{I}_{3},\overline{\mathbf{3}}^\mathrm{C}_{5} )$ ;\rule{0pt}{3ex}  $\beta$: $ \overline{\mathbf{3}}^\mathrm{C}_{4}\overline{\mathbf{3}}^\mathrm{C}_{5}\overline{\mathbf{3}}^\mathrm{C}_{5}$; $\delta$: $ - \,$; \\ [.3ex]
	$\gamma$: $ \overline{\mathbf{2}}^\mathrm{I}_{1}\mathbf{2}^\mathrm{I}_{3}\overline{\mathbf{1}}^\mathrm{(6)}$, $\mathbf{2}^\mathrm{I}_{3}\mathbf{2}^\mathrm{I}_{3}\overline{\mathbf{1}}^\mathrm{(2)}$; \\ [.3ex]
	$\lambda_1$: $ \overline{\mathbf{2}}^\mathrm{I}_{1} $; $\lambda_3$: $- \, $; $\lambda_6$: $\mathbf{2}^\mathrm{I}_{1}$; $\lambda_8$: $ - \, $; $\lambda_{10}$:  $(\overline{\mathbf{3}}^\mathrm{C}_{5})^* $; \\ [.3ex]
	$\lambda_4$:  $( \overline{\mathbf{3}}^\mathrm{C}_{4} , \overline{\mathbf{1}}^\mathrm{(6)} )$; $\lambda_5$:  $( \overline{\mathbf{2}}^\mathrm{I}_{1} , \overline{\mathbf{2}}^\mathrm{I}_{1} )$; \\ [.3ex]
	$\lambda_9$:  $(\overline{\mathbf{3}}^\mathrm{C}_{2}, (\overline{\mathbf{2}}^\mathrm{I}_{1})^*  )$, $(\overline{\mathbf{3}}^\mathrm{C}_{4}, (\mathbf{2}^\mathrm{I}_{3})^*  )$; \\ [.3ex]
	$\lambda_7$:  $\overline{\mathbf{3}}^\mathrm{C}_{2} (\overline{\mathbf{3}}^\mathrm{C}_{5})^* \overline{\mathbf{1}}^\mathrm{(6)}$, $\overline{\mathbf{3}}^\mathrm{C}_{4}(\overline{\mathbf{3}}^\mathrm{C}_{5})^*\overline{\mathbf{1}}^\mathrm{(2)}$; \\ [.3ex]
	$\lambda_2$:  $\overline{\mathbf{3}}^\mathrm{C}_{2}\overline{\mathbf{3}}^\mathrm{C}_{4}\overline{\mathbf{3}}^\mathrm{C}_{5}\overline{\mathbf{1}}^\mathrm{(6)}$, $\overline{\mathbf{3}}^\mathrm{C}_{4}\overline{\mathbf{3}}^\mathrm{C}_{4}\overline{\mathbf{3}}^\mathrm{C}_{5}\overline{\mathbf{1}}^\mathrm{(2)}$
	\end{tabular}
	\\
\hline
\end{longtable}
\end{center}

%% file: outIIIxA.txt
\begin{center}
\begin{longtable}{| l | l |}
\caption[Possible matches for IIIxA]{Possible matches for $\mathrm{III} \times \mathrm{A}$ } \label{tab:match_IIIxA} \\*
\nobreakhline \textit{Matter spectrum} & \textit{Baryon- and Lepton number violation} \\* \nobreakhline
\endfirsthead
\multicolumn{2}{c}{{\bfseries \tablename\ \thetable{}, Possible matches for {\boldmath $\mathrm{III} \times \mathrm{A}$} -- continued from previous page}} \\
\hline
\textit{Matter spectrum} &
\textit{Baryon- and Lepton number violation} \\* \nobreakhline 
\endhead
\multicolumn{2}{|r|}{{Continued on next page}} \\ \hline
\endfoot
\endlastfoot
\hline
\begin{tabular}{@{} l }
	\textit{possibility no. 1} \rule{0pt}{3ex} \\ [1.5ex]
	$ a =\frac{1}{2}, b = -1 ; \, (H_u,H_d) = ( \mathbf{2}^\mathrm{III}_{1} , \overline{\mathbf{2}}^\mathrm{III}_{1} )$ \\[.3ex]
	heavy $(u_R^c , d_R^c)$: $(\overline{\mathbf{3}}^\mathrm{A}_{1} , - \, )$ \\ [.3ex]
	light $u_R^c$: $- \, $; light $d_R^c$: $\overline{\mathbf{3}}^\mathrm{A}_{4}$ \\ [1ex]
	heavy generations of $(L, \nu_R^c, e_R^c)$: \\ [.3ex]$( \overline{\mathbf{2}}^\mathrm{III}_{1} , - \,  , - \,  )$, $( \mathbf{2}^\mathrm{III}_{2} , - \,  , \overline{\mathbf{1}}^\mathrm{(6)} )$\\ [1.5ex] light $\nu_R^c$: $- \, $; light $e_R^c$: $- \, $ \\ [.3ex]
	$\mathbf{1}_\mu$: $- \, $\end{tabular}
	&
	\begin{tabular}{ @{} >{\raggedright\arraybackslash}p{.45\textwidth} @{}}
	$\alpha$: $ ( \mathbf{2}^\mathrm{III}_{2},\overline{\mathbf{3}}^\mathrm{A}_{4} )$ ;\rule{0pt}{3ex}  $\beta$: - \, ; $\delta$: $ - \,$; \\ [.3ex]
	$\gamma$: $ \overline{\mathbf{2}}^\mathrm{III}_{1}\mathbf{2}^\mathrm{III}_{2}\overline{\mathbf{1}}^\mathrm{(6)}$; \\ [.3ex]
	$\lambda_1$: $ \overline{\mathbf{2}}^\mathrm{III}_{1} $; $\lambda_3$: $- \, $; $\lambda_6$: $\mathbf{2}^\mathrm{III}_{1}$; $\lambda_8$: $ - \, $; $\lambda_{10}$:  $(\overline{\mathbf{3}}^\mathrm{A}_{4})^* $; \\ [.3ex]
	$\lambda_4$: $ - \,$ ; $\lambda_5$:  $( \overline{\mathbf{2}}^\mathrm{III}_{1} , \overline{\mathbf{2}}^\mathrm{III}_{1} )$; \\ [.3ex]
	$\lambda_9$:  $(\overline{\mathbf{3}}^\mathrm{A}_{1}, (\overline{\mathbf{2}}^\mathrm{III}_{1})^*  )$; \\ [.3ex]
	$\lambda_7$:  $\overline{\mathbf{3}}^\mathrm{A}_{1} (\overline{\mathbf{3}}^\mathrm{A}_{4})^* \overline{\mathbf{1}}^\mathrm{(6)}$; \\ [.3ex]
	$\lambda_2$: $ - \,$ 
	\end{tabular}
	\\
\hline
\end{longtable}
\end{center}

%% file: outIIIxB.txt
\begin{center}
\begin{longtable}{| l | l |}
\caption[Possible matches for IIIxB]{Possible matches for $\mathrm{III} \times \mathrm{B}$ } \label{tab:match_IIIxB} \\*
\nobreakhline \textit{Matter spectrum} & \textit{Baryon- and Lepton number violation} \\* \nobreakhline
\endfirsthead
\multicolumn{2}{c}{{\bfseries \tablename\ \thetable{}, Possible matches for {\boldmath $\mathrm{III} \times \mathrm{B}$} -- continued from previous page}} \\
\hline
\textit{Matter spectrum} &
\textit{Baryon- and Lepton number violation} \\* \nobreakhline 
\endhead
\multicolumn{2}{|r|}{{Continued on next page}} \\ \hline
\endfoot
\endlastfoot
\hline
\begin{tabular}{@{} l }
	\textit{possibility no. 1} \rule{0pt}{3ex} \\ [1.5ex]
	$ a =-\frac{1}{2}, b = \frac{1}{2} ; \, (H_u,H_d) = ( \overline{\mathbf{2}}^\mathrm{III}_{1} , \mathbf{2}^\mathrm{III}_{1} )$ \\[.3ex]
	heavy $(u_R^c , d_R^c)$: $(- \,  , \overline{\mathbf{3}}^\mathrm{B}_{5})$ \\ [.3ex]
	light $u_R^c$: $\overline{\mathbf{3}}^\mathrm{B}_{2}$; light $d_R^c$: $\overline{\mathbf{3}}^\mathrm{B}_{1}$ \\ [1ex]
	heavy generations of $(L, \nu_R^c, e_R^c)$: \\ [.3ex]$( \mathbf{2}^\mathrm{III}_{1} , - \,  , - \,  )$, $( \overline{\mathbf{2}}^\mathrm{III}_{3} , \mathbf{1}^\mathrm{(4)} , \overline{\mathbf{1}}^\mathrm{(1)} )$\\ [1.5ex] light $\nu_R^c$: $\overline{\mathbf{1}}^\mathrm{(4)}$; light $e_R^c$: $\mathbf{1}^\mathrm{(5)}$ \\ [.3ex]
	$\mathbf{1}_\mu$: $- \, $\end{tabular}
	&
	\begin{tabular}{ @{} >{\raggedright\arraybackslash}p{.45\textwidth} @{}}
	$\alpha$: $ ( \mathbf{2}^\mathrm{III}_{1},\overline{\mathbf{3}}^\mathrm{B}_{5} )$ , $( \overline{\mathbf{2}}^\mathrm{III}_{3},\overline{\mathbf{3}}^\mathrm{B}_{1} )$ ;\rule{0pt}{3ex}  $\beta$: $ \overline{\mathbf{3}}^\mathrm{B}_{2}\overline{\mathbf{3}}^\mathrm{B}_{5}\overline{\mathbf{3}}^\mathrm{B}_{5}$; $\delta$: $ - \,$; \\ [.3ex]
	$\gamma$: $ \mathbf{2}^\mathrm{III}_{1}\overline{\mathbf{2}}^\mathrm{III}_{3}\overline{\mathbf{1}}^\mathrm{(1)}$, $\overline{\mathbf{2}}^\mathrm{III}_{3}\overline{\mathbf{2}}^\mathrm{III}_{3}\mathbf{1}^\mathrm{(5)}$; \\ [.3ex]
	$\lambda_1$: $- \,$; $\lambda_3$: $- \, $; $\lambda_6$: $\mathbf{2}^\mathrm{III}_{1}$; $\lambda_8$: $ - \, $; $\lambda_{10}$: $ - \, $; \\ [.3ex]
	$\lambda_4$:  $( \overline{\mathbf{3}}^\mathrm{B}_{2} , \overline{\mathbf{1}}^\mathrm{(1)} )$; $\lambda_5$:  $( \mathbf{2}^\mathrm{III}_{1} , \mathbf{2}^\mathrm{III}_{1} )$; \\ [.3ex]
	$\lambda_9$:  $(\overline{\mathbf{3}}^\mathrm{B}_{2}, (\overline{\mathbf{2}}^\mathrm{III}_{3})^*  )$; \\ [.3ex]
	$\lambda_7$:  $\overline{\mathbf{3}}^\mathrm{B}_{2} (\overline{\mathbf{3}}^\mathrm{B}_{5})^* \overline{\mathbf{1}}^\mathrm{(1)}$, $\overline{\mathbf{3}}^\mathrm{B}_{2}(\overline{\mathbf{3}}^\mathrm{B}_{1})^*\mathbf{1}^\mathrm{(5)}$; \\ [.3ex]
	$\lambda_2$:  $\overline{\mathbf{3}}^\mathrm{B}_{2}\overline{\mathbf{3}}^\mathrm{B}_{2}\overline{\mathbf{3}}^\mathrm{B}_{5}\overline{\mathbf{1}}^\mathrm{(1)}$
	\end{tabular}
	\\
\hline \hline
\end{longtable}
\end{center}

%% file: f_theory-standard_model-v5.bbl
\begin{thebibliography}{100}
\expandafter\ifx\csname url\endcsname\relax
  \def\url#1{{\tt #1}}\fi
\expandafter\ifx\csname urlprefix\endcsname\relax\def\urlprefix{URL }\fi
\providecommand{\eprint}[2][]{\url{#2}}

\bibitem{Bouchard:2005ag}
V.~Bouchard and R.~Donagi, {\it {An SU(5) heterotic standard model}\/}, {\it
  Phys.Lett.\/} {\bf B633} (2006) 783--791,
  \href{http://www.arxiv.org/abs/hep-th/0512149}{{\tt [hep-th/0512149]}}.

\bibitem{Braun:2005nv}
V.~Braun, Y.-H. He, B.~A. Ovrut and T.~Pantev, {\it {The Exact MSSM spectrum
  from string theory}\/}, {\it JHEP\/} {\bf 0605} (2006) 043,
  \href{http://www.arxiv.org/abs/hep-th/0512177}{{\tt [hep-th/0512177]}}.

\bibitem{Lebedev:2006kn}
O.~Lebedev, H.~P. Nilles, S.~Raby, S.~Ramos-Sanchez, M.~Ratz et~al., {\it {A
  Mini-landscape of exact MSSM spectra in heterotic orbifolds}\/}, {\it
  Phys.Lett.\/} {\bf B645} (2007) 88--94,
  \href{http://www.arxiv.org/abs/hep-th/0611095}{{\tt [hep-th/0611095]}}.

\bibitem{Anderson:2011ns}
L.~B. Anderson, J.~Gray, A.~Lukas and E.~Palti, {\it {Two Hundred Heterotic
  Standard Models on Smooth Calabi-Yau Threefolds}\/}, {\it Phys.Rev.\/} {\bf
  D84} (2011) 106005, \href{http://www.arxiv.org/abs/1106.4804}{{\tt
  [1106.4804]}}.

\bibitem{Donagi:2008ca}
R.~Donagi and M.~Wijnholt, {\it {Model Building with F-Theory}\/}, {\it
  Adv.Theor.Math.Phys.\/} {\bf 15} (2011) 1237--1318,
  \href{http://www.arxiv.org/abs/0802.2969}{{\tt [0802.2969]}}.

\bibitem{oai:arXiv.org:0802.3391}
C.~Beasley, J.~J. Heckman and C.~Vafa, {\it {GUTs and Exceptional Branes in
  F-theory - I}\/}, {\it JHEP\/} {\bf 0901} (2009) 058,
  \href{http://www.arxiv.org/abs/0802.3391}{{\tt [0802.3391]}}.

\bibitem{Beasley:2008kw}
C.~Beasley, J.~J. Heckman and C.~Vafa, {\it {GUTs and Exceptional Branes in
  F-theory - II: Experimental Predictions}\/}, {\it JHEP\/} {\bf 01} (2009)
  059, \href{http://www.arxiv.org/abs/0806.0102}{{\tt [0806.0102]}}.

\bibitem{Donagi:2008kj}
R.~Donagi and M.~Wijnholt, {\it {Breaking GUT Groups in F-Theory}\/}, {\it
  Adv.Theor.Math.Phys.\/} {\bf 15} (2011) 1523--1604,
  \href{http://www.arxiv.org/abs/0808.2223}{{\tt [0808.2223]}}.

\bibitem{Heckman:2010bq}
J.~J. Heckman, {\it {Particle Physics Implications of F-theory}\/}
  \href{http://www.arxiv.org/abs/1001.0577}{{\tt [1001.0577]}}.

\bibitem{Weigand:2010wm}
T.~Weigand, {\it {Lectures on F-theory compactifications and model
  building}\/}, {\it Class. Quant. Grav.\/} {\bf 27} (2010) 214004,
  \href{http://www.arxiv.org/abs/1009.3497}{{\tt [1009.3497]}}.

\bibitem{Maharana:2012tu}
A.~Maharana and E.~Palti, {\it {Models of Particle Physics from Type IIB String
  Theory and F-theory: A Review}\/}, {\it Int.J.Mod.Phys.\/} {\bf A28} (2013)
  1330005, \href{http://www.arxiv.org/abs/1212.0555}{{\tt [1212.0555]}}.

\bibitem{Blumenhagen:2000wh}
R.~Blumenhagen, L.~Goerlich, B.~Kors and D.~Lust, {\it {Noncommutative
  compactifications of type I strings on tori with magnetic background
  flux}\/}, {\it JHEP\/} {\bf 0010} (2000) 006,
  \href{http://www.arxiv.org/abs/hep-th/0007024}{{\tt [hep-th/0007024]}}.

\bibitem{Aldazabal:2000cn}
G.~Aldazabal, S.~Franco, L.~E. Ibanez, R.~Rabadan and A.~Uranga, {\it
  {Intersecting brane worlds}\/}, {\it JHEP\/} {\bf 0102} (2001) 047,
  \href{http://www.arxiv.org/abs/hep-ph/0011132}{{\tt [hep-ph/0011132]}}.

\bibitem{Cvetic:2001nr}
M.~Cvetic, G.~Shiu and A.~M. Uranga, {\it {Chiral four-dimensional N=1
  supersymmetric type 2A orientifolds from intersecting D6 branes}\/}, {\it
  Nucl.Phys.\/} {\bf B615} (2001) 3--32,
  \href{http://www.arxiv.org/abs/hep-th/0107166}{{\tt [hep-th/0107166]}}.

\bibitem{Blumenhagen:2003jy}
R.~Blumenhagen, D.~Lust and S.~Stieberger, {\it {Gauge unification in
  supersymmetric intersecting brane worlds}\/}, {\it JHEP\/} {\bf 0307} (2003)
  036, \href{http://www.arxiv.org/abs/hep-th/0305146}{{\tt [hep-th/0305146]}}.

\bibitem{Lust:2004ks}
D.~Lust, {\it {Intersecting brane worlds: A Path to the standard model?}\/},
  {\it Class.Quant.Grav.\/} {\bf 21} (2004) S1399--1424,
  \href{http://www.arxiv.org/abs/hep-th/0401156}{{\tt [hep-th/0401156]}}.

\bibitem{Blumenhagen:2006ci}
R.~Blumenhagen, B.~Kors, D.~Lust and S.~Stieberger, {\it {Four-dimensional
  String Compactifications with D-Branes, Orientifolds and Fluxes}\/}, {\it
  Phys.Rept.\/} {\bf 445} (2007) 1--193,
  \href{http://www.arxiv.org/abs/hep-th/0610327}{{\tt [hep-th/0610327]}}.

\bibitem{Marchesano:2007de}
F.~Marchesano, {\it {Progress in D-brane model building}\/}, {\it
  Fortsch.Phys.\/} {\bf 55} (2007) 491--518,
  \href{http://www.arxiv.org/abs/hep-th/0702094}{{\tt [hep-th/0702094]}}.

\bibitem{Cvetic:2011vz}
M.~Cvetic and J.~Halverson, {\it {TASI Lectures: Particle Physics from
  Perturbative and Non-perturbative Effects in D-braneworlds}\/} pp. 245--292,
  \href{http://www.arxiv.org/abs/1101.2907}{{\tt [1101.2907]}}.

\bibitem{Ibanez:2012zz}
L.~E. Ibanez and A.~M. Uranga, {\it {String theory and particle physics: An
  introduction to string phenomenology}\/} .

\bibitem{Honecker:2012qr}
G.~Honecker, M.~Ripka and W.~Staessens, {\it {The Importance of Being Rigid:
  D6-Brane Model Building on $T^6/Z_2 x Z_6'$ with Discrete Torsion}\/}, {\it
  Nucl.Phys.\/} {\bf B868} (2013) 156--222,
  \href{http://www.arxiv.org/abs/1209.3010}{{\tt [1209.3010]}}.

\bibitem{Honecker:2012jd}
G.~Honecker and J.~Vanhoof, {\it {Yukawa couplings and masses of non-chiral
  states for the Standard Model on D6-branes on T6/Z6'}\/}, {\it JHEP\/} {\bf
  1204} (2012) 085, \href{http://www.arxiv.org/abs/1201.3604}{{\tt
  [1201.3604]}}.

\bibitem{Dijkstra:2004cc}
T.~Dijkstra, L.~Huiszoon and A.~Schellekens, {\it {Supersymmetric standard
  model spectra from RCFT orientifolds}\/}, {\it Nucl.Phys.\/} {\bf B710}
  (2005) 3--57, \href{http://www.arxiv.org/abs/hep-th/0411129}{{\tt
  [hep-th/0411129]}}.

\bibitem{Anastasopoulos:2006da}
P.~Anastasopoulos, T.~Dijkstra, E.~Kiritsis and A.~Schellekens, {\it
  {Orientifolds, hypercharge embeddings and the Standard Model}\/}, {\it
  Nucl.Phys.\/} {\bf B759} (2006) 83--146,
  \href{http://www.arxiv.org/abs/hep-th/0605226}{{\tt [hep-th/0605226]}}.

\bibitem{Dolan:2011qu}
M.~J. Dolan, S.~Krippendorf and F.~Quevedo, {\it {Towards a Systematic
  Construction of Realistic D-brane Models on a del Pezzo Singularity}\/}, {\it
  JHEP\/} {\bf 1110} (2011) 024, \href{http://www.arxiv.org/abs/1106.6039}{{\tt
  [1106.6039]}}.

\bibitem{Cicoli:2012vw}
M.~Cicoli, S.~Krippendorf, C.~Mayrhofer, F.~Quevedo and R.~Valandro, {\it
  {D-Branes at del Pezzo Singularities: Global Embedding and Moduli
  Stabilisation}\/}, {\it JHEP\/} {\bf 1209} (2012) 019,
  \href{http://www.arxiv.org/abs/1206.5237}{{\tt [1206.5237]}}.

\bibitem{Ibanez:1998qp}
L.~E. Ibanez, R.~Rabadan and A.~Uranga, {\it {Anomalous U(1)'s in type I and
  type IIB D = 4, N=1 string vacua}\/}, {\it Nucl.Phys.\/} {\bf B542} (1999)
  112--138, \href{http://www.arxiv.org/abs/hep-th/9808139}{{\tt
  [hep-th/9808139]}}.

\bibitem{Blumenhagen:2006xt}
R.~Blumenhagen, M.~Cvetic and T.~Weigand, {\it {Spacetime instanton corrections
  in 4D string vacua: The Seesaw mechanism for D-Brane models}\/}, {\it
  Nucl.Phys.\/} {\bf B771} (2007) 113--142,
  \href{http://www.arxiv.org/abs/hep-th/0609191}{{\tt [hep-th/0609191]}}.

\bibitem{Ibanez:2006da}
L.~Ibanez and A.~Uranga, {\it {Neutrino Majorana Masses from String Theory
  Instanton Effects}\/}, {\it JHEP\/} {\bf 0703} (2007) 052,
  \href{http://www.arxiv.org/abs/hep-th/0609213}{{\tt [hep-th/0609213]}}.

\bibitem{Haack:2006cy}
M.~Haack, D.~Krefl, D.~Lust, A.~Van~Proeyen and M.~Zagermann, {\it {Gaugino
  Condensates and D-terms from D7-branes}\/}, {\it JHEP\/} {\bf 0701} (2007)
  078, \href{http://www.arxiv.org/abs/hep-th/0609211}{{\tt [hep-th/0609211]}}.

\bibitem{Florea:2006si}
B.~Florea, S.~Kachru, J.~McGreevy and N.~Saulina, {\it {Stringy Instantons and
  Quiver Gauge Theories}\/}, {\it JHEP\/} {\bf 0705} (2007) 024,
  \href{http://www.arxiv.org/abs/hep-th/0610003}{{\tt [hep-th/0610003]}}.

\bibitem{Blumenhagen:2009qh}
R.~Blumenhagen, M.~Cvetic, S.~Kachru and T.~Weigand, {\it {D-Brane Instantons
  in Type II Orientifolds}\/}, {\it Ann.Rev.Nucl.Part.Sci.\/} {\bf 59} (2009)
  269--296, \href{http://www.arxiv.org/abs/0902.3251}{{\tt [0902.3251]}}.

\bibitem{BerasaluceGonzalez:2011wy}
M.~Berasaluce-Gonzalez, L.~E. Ibanez, P.~Soler and A.~M. Uranga, {\it {Discrete
  gauge symmetries in D-brane models}\/}, {\it JHEP\/} {\bf 1112} (2011) 113,
  \href{http://www.arxiv.org/abs/1106.4169}{{\tt [1106.4169]}}.

\bibitem{BerasaluceGonzalez:2012vb}
M.~Berasaluce-Gonzalez, P.~Camara, F.~Marchesano, D.~Regalado and A.~Uranga,
  {\it {Non-Abelian discrete gauge symmetries in 4d string models}\/}, {\it
  JHEP\/} {\bf 1209} (2012) 059, \href{http://www.arxiv.org/abs/1206.2383}{{\tt
  [1206.2383]}}.

\bibitem{Ibanez:2012wg}
L.~Ibanez, A.~Schellekens and A.~Uranga, {\it {Discrete Gauge Symmetries in
  Discrete MSSM-like Orientifolds}\/}, {\it Nucl.Phys.\/} {\bf B865} (2012)
  509--540, \href{http://www.arxiv.org/abs/1205.5364}{{\tt [1205.5364]}}.

\bibitem{Anastasopoulos:2012zu}
P.~Anastasopoulos, M.~Cvetic, R.~Richter and P.~K. Vaudrevange, {\it {String
  Constraints on Discrete Symmetries in MSSM Type II Quivers}\/}, {\it JHEP\/}
  {\bf 1303} (2013) 011, \href{http://www.arxiv.org/abs/1211.1017}{{\tt
  [1211.1017]}}.

\bibitem{Honecker:2013hda}
G.~Honecker and W.~Staessens, {\it {To Tilt or Not To Tilt: Discrete Gauge
  Symmetries in Global Intersecting D-Brane Models}\/}, {\it JHEP\/} {\bf 1310}
  (2013) 146, \href{http://www.arxiv.org/abs/1303.4415}{{\tt [1303.4415]}}.

\bibitem{Ibanez:2008my}
L.~Ibanez and R.~Richter, {\it {Stringy Instantons and Yukawa Couplings in
  MSSM-like Orientifold Models}\/}, {\it JHEP\/} {\bf 0903} (2009) 090,
  \href{http://www.arxiv.org/abs/0811.1583}{{\tt [0811.1583]}}.

\bibitem{Cvetic:2009yh}
M.~Cvetic, J.~Halverson and R.~Richter, {\it {Realistic Yukawa structures from
  orientifold compactifications}\/}, {\it JHEP\/} {\bf 0912} (2009) 063,
  \href{http://www.arxiv.org/abs/0905.3379}{{\tt [0905.3379]}}.

\bibitem{Cvetic:2009ez}
M.~Cvetic, J.~Halverson and R.~Richter, {\it {Mass Hierarchies from MSSM
  Orientifold Compactifications}\/}, {\it JHEP\/} {\bf 1007} (2010) 005,
  \href{http://www.arxiv.org/abs/0909.4292}{{\tt [0909.4292]}}.

\bibitem{Blumenhagen:2001te}
R.~Blumenhagen, B.~Kors, D.~Lust and T.~Ott, {\it {The standard model from
  stable intersecting brane world orbifolds}\/}, {\it Nucl.Phys.\/} {\bf B616}
  (2001) 3--33, \href{http://www.arxiv.org/abs/hep-th/0107138}{{\tt
  [hep-th/0107138]}}.

\bibitem{Blumenhagen:2007zk}
R.~Blumenhagen, M.~Cvetic, D.~Lust, R.~Richter and T.~Weigand, {\it
  {Non-perturbative Yukawa Couplings from String Instantons}\/}, {\it
  Phys.Rev.Lett.\/} {\bf 100} (2008) 061602,
  \href{http://www.arxiv.org/abs/0707.1871}{{\tt [0707.1871]}}.

\bibitem{Donagi:2009ra}
R.~Donagi and M.~Wijnholt, {\it {Higgs Bundles and UV Completion in
  F-Theory}\/}, {\it Commun.Math.Phys.\/} {\bf 326} (2014) 287--327,
  \href{http://www.arxiv.org/abs/0904.1218}{{\tt [0904.1218]}}.

\bibitem{Krause:2012yh}
S.~Krause, C.~Mayrhofer and T.~Weigand, {\it {Gauge Fluxes in F-theory and Type
  IIB Orientifolds}\/}, {\it JHEP\/} {\bf 1208} (2012) 119,
  \href{http://www.arxiv.org/abs/1202.3138}{{\tt [1202.3138]}}.

\bibitem{Grimm:2010ez}
T.~W. Grimm and T.~Weigand, {\it {On Abelian Gauge Symmetries and Proton Decay
  in Global F-theory GUTs}\/}, {\it Phys.Rev.\/} {\bf D82} (2010) 086009,
  \href{http://www.arxiv.org/abs/1006.0226}{{\tt [1006.0226]}}.

\bibitem{Grimm:2011tb}
T.~W. Grimm, M.~Kerstan, E.~Palti and T.~Weigand, {\it {Massive Abelian Gauge
  Symmetries and Fluxes in F-theory}\/}, {\it JHEP\/} {\bf 1112} (2011) 004,
  \href{http://www.arxiv.org/abs/1107.3842}{{\tt [1107.3842]}}.

\bibitem{Braun:2014nva}
A.~P. Braun, A.~Collinucci and R.~Valandro, {\it {The fate of U(1)'s at strong
  coupling in F-theory}\/} \href{http://www.arxiv.org/abs/1402.4054}{{\tt
  [1402.4054]}}.

\bibitem{Anderson:2014yva}
L.~B. Anderson, I.~Garca-Etxebarria, T.~W. Grimm and J.~Keitel, {\it {Physics
  of F-theory compactifications without section}\/}
  \href{http://www.arxiv.org/abs/1406.5180}{{\tt [1406.5180]}}.

\bibitem{Braun:2011zm}
A.~P. Braun, A.~Collinucci and R.~Valandro, {\it {G-flux in F-theory and
  algebraic cycles}\/}, {\it Nucl.Phys.\/} {\bf B856} (2012) 129--179,
  \href{http://www.arxiv.org/abs/1107.5337}{{\tt [1107.5337]}}.

\bibitem{Krause:2011xj}
S.~Krause, C.~Mayrhofer and T.~Weigand, {\it {$G_4$ flux, chiral matter and
  singularity resolution in F-theory compactifications}\/}, {\it Nucl.Phys.\/}
  {\bf B858} (2012) 1--47, \href{http://www.arxiv.org/abs/1109.3454}{{\tt
  [1109.3454]}}.

\bibitem{Grimm:2011fx}
T.~W. Grimm and H.~Hayashi, {\it {F-theory fluxes, Chirality and Chern-Simons
  theories}\/}, {\it JHEP\/} {\bf 1203} (2012) 027,
  \href{http://www.arxiv.org/abs/1111.1232}{{\tt [1111.1232]}}.

\bibitem{oai:arXiv.org:1202.3138}
S.~Krause, C.~Mayrhofer and T.~Weigand, {\it {Gauge Fluxes in F-theory and Type
  IIB Orientifolds}\/}, {\it JHEP\/} {\bf 1208} (2012) 119,
  \href{http://www.arxiv.org/abs/1202.3138}{{\tt [1202.3138]}}.

\bibitem{Morrison:2012ei}
D.~R. Morrison and D.~S. Park, {\it {F-Theory and the Mordell-Weil Group of
  Elliptically-Fibered Calabi-Yau Threefolds}\/}, {\it JHEP\/} {\bf 1210}
  (2012) 128, \href{http://www.arxiv.org/abs/1208.2695}{{\tt [1208.2695]}}.

\bibitem{oai:arXiv.org:1210.6034}
M.~Cvetic, T.~W. Grimm and D.~Klevers, {\it {Anomaly Cancellation And Abelian
  Gauge Symmetries In F-theory}\/}, {\it JHEP\/} {\bf 1302} (2013) 101,
  \href{http://www.arxiv.org/abs/1210.6034}{{\tt [1210.6034]}}.

\bibitem{Mayrhofer:2012zy}
C.~Mayrhofer, E.~Palti and T.~Weigand, {\it {U(1) symmetries in F-theory GUTs
  with multiple sections}\/}, {\it JHEP\/} {\bf 1303} (2013) 098,
  \href{http://www.arxiv.org/abs/1211.6742}{{\tt [1211.6742]}}.

\bibitem{Braun:2013yti}
V.~Braun, T.~W. Grimm and J.~Keitel, {\it {New Global F-theory GUTs with U(1)
  symmetries}\/}, {\it JHEP\/} {\bf 1309} (2013) 154,
  \href{http://www.arxiv.org/abs/1302.1854}{{\tt [1302.1854]}}.

\bibitem{Borchmann:2013jwa}
J.~Borchmann, C.~Mayrhofer, E.~Palti and T.~Weigand, {\it {Elliptic fibrations
  for $SU(5)\times U(1)\times U(1)$ F-theory vacua}\/}, {\it Phys.Rev.\/} {\bf
  D88}, no.~4 (2013) 046005, \href{http://www.arxiv.org/abs/1303.5054}{{\tt
  [1303.5054]}}.

\bibitem{Cvetic:2013nia}
M.~Cvetic, D.~Klevers and H.~Piragua, {\it {F-Theory Compactifications with
  Multiple U(1)-Factors: Constructing Elliptic Fibrations with Rational
  Sections}\/}, {\it JHEP\/} {\bf 1306} (2013) 067,
  \href{http://www.arxiv.org/abs/1303.6970}{{\tt [1303.6970]}}.

\bibitem{Braun:2013nqa}
V.~Braun, T.~W. Grimm and J.~Keitel, {\it {Geometric Engineering in Toric
  F-Theory and GUTs with U(1) Gauge Factors}\/}
  \href{http://www.arxiv.org/abs/1306.0577}{{\tt [1306.0577]}}.

\bibitem{Cvetic:2013uta}
M.~Cveti\v{c}, A.~Grassi, D.~Klevers and H.~Piragua, {\it {Chiral
  Four-Dimensional F-Theory Compactifications With SU(5) and Multiple
  U(1)-Factors}\/} \href{http://www.arxiv.org/abs/1306.3987}{{\tt
  [1306.3987]}}.

\bibitem{Borchmann:2013hta}
J.~Borchmann, C.~Mayrhofer, E.~Palti and T.~Weigand, {\it {SU(5) Tops with
  Multiple U(1)s in F-theory}\/}, {\it Nucl.Phys.\/} {\bf B882} (2014) 1--69,
  \href{http://www.arxiv.org/abs/1307.2902}{{\tt [1307.2902]}}.

\bibitem{Cvetic:2013jta}
M.~Cveti{\v c}, D.~Klevers and H.~Piragua, {\it {F-Theory Compactifications
  with Multiple U(1)-Factors: Addendum}\/}, {\it JHEP\/} {\bf 1312} (2013) 056,
  \href{http://www.arxiv.org/abs/1307.6425}{{\tt [1307.6425]}}.

\bibitem{Cvetic:2013qsa}
M.~Cvetic, D.~Klevers, H.~Piragua and P.~Song, {\it {Elliptic Fibrations with
  Rank Three Mordell-Weil Group: F-theory with U(1) x U(1) x U(1) Gauge
  Symmetry}\/} \href{http://www.arxiv.org/abs/1310.0463}{{\tt [1310.0463]}}.

\bibitem{Morrison:2014era}
D.~R. Morrison and W.~Taylor, {\it {Sections, multisections, and U(1) fields in
  F-theory}\/} \href{http://www.arxiv.org/abs/1404.1527}{{\tt [1404.1527]}}.

\bibitem{Martini:2014iza}
G.~Martini and W.~Taylor, {\it {6D F-theory models and elliptically fibered
  Calabi-Yau threefolds over semi-toric base surfaces}\/}
  \href{http://www.arxiv.org/abs/1404.6300}{{\tt [1404.6300]}}.

\bibitem{Bizet:2014uua}
N.~C. Bizet, A.~Klemm and D.~V. Lopes, {\it {Landscaping with fluxes and the E8
  Yukawa Point in F-theory}\/} \href{http://www.arxiv.org/abs/1404.7645}{{\tt
  [1404.7645]}}.

\bibitem{Marsano:2009wr}
J.~Marsano, N.~Saulina and S.~Schafer-Nameki, {\it {Compact F-theory GUTs with
  U(1)(PQ)}\/}, {\it JHEP\/} {\bf 1004} (2010) 095,
  \href{http://www.arxiv.org/abs/0912.0272}{{\tt [0912.0272]}}.

\bibitem{Hayashi:2010zp}
H.~Hayashi, T.~Kawano, Y.~Tsuchiya and T.~Watari, {\it {More on Dimension-4
  Proton Decay Problem in F-theory -- Spectral Surface, Discriminant Locus and
  Monodromy}\/}, {\it Nucl.Phys.\/} {\bf B840} (2010) 304--348,
  \href{http://www.arxiv.org/abs/1004.3870}{{\tt [1004.3870]}}.

\bibitem{Dolan:2011iu}
M.~J. Dolan, J.~Marsano, N.~Saulina and S.~Schafer-Nameki, {\it {F-theory GUTs
  with U(1) Symmetries: Generalities and Survey}\/}, {\it Phys.Rev.\/} {\bf
  D84} (2011) 066008, \href{http://www.arxiv.org/abs/1102.0290}{{\tt
  [1102.0290]}}.

\bibitem{Dolan:2011aq}
M.~J. Dolan, J.~Marsano and S.~Schafer-Nameki, {\it {Unification and
  Phenomenology of F-Theory GUTs with U(1)PQ}\/}, {\it JHEP\/} {\bf 1112}
  (2011) 032, \href{http://www.arxiv.org/abs/1109.4958}{{\tt [1109.4958]}}.

\bibitem{Marsano:2011nn}
J.~Marsano, N.~Saulina and S.~Schafer-Nameki, {\it {On G-flux, M5 instantons,
  and U(1)s in F-theory}\/} \href{http://www.arxiv.org/abs/1107.1718}{{\tt
  [1107.1718]}}.

\bibitem{Krippendorf:2014xba}
S.~Krippendorf, D.~K.~M. Pena, P.-K. Oehlmann and F.~Ruehle, {\it {Rational
  F-Theory GUTs without exotics}\/}
  \href{http://www.arxiv.org/abs/1401.5084}{{\tt [1401.5084]}}.

\bibitem{Candelas:1996su}
P.~Candelas and A.~Font, {\it {Duality between the webs of heterotic and type
  II vacua}\/}, {\it Nucl.Phys.\/} {\bf B511} (1998) 295--325,
  \href{http://www.arxiv.org/abs/hep-th/9603170}{{\tt [hep-th/9603170]}}.

\bibitem{Candelas:1997eh}
P.~Candelas, E.~Perevalov and G.~Rajesh, {\it {Toric geometry and enhanced
  gauge symmetry of F theory / heterotic vacua}\/}, {\it Nucl.Phys.\/} {\bf
  B507} (1997) 445--474, \href{http://www.arxiv.org/abs/hep-th/9704097}{{\tt
  [hep-th/9704097]}}.

\bibitem{Choi:2013hua}
K.-S. Choi, {\it {On the Standard Model Group in F-theory}\/}
  \href{http://www.arxiv.org/abs/1309.7297}{{\tt [1309.7297]}}.

\bibitem{Choi:2010su}
K.-S. Choi and T.~Kobayashi, {\it {Towards the MSSM from F-theory}\/}, {\it
  Phys.Lett.\/} {\bf B693} (2010) 330--333,
  \href{http://www.arxiv.org/abs/1003.2126}{{\tt [1003.2126]}}.

\bibitem{Choi:2010nf}
K.-S. Choi, {\it {SU(3) x SU(2) x U(1) Vacua in F-Theory}\/}, {\it
  Nucl.Phys.\/} {\bf B842} (2011) 1--32,
  \href{http://www.arxiv.org/abs/1007.3843}{{\tt [1007.3843]}}.

\bibitem{Bouchard:2003bu}
V.~Bouchard and H.~Skarke, {\it {Affine Kac-Moody algebras, CHL strings and the
  classification of tops}\/}, {\it Adv.Theor.Math.Phys.\/} {\bf 7} (2003)
  205--232, \href{http://www.arxiv.org/abs/hep-th/0303218}{{\tt
  [hep-th/0303218]}}.

\bibitem{Klemm:1996hh}
A.~Klemm, P.~Mayr and C.~Vafa, {\it {BPS states of exceptional noncritical
  strings}\/} \href{http://www.arxiv.org/abs/hep-th/9607139}{{\tt
  [hep-th/9607139]}}.

\bibitem{Grimm:2013oga}
T.~W. Grimm, A.~Kapfer and J.~Keitel, {\it {Effective action of 6D F-Theory
  with U(1) factors: Rational sections make Chern-Simons terms jump}\/}, {\it
  JHEP\/} {\bf 1307} (2013) 115, \href{http://www.arxiv.org/abs/1305.1929}{{\tt
  [1305.1929]}}.

\bibitem{Shioda:1989}
T.~Shioda, {\it {Mordell-Weil Lattices and Galois Representation. I}\/}, {\it
  Proc. Japan Acad.\/} {\bf A65} (1989) 268--271.

\bibitem{Wazir:2001}
R.~Wazir, {\it Arithmetic on elliptic threefolds\/}, {\it Compos.Math.\/} {\bf
  140} (2001) 567--580, \href{http://www.arxiv.org/abs/math.NT/0112259}{{\tt
  [math.NT/0112259]}}.

\bibitem{Park:2011ji}
D.~S. Park, {\it {Anomaly Equations and Intersection Theory}\/}, {\it JHEP\/}
  {\bf 01} (2012) 093, \href{http://www.arxiv.org/abs/1111.2351}{{\tt
  [1111.2351]}}.

\bibitem{hartshorne:alggeo}
R.~Hartshorne, {\it Algebraic Geometry\/}, Graduate Texts in Mathematics,
  Springer, 2010, ISBN 9781441928078,
  \urlprefix\url{http://books.google.de/books?id=zO9wcgAACAAJ}.

\bibitem{cox:alggeo}
D.~A. Cox, J.~Little and D.~O'Shea, {\it Using algebraic geometry\/}, number
  185 in Graduate texts in mathematics ; 185 ; Graduate texts in mathematics,
  New York, NY: Springer, 2. ed. edition, 2005, ISBN 0-387-20706-6 ;
  0-387-20733-3 ; 978-0-387-20733-9 ; 978-0-387-20706-3, XII, 572 S. pp.,
  includes bibliographical references and index.

\bibitem{cox:ideals}
D.~A. Cox, J.~N. Little and D.~O'Shea, {\it Ideals, Varieties, and
  Algorithms\/}, Undergraduate Texts in Mathematics, New York, NY: Springer New
  York, 3., rd ed. 2007. corr. 2nd printing. softcover version of original
  hardcover edition 2007 edition, 2010, ISBN 978-1-441-92257-1 ; 1-441-92257-1
  ; 978-1-4419-2257-1, XVI, 551 S. pp.

\bibitem{Piragua}
H.~Piragua, {\it Aspects of elliptic fibrations with higher rank Mordell-Weil
  groups\/},
  \url{http://www.match.uni-heidelberg.de/GPF/Talks/poster-Hernan_Piragua.pdf}.

\bibitem{singular}
W.~Decker, G.-M. Greuel, G.~Pfister and H.~Sch\"onemann, {\it {\sc Singular}
  {3-1-6} --- {A} computer algebra system for polynomial computations\/},
  \url{http://www.singular.uni-kl.de}, 2012.

\bibitem{philipp:BA}
P.~Arras, {\it T.B.A. (work in progress)\/}, Bachelor thesis, University of
  Heidelberg, 2014.

\bibitem{Kuntzler:2014ila}
M.~Kuntzler and S.~Schafer-Nameki, {\it {Tate Trees for Elliptic Fibrations
  with Rank one Mordell-Weil group}\/}
  \href{http://www.arxiv.org/abs/1406.5174}{{\tt [1406.5174]}}.

\bibitem{Grassi:2013kha}
A.~Grassi, J.~Halverson and J.~L. Shaneson, {\it {Matter From Geometry Without
  Resolution}\/} \href{http://www.arxiv.org/abs/1306.1832}{{\tt [1306.1832]}}.

\bibitem{Grassi:2014sda}
A.~Grassi, J.~Halverson and J.~L. Shaneson, {\it {Non-Abelian Gauge Symmetry
  and the Higgs Mechanism in F-theory}\/}
  \href{http://www.arxiv.org/abs/1402.5962}{{\tt [1402.5962]}}.

\bibitem{Grassi:2011hq}
A.~Grassi and D.~R. Morrison, {\it {Anomalies and the Euler characteristic of
  elliptic Calabi-Yau threefolds}\/}, {\it Commun.Num.Theor.Phys.\/} {\bf 6}
  (2012) 51--127, \href{http://www.arxiv.org/abs/1109.0042}{{\tt [1109.0042]}}.

\bibitem{Ellwanger:2009dp}
U.~Ellwanger, C.~Hugonie and A.~M. Teixeira, {\it {The Next-to-Minimal
  Supersymmetric Standard Model}\/}, {\it Phys.Rept.\/} {\bf 496} (2010) 1--77,
  \href{http://www.arxiv.org/abs/0910.1785}{{\tt [0910.1785]}}.

\bibitem{Cvetic:2010dz}
M.~Cvetic, J.~Halverson and P.~Langacker, {\it {Singlet Extensions of the MSSM
  in the Quiver Landscape}\/}, {\it JHEP\/} {\bf 1009} (2010) 076,
  \href{http://www.arxiv.org/abs/1006.3341}{{\tt [1006.3341]}}.

\bibitem{Allanach:1999ic}
B.~Allanach, A.~Dedes and H.~K. Dreiner, {\it {Bounds on R-parity violating
  couplings at the weak scale and at the GUT scale}\/}, {\it Phys.Rev.\/} {\bf
  D60} (1999) 075014, \href{http://www.arxiv.org/abs/hep-ph/9906209}{{\tt
  [hep-ph/9906209]}}.

\bibitem{Nath:2006ut}
P.~Nath and P.~Fileviez~Perez, {\it {Proton stability in grand unified
  theories, in strings and in branes}\/}, {\it Phys.Rept.\/} {\bf 441} (2007)
  191--317, \href{http://www.arxiv.org/abs/hep-ph/0601023}{{\tt
  [hep-ph/0601023]}}.

\bibitem{Allanach:2003eb}
B.~Allanach, A.~Dedes and H.~Dreiner, {\it {R parity violating minimal
  supergravity model}\/}, {\it Phys.Rev.\/} {\bf D69} (2004) 115002,
  \href{http://www.arxiv.org/abs/hep-ph/0309196}{{\tt [hep-ph/0309196]}}.

\bibitem{Hebecker:2012qp}
A.~Hebecker, A.~K. Knochel and T.~Weigand, {\it {A Shift Symmetry in the Higgs
  Sector: Experimental Hints and Stringy Realizations}\/}, {\it JHEP\/} {\bf
  1206} (2012) 093, \href{http://www.arxiv.org/abs/1204.2551}{{\tt
  [1204.2551]}}.

\bibitem{Ibanez:2012zg}
L.~E. Ibanez, F.~Marchesano, D.~Regalado and I.~Valenzuela, {\it {The
  Intermediate Scale MSSM, the Higgs Mass and F-theory Unification}\/}, {\it
  JHEP\/} {\bf 1207} (2012) 195, \href{http://www.arxiv.org/abs/1206.2655}{{\tt
  [1206.2655]}}.

\bibitem{Hebecker:2013lha}
A.~Hebecker, A.~K. Knochel and T.~Weigand, {\it {The Higgs mass from a
  String-Theoretic Perspective}\/}, {\it Nucl.Phys.\/} {\bf B874} (2013) 1--35,
  \href{http://www.arxiv.org/abs/1304.2767}{{\tt [1304.2767]}}.

\bibitem{Ibanez:2013gf}
L.~E. Ibanez and I.~Valenzuela, {\it {The Higgs Mass as a Signature of Heavy
  SUSY}\/}, {\it JHEP\/} {\bf 1305} (2013) 064,
  \href{http://www.arxiv.org/abs/1301.5167}{{\tt [1301.5167]}}.

\bibitem{Hebecker:2014uaa}
A.~Hebecker and J.~Unwin, {\it {Precision Unification and Proton Decay in
  F-Theory GUTs with High Scale Supersymmetry}\/}
  \href{http://www.arxiv.org/abs/1405.2930}{{\tt [1405.2930]}}.

\bibitem{Chatzistavrakidis:2012bb}
A.~Chatzistavrakidis, E.~Erfani, H.~P. Nilles and I.~Zavala, {\it
  {Axiology}\/}, {\it JCAP\/} {\bf 1209} (2012) 006,
  \href{http://www.arxiv.org/abs/1207.1128}{{\tt [1207.1128]}}.

\bibitem{Hall:2013eko}
L.~J. Hall and Y.~Nomura, {\it {Grand Unification and Intermediate Scale
  Supersymmetry}\/}, {\it JHEP\/} {\bf 1402} (2014) 129,
  \href{http://www.arxiv.org/abs/1312.6695}{{\tt [1312.6695]}}.

\bibitem{Ibanez:2014zsa}
L.~E. Ibanez and I.~Valenzuela, {\it {BICEP2, the Higgs Mass and the
  SUSY-breaking Scale}\/} \href{http://www.arxiv.org/abs/1403.6081}{{\tt
  [1403.6081]}}.

\bibitem{Hall:2014vga}
L.~J. Hall, Y.~Nomura and S.~Shirai, {\it {Grand Unification, Axion, and
  Inflation in Intermediate Scale Supersymmetry}\/}
  \href{http://www.arxiv.org/abs/1403.8138}{{\tt [1403.8138]}}.

\bibitem{Heckman:2008qa}
J.~J. Heckman and C.~Vafa, {\it {Flavor Hierarchy From F-theory}\/}, {\it
  Nucl.Phys.\/} {\bf B837} (2010) 137--151,
  \href{http://www.arxiv.org/abs/0811.2417}{{\tt [0811.2417]}}.

\bibitem{Font:2009gq}
A.~Font and L.~Ibanez, {\it {Matter wave functions and Yukawa couplings in
  F-theory Grand Unification}\/}, {\it JHEP\/} {\bf 0909} (2009) 036,
  \href{http://www.arxiv.org/abs/0907.4895}{{\tt [0907.4895]}}.

\bibitem{Cecotti:2009zf}
S.~Cecotti, M.~C. Cheng, J.~J. Heckman and C.~Vafa, {\it {Yukawa Couplings in
  F-theory and Non-Commutative Geometry}\/}
  \href{http://www.arxiv.org/abs/0910.0477}{{\tt [0910.0477]}}.

\bibitem{Conlon:2009qq}
J.~P. Conlon and E.~Palti, {\it {Aspects of Flavour and Supersymmetry in
  F-theory GUTs}\/}, {\it JHEP\/} {\bf 1001} (2010) 029,
  \href{http://www.arxiv.org/abs/0910.2413}{{\tt [0910.2413]}}.

\bibitem{Marchesano:2009rz}
F.~Marchesano and L.~Martucci, {\it {Non-perturbative effects on seven-brane
  Yukawa couplings}\/}, {\it Phys.Rev.Lett.\/} {\bf 104} (2010) 231601,
  \href{http://www.arxiv.org/abs/0910.5496}{{\tt [0910.5496]}}.

\bibitem{Aparicio:2011jx}
L.~Aparicio, A.~Font, L.~E. Ibanez and F.~Marchesano, {\it {Flux and Instanton
  Effects in Local F-theory Models and Hierarchical Fermion Masses}\/}, {\it
  JHEP\/} {\bf 1108} (2011) 152, \href{http://www.arxiv.org/abs/1104.2609}{{\tt
  [1104.2609]}}.

\bibitem{Font:2012wq}
A.~Font, L.~E. Ibanez, F.~Marchesano and D.~Regalado, {\it {Non-perturbative
  effects and Yukawa hierarchies in F-theory SU(5) Unification}\/}, {\it
  JHEP\/} {\bf 1303} (2013) 140, \href{http://www.arxiv.org/abs/1211.6529}{{\tt
  [1211.6529]}}.

\bibitem{Font:2013ida}
A.~Font, F.~Marchesano, D.~Regalado and G.~Zoccarato, {\it {Up-type quark
  masses in SU(5) F-theory models}\/}, {\it JHEP\/} {\bf 1311} (2013) 125,
  \href{http://www.arxiv.org/abs/1307.8089}{{\tt [1307.8089]}}.

\bibitem{Marsano:2008py}
J.~Marsano, N.~Saulina and S.~Schafer-Nameki, {\it {An Instanton Toolbox for
  F-Theory Model Building}\/}, {\it JHEP\/} {\bf 1001} (2010) 128,
  \href{http://www.arxiv.org/abs/0808.2450}{{\tt [0808.2450]}}.

\bibitem{Blumenhagen:2010ja}
R.~Blumenhagen, A.~Collinucci and B.~Jurke, {\it {On Instanton Effects in
  F-theory}\/}, {\it JHEP\/} {\bf 1008} (2010) 079,
  \href{http://www.arxiv.org/abs/1002.1894}{{\tt [1002.1894]}}.

\bibitem{Donagi:2010pd}
R.~Donagi and M.~Wijnholt, {\it {MSW Instantons}\/}, {\it JHEP\/} {\bf 1306}
  (2013) 050, \href{http://www.arxiv.org/abs/1005.5391}{{\tt [1005.5391]}}.

\bibitem{Cvetic:2011gp}
M.~Cvetic, I.~Garcia~Etxebarria and J.~Halverson, {\it {Three Looks at
  Instantons in F-theory -- New Insights from Anomaly Inflow, String Junctions
  and Heterotic Duality}\/}, {\it JHEP\/} {\bf 1111} (2011) 101,
  \href{http://www.arxiv.org/abs/1107.2388}{{\tt [1107.2388]}}.

\bibitem{Grimm:2011dj}
T.~W. Grimm, M.~Kerstan, E.~Palti and T.~Weigand, {\it {On Fluxed Instantons
  and Moduli Stabilisation in IIB Orientifolds and F-theory}\/}, {\it
  Phys.Rev.\/} {\bf D84} (2011) 066001,
  \href{http://www.arxiv.org/abs/1105.3193}{{\tt [1105.3193]}}.

\bibitem{Kerstan:2012cy}
M.~Kerstan and T.~Weigand, {\it {Fluxed M5-instantons in F-theory}\/}, {\it
  Nucl.Phys.\/} {\bf B864} (2012) 597--639,
  \href{http://www.arxiv.org/abs/1205.4720}{{\tt [1205.4720]}}.

\bibitem{Cvetic:2009ah}
M.~Cvetic, I.~Garcia-Etxebarria and R.~Richter, {\it {Branes and instantons at
  angles and the F-theory lift of O(1) instantons}\/}, {\it AIP Conf.Proc.\/}
  {\bf 1200} (2010) 246--260, \href{http://www.arxiv.org/abs/0911.0012}{{\tt
  [0911.0012]}}.

\bibitem{Cvetic:2010rq}
M.~Cvetic, I.~Garcia-Etxebarria and J.~Halverson, {\it {Global F-theory Models:
  Instantons and Gauge Dynamics}\/}, {\it JHEP\/} {\bf 1101} (2011) 073,
  \href{http://www.arxiv.org/abs/1003.5337}{{\tt [1003.5337]}}.

\bibitem{Bianchi:2011qh}
M.~Bianchi, A.~Collinucci and L.~Martucci, {\it {Magnetized E3-brane instantons
  in F-theory}\/}, {\it JHEP\/} {\bf 1112} (2011) 045,
  \href{http://www.arxiv.org/abs/1107.3732}{{\tt [1107.3732]}}.

\bibitem{Bianchi:2012kt}
M.~Bianchi, G.~Inverso and L.~Martucci, {\it {Brane instantons and fluxes in
  F-theory}\/}, {\it JHEP\/} {\bf 1307} (2013) 037,
  \href{http://www.arxiv.org/abs/1212.0024}{{\tt [1212.0024]}}.

\bibitem{Cvetic:2012ts}
M.~Cvetic, R.~Donagi, J.~Halverson and J.~Marsano, {\it {On Seven-Brane
  Dependent Instanton Prefactors in F-theory}\/}, {\it JHEP\/} {\bf 1211}
  (2012) 004, \href{http://www.arxiv.org/abs/1209.4906}{{\tt [1209.4906]}}.

\bibitem{Martucci:2014ema}
L.~Martucci, {\it {Topological duality twist and brane instantons in
  F-theory}\/} \href{http://www.arxiv.org/abs/1403.2530}{{\tt [1403.2530]}}.

\bibitem{Bies:2014sra}
M.~Bies, C.~Mayrhofer, C.~Pehle and T.~Weigand, {\it {Chow groups, Deligne
  cohomology and massless matter in F-theory}\/}
  \href{http://www.arxiv.org/abs/1402.5144}{{\tt [1402.5144]}}.

\bibitem{Kiritsis:2009sf}
E.~Kiritsis, M.~Lennek and B.~Schellekens, {\it {SU(5) orientifolds, Yukawa
  couplings, Stringy Instantons and Proton Decay}\/}, {\it Nucl.Phys.\/} {\bf
  B829} (2010) 298--324, \href{http://www.arxiv.org/abs/0909.0271}{{\tt
  [0909.0271]}}.

\bibitem{Cvetic:2008hi}
M.~Cvetic and P.~Langacker, {\it {D-Instanton Generated Dirac Neutrino
  Masses}\/}, {\it Phys.Rev.\/} {\bf D78} (2008) 066012,
  \href{http://www.arxiv.org/abs/0803.2876}{{\tt [0803.2876]}}.

\bibitem{oai:arXiv.org:hep-th/9908088}
K.~Dasgupta, G.~Rajesh and S.~Sethi, {\it {M theory, orientifolds and G -
  flux}\/}, {\it JHEP\/} {\bf 9908} (1999) 023,
  \href{http://www.arxiv.org/abs/hep-th/9908088}{{\tt [hep-th/9908088]}}.

\bibitem{oai:arXiv.org:hep-th/9609122}
E.~Witten, {\it {On flux quantization in M theory and the effective action}\/},
  {\it J.Geom.Phys.\/} {\bf 22} (1997) 1--13,
  \href{http://www.arxiv.org/abs/hep-th/9609122}{{\tt [hep-th/9609122]}}.

\bibitem{oai:arXiv.org:1011.6388}
A.~Collinucci and R.~Savelli, {\it {On Flux Quantization in F-Theory}\/}, {\it
  JHEP\/} {\bf 1202} (2012) 015, \href{http://www.arxiv.org/abs/1011.6388}{{\tt
  [1011.6388]}}.

\bibitem{oai:arXiv.org:1203.4542}
A.~Collinucci and R.~Savelli, {\it {On Flux Quantization in F-Theory II:
  Unitary and Symplectic Gauge Groups}\/}, {\it JHEP\/} {\bf 1208} (2012) 094,
  \href{http://www.arxiv.org/abs/1203.4542}{{\tt [1203.4542]}}.

\bibitem{Grimm:2010ks}
T.~W. Grimm, {\it {The N=1 effective action of F-theory compactifications}\/},
  {\it Nucl.Phys.\/} {\bf B845} (2011) 48--92,
  \href{http://www.arxiv.org/abs/1008.4133}{{\tt [1008.4133]}}.

\bibitem{oai:arXiv.org:0802.2969}
R.~Donagi and M.~Wijnholt, {\it {Model Building with F-Theory}\/}, {\it
  Adv.Theor.Math.Phys.\/} {\bf 15} (2011) 1237--1318,
  \href{http://www.arxiv.org/abs/0802.2969}{{\tt [0802.2969]}}.

\bibitem{oai:arXiv.org:0904.1218}
R.~Donagi and M.~Wijnholt, {\it {Higgs Bundles and UV Completion in
  F-Theory}\/} \href{http://www.arxiv.org/abs/0904.1218}{{\tt [0904.1218]}}.

\bibitem{oai:arXiv.org:1108.1794}
J.~Marsano and S.~Schafer-Nameki, {\it {Yukawas, G-flux, and Spectral Covers
  from Resolved Calabi-Yau's}\/}, {\it JHEP\/} {\bf 1111} (2011) 098,
  \href{http://www.arxiv.org/abs/1108.1794}{{\tt [1108.1794]}}.

\bibitem{oai:arXiv.org:1111.1232}
T.~W. Grimm and H.~Hayashi, {\it {F-theory fluxes, Chirality and Chern-Simons
  theories}\/}, {\it JHEP\/} {\bf 1203} (2012) 027,
  \href{http://www.arxiv.org/abs/1111.1232}{{\tt [1111.1232]}}.

\bibitem{oai:arXiv.org:1203.6662}
K.~Intriligator, H.~Jockers, P.~Mayr, D.~R. Morrison and M.~R. Plesser, {\it
  {Conifold Transitions in M-theory on Calabi-Yau Fourfolds with Background
  Fluxes}\/} \href{http://www.arxiv.org/abs/1203.6662}{{\tt [1203.6662]}}.

\bibitem{Greene:1993vm}
B.~R. Greene, D.~R. Morrison and M.~Plesser, {\it {Mirror manifolds in higher
  dimension}\/}, {\it Commun.Math.Phys.\/} {\bf 173} (1995) 559--598,
  \href{http://www.arxiv.org/abs/hep-th/9402119}{{\tt [hep-th/9402119]}}.

\bibitem{Blumenhagen:2009yv}
R.~Blumenhagen, T.~W. Grimm, B.~Jurke and T.~Weigand, {\it {Global F-theory
  GUTs}\/}, {\it Nucl.Phys.\/} {\bf B829} (2010) 325--369,
  \href{http://www.arxiv.org/abs/0908.1784}{{\tt [0908.1784]}}.

\bibitem{Grimm:2009yu}
T.~W. Grimm, S.~Krause and T.~Weigand, {\it {F-Theory GUT Vacua on Compact
  Calabi-Yau Fourfolds}\/}, {\it JHEP\/} {\bf 1007} (2010) 037,
  \href{http://www.arxiv.org/abs/0912.3524}{{\tt [0912.3524]}}.

\bibitem{Chen:2010ts}
C.-M. Chen, J.~Knapp, M.~Kreuzer and C.~Mayrhofer, {\it {Global SO(10) F-theory
  GUTs}\/}, {\it JHEP\/} {\bf 1010} (2010) 057,
  \href{http://www.arxiv.org/abs/1005.5735}{{\tt [1005.5735]}}.

\bibitem{Knapp:2011wk}
J.~Knapp, M.~Kreuzer, C.~Mayrhofer and N.-O. Walliser, {\it {Toric Construction
  of Global F-Theory GUTs}\/}, {\it JHEP\/} {\bf 1103} (2011) 138,
  \href{http://www.arxiv.org/abs/1101.4908}{{\tt [1101.4908]}}.

\bibitem{Blumenhagen:2008aw}
R.~Blumenhagen, {\it {Gauge Coupling Unification in F-Theory Grand Unified
  Theories}\/}, {\it Phys. Rev. Lett.\/} {\bf 102} (2009) 071601,
  \href{http://www.arxiv.org/abs/0812.0248}{{\tt [0812.0248]}}.

\bibitem{Conlon:2009qa}
J.~P. Conlon and E.~Palti, {\it {On Gauge Threshold Corrections for Local
  IIB/F-theory GUTs}\/}, {\it Phys.Rev.\/} {\bf D80} (2009) 106004,
  \href{http://www.arxiv.org/abs/0907.1362}{{\tt [0907.1362]}}.

\bibitem{Mayrhofer:2013ara}
C.~Mayrhofer, E.~Palti and T.~Weigand, {\it {Hypercharge Flux in IIB and
  F-theory: Anomalies and Gauge Coupling Unification}\/}
  \href{http://www.arxiv.org/abs/1303.3589}{{\tt [1303.3589]}}.

\bibitem{Cvetic:2011iq}
M.~Cvetic, J.~Halverson and P.~Langacker, {\it {Implications of String
  Constraints for Exotic Matter and Z' s Beyond the Standard Model}\/}, {\it
  JHEP\/} {\bf 1111} (2011) 058, \href{http://www.arxiv.org/abs/1108.5187}{{\tt
  [1108.5187]}}.

\bibitem{Halverson:2013ska}
J.~Halverson, {\it {Anomaly Nucleation Constrains SU(2) Gauge Theories}\/},
  {\it Phys.Rev.Lett.\/} {\bf 111} (2013) 261601,
  \href{http://www.arxiv.org/abs/1310.1091}{{\tt [1310.1091]}}.

\bibitem{Donagi:2011jy}
R.~Donagi and M.~Wijnholt, {\it {Gluing Branes, I}\/}, {\it JHEP\/} {\bf 1305}
  (2013) 068, \href{http://www.arxiv.org/abs/1104.2610}{{\tt [1104.2610]}}.

\bibitem{Donagi:2011dv}
R.~Donagi and M.~Wijnholt, {\it {Gluing Branes II: Flavour Physics and String
  Duality}\/}, {\it JHEP\/} {\bf 1305} (2013) 092,
  \href{http://www.arxiv.org/abs/1112.4854}{{\tt [1112.4854]}}.

\end{thebibliography}
